\documentclass[a4paper,fleqn]{cas-dc}

\usepackage[numbers,sort&compress]{natbib}
\usepackage{subfigure}
\usepackage{tikz}
\usepackage{adjustbox}
\usepackage{lipsum}
\usepackage{algorithm2e}
\usepackage{threeparttable, tablefootnote}
\usepackage{subfigure}
\usepackage{wrapfig}
\usepackage{adjustbox}
\usepackage{float}

\def\tsc#1{\csdef{#1}{\textsc{\lowercase{#1}}\xspace}}
\tsc{WGM}
\tsc{QE}
\tsc{EP}
\tsc{PMS}
\tsc{BEC}
\tsc{DE}

\begin{document}
\let\WriteBookmarks\relax
\def\floatpagepagefraction{1}
\def\textpagefraction{.001}
\shorttitle{ }
\shortauthors{Islas et~al.}

\title [mode = title]{Computational Assessment of Biomass Dust Explosions in the 20L Sphere}                      



\author[1]{{Alain Islas}}[]


\address[1]{Department of Energy, University of Oviedo - 33203 Gijón, Asturias, Spain}

\author[1]{{Andrés Rodríguez Fernández}}[]

\author[2]{{Covadonga Betegón}}[%
   ]


\address[2]{Department of Construction and Manufacturing Engineering, University of Oviedo - 33203 Gijón, Asturias, Spain}

\author[3]{{Emilio Martínez-Pañeda}}[]

\address[3]{Department of Civil and Environmental Engineering, Imperial College London - London, SW7 2AZ, United Kingdom}

\author[1]{{Adrián Pandal}}[orcid=0000-0001-6006-2199]
\cormark[1]

\cortext[cor1]{Corresponding author:}

\begin{abstract}
Determination of the explosion severity parameters of biomass is crucial for the safety management and dust explosion risk assessment of biomass-processing industries. These are commonly determined following experimental tests in the 20L sphere according to the international standards. Recently, CFD simulations have emerged as a reliable alternative to predict the explosion behavior with good accuracy and reduced labor and capital. In this work, numerical simulations of biomass dust explosions are conducted with the open-source CFD code OpenFOAM. The multi-phase (gas-solid) flow is treated in an Eulerian-Lagrangian framework, using a two-way coupling regime and considering the reactions of biomass conversion (moisture evaporation, devolatilization, and char oxidation), the combustion of volatile gases, and convective and radiative heat transfer. The model is validated with pressure-time and concentration-dependent experimental measurements of two biomass samples. 
Results suggest that the characteristics of the cold-flow (\i.e. turbulence levels, actual dust concentration, spatial distribution of the dust cloud, and turbophoresis effect) govern the course of the explosion process, and depend strongly on particle size, dust concentration, and ignition delay time effects. These findings may be relevant in the design of better dust explosion testing devices and to the reexamination of the guidelines for the operation of the experiment. Finally, a thorough discussion on the explosion pressures, degree of biomass conversion, flame temperature, flame propagation patterns, and the dust agglomeration effect is presented.

\end{abstract}

\begin{keywords}
Dust explosions \sep Biomass \sep CFD \sep  OpenFOAM
\end{keywords}

\maketitle

\section{Introduction}
\label{Section:Introduction}

Dust explosions are an ever-present threat wherever bulk powders are handled in the process industries \cite{AMYOTTE201015}. Since the first reported accident in 1785 \cite{eckhoff2003dust}, dust explosions have become a serious concern due to their inherent destructive power and high occurrence. Yuan et al. \cite{YUAN201557} reported that more than 2000 dust explosion disasters occurred worldwide between 1785 and 2012, being China and the USA the countries with most incidence. More recently, in 2020 a total of 60 dust explosions were reported worldwide \cite{DSS_2020}, from which 72\% were caused by wood and food products. Among these combustible dust incidents (fires and explosions), storage silos demonstrated the highest percentage of recurrence. Consequently, organic dust explosions should be considered as a serious hazard in the process industries (e.g. biomass or agricultural), being operational and dynamic risk assessments required to better comprehend the probability of occurrence of dust explosions and its potential severity \cite{KHAN2015116,ZHOU2019144,LIN201948, AMYOTTE2014292}.\\

A first step in the dust explosion risk assessment is to (1) identify dust hazards and determine the likelihood of explosion of dust clouds. Following the ASTM E1226 \cite{ASTME1226}, ISO 6184 \cite{iso1985explosion} or EN 14034 \cite{EN14034} standards, closed vessel testing is used to determine whether or not a dust cloud in suspension is capable of initiating and sustaining an explosion in the presence of an ignition source. The explosion parameters of interest are: (a) the maximum explosion pressure $P_{max}$, (b) the deflagration index $K_{st}$, (c) the limiting oxygen concentration (LOC), and (d) the minimum explosive concentration (MEC). The other common steps in a dust explosion risk assessment include: (2) evaluation of the dust explosion hazards (e.g., flash fires, secondary explosions), (3) identification of the risks (e.g., injuries to personnel, fatalities, plant damages), (4) dust hazard management (i.e., risk reduction and explosion prevention \& protection measures) and (5) data collection, documentation and training \& competence development. The explosion parameters are particularly useful to classify the dust hazards according to explosion risk levels \cite{ogle2019dust} and serve as the base input for designing explosion protection system: venting panels \cite{ADDAI201572,LI2017489,HOLBROW2013183}, isolation valves \cite{TAVEAU2017348,ajrash2017experimental}, and suppression systems \cite{amyotte2009application,jiang2020suppression}. \\

Formerly, when compared to coal or metal dust \cite{YANG202183,YANG202172}, reporting explosivity test results of biomass samples was of secondary interest. First data was published in the late 1990's \cite{WILEN,GARCIATORRENT19981093} and early 2000's \cite{CALLE2005144}. However, with the advent of biomass as a CO\textsubscript{2} neutral, renewable energy source for power generation, and because wood represents one of the largest biomass energy resources today \cite{clark2017sustainable}, the demand for conducting dust explosion risk assessments has increased considerably. What is more, with the progress of technology and growth of large-scale storage equipment, safe dimensioning of mitigating measures requires adequate knowledge about the burning rate of dust clouds in actual process situations \cite{SKJOLD2005151}. For this reason, in the last decade the number of experimental studies on dust explosion testing raised significantly and focused mainly on describing the effects of dust concentration \cite{Lee2016}, calorific values \cite{HUESCARMEDINA2015116,HUESCARMEDINA201591,HUESCARMEDINA2015287}, burnt mass \cite{SLATTER2015318}, particle size \cite{GUO2019}, and volatile matter content \cite{liu2019explosion, LIU2021384, JIANG201845} on the explosion severity parameters.\\

Along with experimental research, the increasing computational capabilities have demonstrated that numerical models can be an effective tool to predict the hazardous explosion potential of dust clouds \cite{SKJOLD2005151,LI2018360}. 
These vary from simple mathematical models \cite{FUMAGALLI201767,Copelli2019329,SCOTTON2020104218,portarapillo2021ignit} to more complex CFD simulations \cite{RANI201514,ABUSWER201649,CLONEY2018215}. CFD methods are especially well suited for understanding
deflagration development and propagation inside equipment or through complex structures \cite{ogle2019dust}. To calibrate these models, a frequent practice is to first perform CFD simulations of dust dispersion and explosion experiments in the standardized apparatus: the Hartmann tube \cite{MURILLO2013103,CHAUDHARI2019192}, the 20L Siwek sphere \cite{di2014cfd,CAO2014456,li2020influence}, or the 1m\textsuperscript{3} ISO vessel \cite{portarapillo2020cfd_1m3vessel}. These models can reduce the time consuming labor and expensive costs of experimental testing. Furthermore, numerical simulations can unveil a broader understanding of the flow phenomena that are not accessible through experiments.  \\

Due to its reduced size and quicker testing times, the 20L sphere is often preferred over the 1m\textsuperscript{3} vessel, being the latter mostly used when spurious data appears in or a double check of the results is necessary. Notably, CFD studies on biomass dust explosions in the 20L sphere are still scarce, with only a couple of works published in the literature \cite{LI2020cornstarch,PICO2020analysis,PICO2020cfddpm}. Yet better methods for predicting real dust cloud generation, ignition, devolatilization, combustion, and heat transfer processes are needed. The present paper accounts for a subsequent step to our first work \cite{ISLAS2022117033} with the long-term objective of constructing an accurate engineering tool for the simulation of large-scale dust explosions in specific industrial geometries. Therefore, dust explosions are simulated in the standard 20L sphere equipped with the rebound nozzle and proceeding according to the ASTM E1226 standard. Specifically, two different biomass dust samples are evaluated, accounting for significantly different particle size distributions. The model features detailed calculations of the radiative properties of the gas-solid mixture and devolatilization kinetics, and it is constructed in the open-source CFD code OpenFOAM 8. The CFD model is initially validated with pressure-time evolution measurements and then, the performance to capture the maximum explosion pressures among different dust concentrations is evaluated. Finally, the CFD model is used to assess the role of dust concentration and ignition delay time on the maximum explosion pressures, aiming to promote the knowledge of the key aspects of dust explosions and the development of the CFD tools towards this end. \\ 

\section{Test samples}
\label{Section:Test_samples}

Two woody fuel samples are considered in this study, namely biomass 1 (\textit{Pellets Asturias}) and biomass 2 (\textit{Cupressus Funebris}. The former is a Spanish biomass sample from a pellet manufacturer in the autonomous community of Asturias and is comprised of natural wood sub-products of the 1\textsuperscript{st} wood processing industry (saw dust, wood chips and debarked wood). The explosion parameters, including the pressure time evolution, chemical composition, and particle size distribution (PSD) of this sample were provided by our industrial third-party PHB Weserhütte S.A. 
Contrarily, biomass 2 is a Chinese biomass sample, whose explosion parameters, chemical composition and PSD were taken exclusively from the literature \cite{liu2019explosion,shen2014emission}. The purpose of the two samples is to: (1) perform a pressure-time validation with the explosion curve of biomass 1, and (2) use biomass 2 to evaluate the performance of the model predicting the explosion pressures when the dust concentration is varied. The corresponding ultimate and proximate analyses of both samples are presented in Table \ref{Table1:Proximate_Ultimate_analysis}.\\

\subsection{Biomass composition}
\label{Subsection:Biomass_composition}

The chemical equilibrium method adopted in this study is based on the representation of biomass as a postulate substance, e.g. C\textsubscript{x}H\textsubscript{y}O\textsubscript{z}N\textsubscript{p}, whose subscripts can be determined from the ultimate and proximate analyses. However, as the nitrogen content is negligible, it is convenient to represent the biomass molecule as C\textsubscript{x}H\textsubscript{y}O\textsubscript{z} only.\\

\begin{table}[H]
\begin{threeparttable}
\caption{Ultimate and proximate analyses of the biomass samples.}
\label{Table1:Proximate_Ultimate_analysis}
\begin{tabular}{@{}lll@{}}
\toprule
Label                             & Biomass 1        & Biomass 2          \\ \midrule
Sample                            & \begin{tabular}[c]{@{}l@{}}Pellets\\ Asturias\tnote{a}\end{tabular} & \begin{tabular}[c]{@{}l@{}}Cupressus\\ Funebris\tnote{b}\end{tabular} \\
                                  &                  &                    \\
\textit{Proximate analysis (wt. \% ar)} &                  &                    \\
Fixed carbon                      & 14.16            & 19.14              \\
Volatile matter                   & 77.04            & 66.86              \\
Moisture                          & 8.33             & 12.71              \\
Ash                               & 0.47             & 1.29               \\
                                  &                  &                    \\
\textit{Ultimate analysis (wt. \% daf)}&                  &                    \\
C                                 & 50.25            & 50.13              \\
H                                 & 6.02             & 6.02               \\
O                                 & 43.45            & 43.49              \\
N                                 & 0.28             & 0.36               \\
Lower calorific value (MJ/kg)     & 18.83            & 18.80\tnote{c}  \\ \bottomrule
\end{tabular}
\begin{tablenotes}\footnotesize
\item [a] Composition measured by a third-party lab.
\item [b] Composition reported by Shen \cite{shen2014emission}.
\item [c] Estimated via empirical correlations.
\end{tablenotes}
\end{threeparttable}
\end{table}

The composition of the volatile gases is determined from mass and energy balances. Based on the principle that "\textit{the total heat produced by a compound is little different from the sum of the heats which would be produced by a separate combustion of its elements}" \cite{given1986calculation}, the lower calorific value (LCV) of biomass can be split into the LCV of volatile matter (VM) and fixed carbon (FC) as:

\begin{equation}
    \text{LCV}_\text{biomass} = Y_\text{VM}^{\text{daf}}\times\text{LCV}_\text{VM} + Y_\text{FC}^{\text{daf}}\times\text{LCV}_\text{FC}
    \label{Eqn:LCV_biomass}
\end{equation}

with Y\textsubscript{VM}\textsuperscript{daf}+Y\textsubscript{FC}\textsuperscript{daf} = 1. LCV\textsubscript{biomass} can be measured directly following the EN 14918 or ISO 18125 standards. Alternatively, it can be estimated using empirical correlations based on the ultimate analysis \cite{sheng2005estimating,garcia2014spanish}.\\

With the above considerations, the thermal breakdown of the postulate substance into gaseous species is modeled as \cite{li2021detailed}:

\begin{align}
    \text{C}\textsubscript{x}\text{H}\textsubscript{y}\text{O}\textsubscript{z}& \xrightarrow[]{k_{v}} \nu_{1}^{\text{"}} \text{CO} + \nu_{2}^{\text{"}} \text{CO\textsubscript{2}} + \nu_{3}^{\text{"}} \text{CH\textsubscript{4}} + \nu_{4}^{\text{"}} \text{H\textsubscript{2}} 
    \label{Eqn:Devolatilization_reaction} \\
    \text{LCV}\textsubscript{VM} &= \sum_{i=1}^{4}Y_{i}\times\Delta H_{R,i}
\end{align}

 where LCV\textsubscript{VM} is found from Eq. (\ref{Eqn:LCV_biomass}) and $\Delta H_{R,i}$ is the enthalpy of combustion of the corresponding volatile component \cite{ansys2011ansys}. Table (\ref{Table2:Volatile_composition}) presents the calculated mass fractions of the volatile species in each sample. \\

\begin{table}[H]
\caption{Calculated volatile gas composition of the biomass samples}
\label{Table2:Volatile_composition}
\begin{tabular}{@{}llll@{}}
\toprule
Label   &   & Biomass 1 & Biomass 2 \\ \midrule
\begin{tabular}[l]{@{}l@{}}Chemical \\ molecule\end{tabular}                                &                         & C\textsubscript{1.03}H\textsubscript{2.13}O\textsubscript{0.97}                                     & C\textsubscript{0.90}H\textsubscript{2.31}O\textsubscript{1.05}                                     \\
\multirow{4}{*}{\begin{tabular}[l]{@{}l@{}}Volatile\\ composition\\ (wt. \%)\end{tabular}} & \multicolumn{1}{l}{CO}  & 0.066                                               & 0                                                   \\
                                                                                            & \multicolumn{1}{l}{CO\textsubscript{2}} & 0.657                                               & 0.778                                               \\
                                                                                            & \multicolumn{1}{l}{CH\textsubscript{4}} & 0.274                                               & 0.194                                               \\
                                                                                            & \multicolumn{1}{l}{H\textsubscript{2}}  & 0.003                                               & 0.028                                               \\
\begin{tabular}[l]{@{}l@{}}LCV\textsubscript{VM}\\ (MJ/kg)\end{tabular}                                   &                         & 16.24                                               & 14.76                                               \\ \bottomrule
\end{tabular}
\end{table}

\subsection{Particle size distribution}
\label{Subsection:Particle_size_distribution}

When testing combustible dusts in the 20L sphere, the standard test procedures provide recommendations on the particle fineness of the dust sample. As per the EN 14034 code, the particle diameter should not exceed 500 $\mu$m \cite{EN14034}. The ASTM E1226 standard is more strict, as the particle diameter should be limited to 95\% minus 200 mesh (75 $\mu$m) \cite{ASTME1226}. However, very often particle size distributions under such conditions do not represent a sample that can be collected from a typical industrial process. Sometimes it is desirable to run tests on an as-received sample.\\

\begin{figure}[h]
    \centering
    \includegraphics[width=0.43\textwidth]{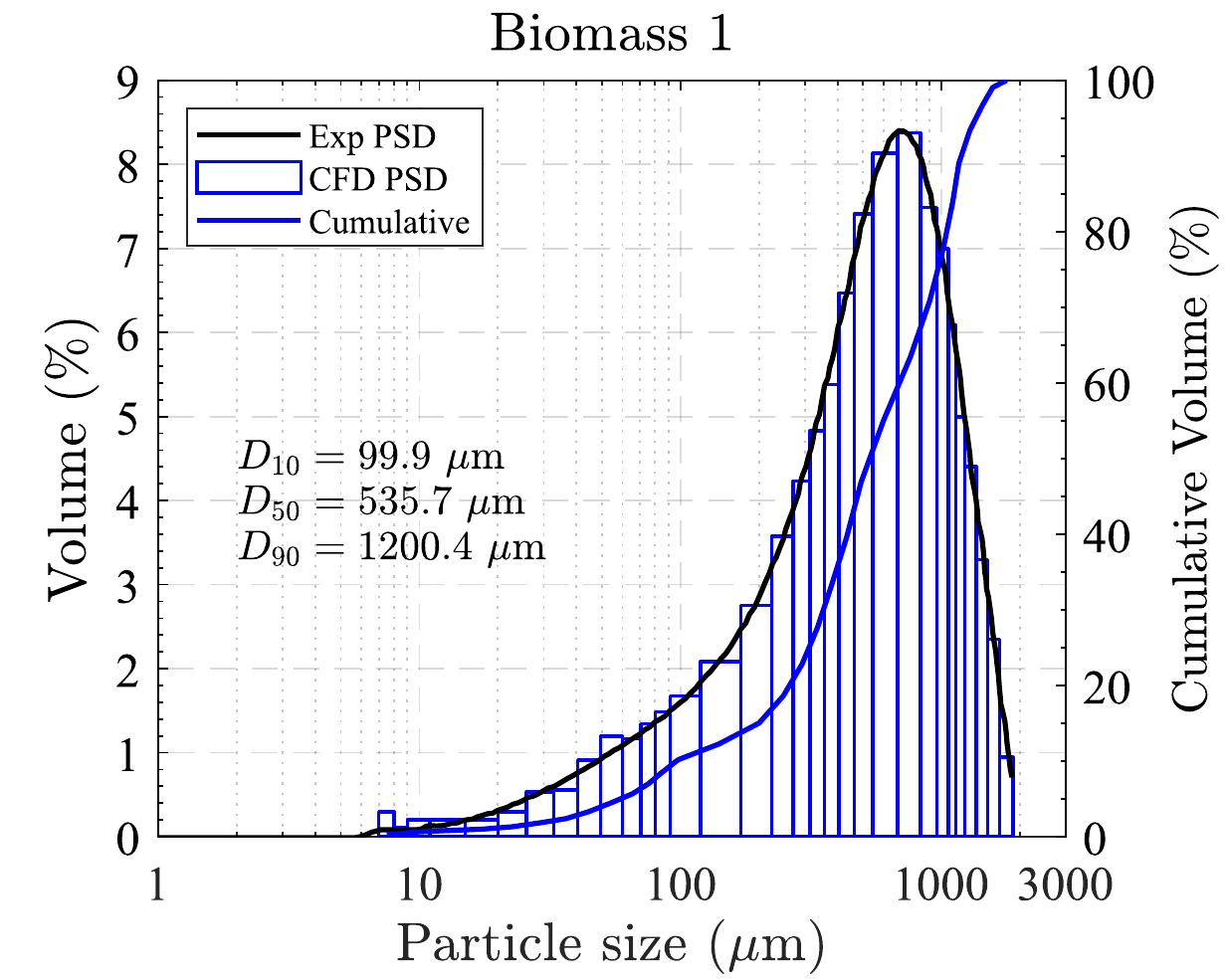}    
    \centering \caption{Particle size distribution of \textit{Pellets Asturias}.}
    \label{Figure:PSD_Pellets_Asturias}
\end{figure}

\begin{figure}[h]
    \centering
    \includegraphics[width=0.43\textwidth]{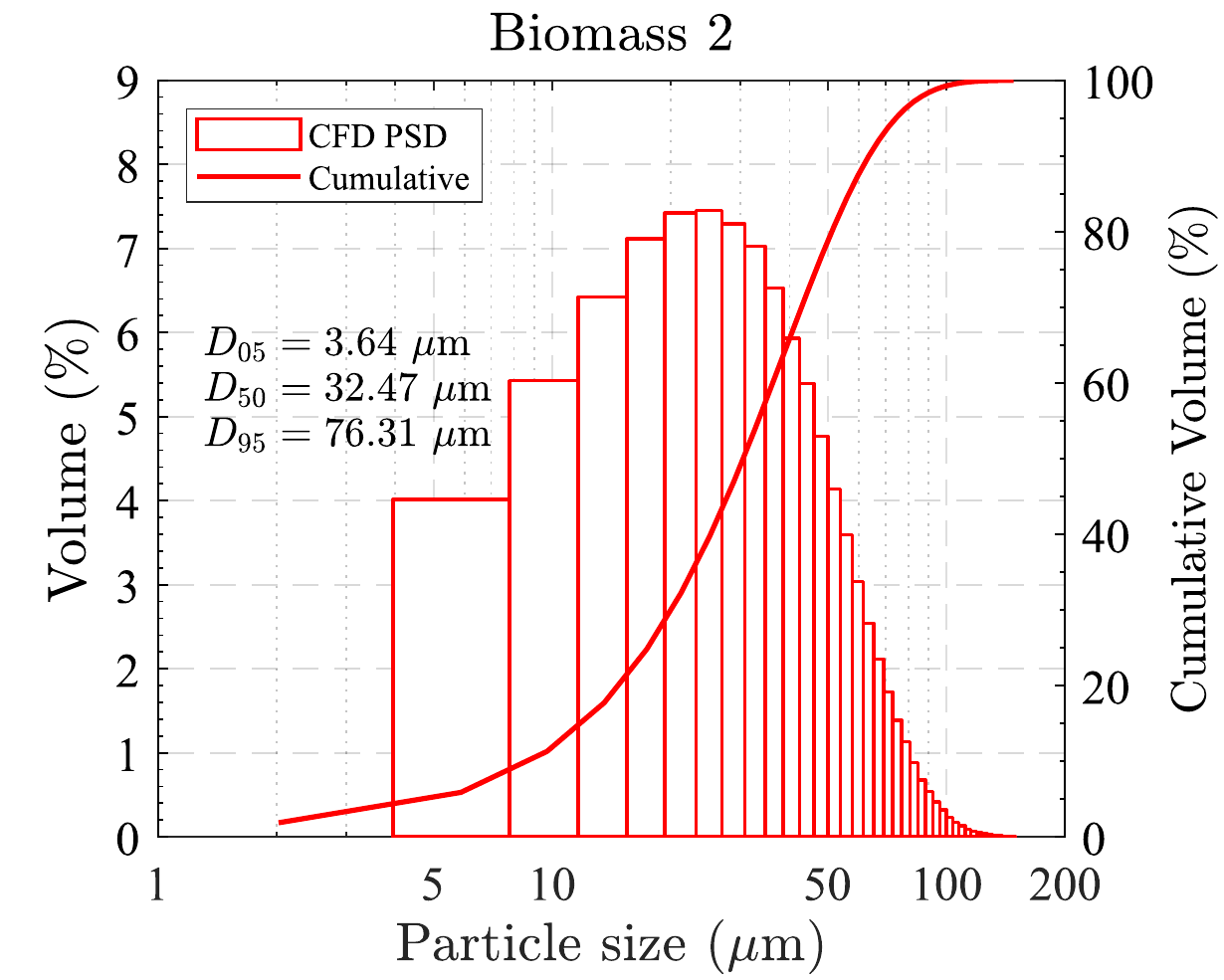}    
    \centering \caption{Particle size distribution of \textit{Cupressus Funebris}.}
    \label{Figure:PSD_Cupressus_Funebris}
\end{figure}

This is the case of biomass 1, a sample that was collected from a pellet storage silo at a power plant and whose PSD is the result of wearing during the conveying and filling operations of the pellets into the silo. The as-received PSD was measured by a third-party lab via laser diffraction (LD). Such size distribution was digitized and given to the CFD code as a {\fontfamily{qcr}\selectfont generalDistribution} which allows one to define an arbitrary probability density function, see Fig. \ref{Figure:PSD_Pellets_Asturias}. In contrast,
since no other data except than the $D_{05}$, $D_{50}$, and $D_{95}$ values were reported by Liu et al. \cite{liu2019explosion}, the PSD of biomass 2 was assumed to follow a Rosin-Rammler distribution whose parameters were calculated by fitting above values to the corresponding probability density function, see Fig. \ref{Figure:PSD_Cupressus_Funebris}.\\

Note that in case of biomass 2, the PSD is about one order of magnitude smaller than biomass 1. This is because, as explicitly mentioned by Liu et al. \cite{liu2019explosion}, the \textit{Cupressus Funebris} sample was broken into smaller pieces and sieved with a mesh size of 50 $\mu$m in order to satisfy the recommendations of the standards.

\section{Physical considerations and modeling}
\label{Section:Physical_considerations_and_modeling}

The numerical simulations are conducted with the {\fontfamily{qcr}\selectfont coalChemistryFoam} solver following a Eulerian-Lagrangian framework in a two-way coupling regime.\\

\subsection{Gas-phase governing equations}
\label{Subsection:Gas-phase_governing_equations}

The reactive flow is described by the compressible form of the Reynolds averaged Navier-Stokes equations (RANS), Eqs. (\ref{Equation:mass_transport}-\ref{Equation:energy_transport}) with source terms $\Gamma_{i}$, $\Lambda_{i}$, $\Theta_{i}$ accounting for the mass, momentum and energy transfer from the dispersed to the gas phase, respectively. 

\begin{equation}
    \frac{\partial \bar{\rho}}{\partial t}+\frac{\partial}{\partial x_i}\left(\bar{\rho} \tilde{u}_{i}\right)=\Gamma_{i}
    \label{Equation:mass_transport}
\end{equation}

\begin{multline}
    \frac{\partial}{\partial t}\left(\bar{\rho} \tilde{u}_{i}\right) + \frac{\partial}{\partial x_j}\left(\bar{\rho} \tilde{u}_{i} \tilde{u}_{j}\right) = - \frac{\partial \bar{p}}{\partial x_j} + \frac{\partial \bar{\tau}^{ij}}{\partial x_j}+\frac{\partial}{\partial x_j}\left(-\bar{\rho} \widetilde{u_i^{'} u_j^{'}}\right)\\
    +\bar{\rho} g_i + \Lambda_{i}
    \label{Equation:momentum_transport}
\end{multline}

\begin{equation}
    \frac{\partial}{\partial t}\left(\bar{\rho} \widetilde{h}_{0}\right) + \frac{\partial}{\partial x_i}\left(\bar{\rho} \tilde{u}_{i} \widetilde{h}_{0}\right) = \frac{D \bar{p}}{D t} - \frac{\partial \bar{q_i}}{\partial x_i} + \overline{\tau^{ij}\frac{\partial u_i}{\partial x_j}}+\Theta_{i}
    \label{Equation:energy_transport}
\end{equation}

The mass source term results from the evaporation, devolatilization and char conversion of the biomass particles. Momentum source term originates from the two-way coupling effect, while the source term in the energy equation, Eq. (\ref{Equation:energy_transport}), includes contributions from the homogeneous gas-phase reactions, heterogeneous combustion of biomass, and combined effect of particle and gas-phase radiation.\\

The production and consumption of chemical species is solved by individual species transport equations Eq. (\ref{Equation:species_transport}) for CO, CO\textsubscript{2}, CH\textsubscript{4}, H\textsubscript{2}, H\textsubscript{2}O, O\textsubscript{2}, and N\textsubscript{2} as bulk gas:

\begin{equation}
    \frac{\partial}{\partial t} \left(\bar{\rho} \widetilde{Y}_{k}\right) + \frac{\partial}{\partial x_{i}} \left(\bar{\rho} \tilde{u_{i}}\widetilde{Y}_{k}\right) = \frac{\partial}{\partial x_{i}}\left(\bar{\rho} \overline{D}_{k}\frac{\partial \widetilde{Y}_{k}}{\partial x_{i}}\right) + \overline{\dot{\omega}}_{k} + \Phi_{k}
    \label{Equation:species_transport}
\end{equation}

The turbulence-chemistry interaction is modeled with the Partially Stirred-Reaction (PaSR) model \cite{chomiak1996flame}, which computes the chemical reaction rate $\overline{\dot{\omega}}_{k}$ as:

\begin{equation}
    \overline{\dot{\omega}}_{k}=\kappa\dot{\omega_{k}}(\widetilde{Y}_{i}, \widetilde{T})
    \label{Equation:PaSR}
\end{equation}

where $\kappa$ is the reactive volume fraction $\kappa = \tfrac{\tau_{c}}{\tau_{c}+\tau_{m}}$ and $\dot{\omega_{k}}(\widetilde{Y}_{i}, \widetilde{T})$ is the formation rate of species $j$. The chemical $\tau_{c}$, and mixing $\tau_{m}$ time scales are calculated as:

\begin{align}
\frac{1}{\tau_{c}}&=\max{\lbrace-\frac{\overline{\dot{\omega}}_{f}}{Y_{f}}, -\frac{\overline{\dot{\omega}}_{o}}{Y_{o}}\rbrace} & \tau_{m}=\sqrt{\frac{k}{\varepsilon}\left(\frac{\nu}{\varepsilon}\right)^{1/2}}
    \label{Equation:Time_scales}
\end{align}

where subscripts \textit{f} and \textit{o} denote the fuel and oxidizer species, respectively. Due to wide range of turbulence scales existing in the flow, in above expression the mixing time scale $\tau_{m}$ is taken as the geometric mean of the integral and Kolmogorov time scales.\\

\subsubsection{Ignition mechanism}
\label{SubSubsection:Ignition_mechanism}

In laboratory tests of dust explosions there are different methods to ignite the dust clouds: electric sparks \cite{Hertzberg19851,Hertzberg19852,Eckhoff+2017+1683+1707}, fuse wires \cite{scheid201301,Scheid201311}, and pyrotechnic ignitors \cite{bartknecht1989dust}. However, given that dust-air mixtures are inherently more difficult to be ignited than gas-air mixtures \cite{GOING2000209}, the energetic pyrotechnic ignitors are usually employed in the ASTM E1226 and EN 14034 standards. Their role is to generate the initial flame which induces dust particles to produce a flame kernel able to allow self-propagation \cite{yuan2014}. As per the standards, two pyrotechnic ignitors with energy of 5 kJ each shall be used, firing horizontally in opposite directions supplying the system with a total ignition energy of 10 kJ.\\

A common practice in previous CFD studies of the 20L sphere, was to represent the ignition source by simply patching a high temperature region at the center of the chamber \cite{Wang2020cornstarch,OGUNGBEMIDE2021104398}. However, this method does not represent accurately the actual behavior of the pyrotechnic ignitors:

\begin{itemize}
    \item The pyrotechnic ignitors produce an acute pressure over-driving in the 20L vessel experiment \cite{GOING2000209,TAVEAU2017348,KUAI2011302}. According to data collected from blank test experiments (i.e. dust-free flows), the pressure increase due to the sole activation of the ignitors can vary between 0.8 and 1.6 bar \cite{ZHAO2020116401,FUMAGALLI201893,portarapillo2021ignit}.
    
    \item When using a single 5 kJ ignitor, the ratio of the volume occupied by the ignition fireball to the volume of the 20L sphere is about 35\% \cite{ZHEN1997317}, while for a 10kJ ignition energy, is above 77\%, almost filling the entire vessel \cite{portarapillo2021ignit,Krietsch2011TestsOS}. 
    
    \item Based on thermal images of Scheid et al. \cite{scheid2013new} and their own experimental work when using pyrotechnical ignitors, Taveau et al. \cite{TAVEAU2017348} assured that temperatures in excess of 923K can be reached within a significant volume in the sphere.
    
    \item Additional experimental studies \cite{HERTZBERG1988303,cashdollar1993,GOING2000209} advocate that the time that elapses between the ignitors are triggered until their effect is extinguished lasts between 10 to 50 ms, being this latter value the one reported in the ASTM E1226 standard \cite{ASTME1226}.
\end{itemize}

Therefore, in the present work the ignition mechanism is simulated by means of a time-dependent {\fontfamily{qcr}\selectfont semiImplicitSource} added to the energy equation, Eq. (\ref{Equation:energy_transport}), as a source term via the {\fontfamily{qcr}\selectfont fvOptions} dictionary. This source term releases a total energy of 10kJ that is distributed over a kernel sphere of 13 cm. Hence, the volume filled by this source is equal to the 77\% of the full sphere one, which is the same value considered by Portarapillo et al.  \cite{portarapillo2021ignit}, who applied a thin-flame model to quantify the flame radius from experimental data. The duration of the source term is taken as an intermediate value from above-mentioned experimental studies, namely 25 ms, and its transient evolution was calibrated with curves from blank test experiments performed in the 20L sphere and reported by Cesana-AG in the 2011 and 2013 Calibration Round-Robin (CaRo) tests \cite{cesana2011}.

\subsubsection{Homogeneous reactions}
\label{SubSubSection:Homogeneous_reactions}

In biomass conversion, the combustion of volatile gases represents about 70\%-80\% of the energy release \cite{sami2001co}. Although the composition of the volatile gases is quite diverse and depends on various factors such as particle temperature, heating rate, residence time or particle size \cite{lu2008comprehensive,yin2010co}, many authors concur that permanent gas composition of volatiles includes CO, CO\textsubscript{2}, CH\textsubscript{4} and H\textsubscript{2} \cite{JIANG201845,sami2001co,di2008modeling,neves2011characterization,li2021detailed,ku2014eulerian}. \\

In various CFD studies of biomass combustion \cite{yin2012towards, yin2010co, marangwanda2021modelling, tabet2015review}, a reaction mechanism that has been applied successfully to the combustion of these volatile gases is the 4-step global mechanism of Jones and Lindstedt \cite{joneslindstedt1988}, which is also adopted in this study:

\begin{align*}
    &\text{CH}\textsubscript{4}+0.5\text{O}\textsubscript{2}\rightarrow{}\text{CO}+2\text{H}\textsubscript{2} \tag{R1} \\
    &\text{CH}\textsubscript{4}+\text{H}\textsubscript{2}\text{O}\rightarrow{}\text{CO}+3\text{H}\textsubscript{2} \tag{R2} \\
    &\text{CO}+\text{H}\textsubscript{2}\text{O}\leftrightarrow{}\text{CO}\textsubscript{2} +\text{H}\textsubscript{2} \tag{R3} \\
    &\text{H}\textsubscript{2}+0.5\text{O}\textsubscript{2}\leftrightarrow{}\text{H}\textsubscript{2}\text{O} \tag{R4}
\end{align*}

The first two reactions describe the breakdown of methane, where (R1) is dominant in fuel lean mixtures, whereas (R2) in fuel rich mixtures \cite{kim2008comparison,wang2012comparison}. (R3) describes the water-gas shift reaction and (R4) is the oxidation of hydrogen. To reduce the computational time of chemical equilibrium calculations, in this global mechanism all reactions are treated as irreversible. A modified rate of (R4) based on the work of Marinov \cite{marinov1996detailed} has demonstrated good agreement with experiments of pulverized fuel combustion \cite{yin2011chemistry}, so it is maintained in this work. The reverse rates of (R3) and (R4) were taken from Wang et al. \cite{wang2018refined}. The corresponding kinetic rates are presented in Table (\ref{Table:homogeneous_reactions}).

\begin{table}[h]
\caption{Kinetic rates of the homogeneous reactions}
\label{Table:homogeneous_reactions}
\adjustbox{width=0.48\textwidth}{%
\begin{tabular}{@{}lll@{}}
\toprule
Reaction & Kinetic rate (kmol/m\textsuperscript{3}/s) & Ref. \\ \midrule
(R1)       & $r_{1}=4.40\times10^{11}\exp{\left(-15154/T_{g}\right)}\left[\text{CH}\textsubscript{4}\right]^{0.5}\left[\text{O}\textsubscript{2}\right]^{1.25}$ & \cite{joneslindstedt1988}     \\
(R2)       & $r_{2}=3.00\times10^{8}\exp{\left(-15154/T_{g}\right)}\left[\text{CH}\textsubscript{4}\right]\left[\text{H}\textsubscript{2}\text{O}\right]$ & \cite{joneslindstedt1988}     \\
(R3)       & $r_{3}=2.75\times10^{9}\exp{\left(-10067/T_{g}\right)}\left[\text{CO}\right]\left[\text{H}\textsubscript{2}\text{O}\right]$ & \cite{joneslindstedt1988}     \\
\textit{rev.}       & $r_{3r}=6.46\times10^{10}\exp{\left(-13590/T_{g}\right)}\left[\text{CO}\textsubscript{2}\right]\left[\text{H}\textsubscript{2}\right]$ & \cite{wang2018refined}      \\
(R4)       & $r_{4}=5.69\times10^{11}\exp{\left(-17560/T_{g}\right)}\left[\text{H}\textsubscript{2}\right]\left[\text{O}\textsubscript{2}\right]^{0.5}$ & \cite{marinov1996detailed}     \\
\textit{rev.}       & $r_{4r}=2.83\times10^{13}\exp{\left(-46906/T_{g}\right)}\left[\text{H}\textsubscript{2}\text{O}\right]$ & \cite{wang2018refined}     \\ \bottomrule
\end{tabular}}
\end{table}

\subsubsection{Radiation modeling}
\label{SubSubSection:Radiation_modeling}

Since thermal radiation contributes strongly to the heat transfer mechanism of biomass combustion, modeling the radiation properties of the combustion gases and the particulates is essential  \cite{modest2021radiative}. \\

In solid fuel combustion CFD, the radiative transfer equation (RTE) is commonly solved by the P1 and discrete ordinates {\fontfamily{qcr}\selectfont fvDOM} models \cite{yin2019coal}. Although the former is computationally cheaper than latter, {\fontfamily{qcr}\selectfont fvDOM} is preferred in this work because is applicable to all the optical thicknesses resulting from the wide range of dust concentrations considered in this study. \\

In OpenFOAM, {\fontfamily{qcr}\selectfont fvDOM} solves the RTE for an absorbing-emitting and non-scattering medium, Eq. (\ref{Eqn:fvDOM}): 

\begin{equation}
    \bold{\hat{s}} \cdot \nabla I\left(\bold{r}, \bold{\hat{s}}\right) = \alpha_{g} I_{b} -\left(\alpha_{g} + \alpha_{p} + \sigma_{p}\right)I
    \label{Eqn:fvDOM}
\end{equation}

in which $I(\bold{r},\bold{\hat{s}})$, $I_{b}$, $\alpha_{g}$, $\alpha_{p}$, and $\sigma_{p}$ represent the radiative intensity at position $\bold{r}$ in direction $\bold{\hat{s}}$, the black body intensity, the absorption coefficient of the gaseous mixture, the particle absorption coefficient, and the particle scattering coefficient, respectively. In {\fontfamily{qcr}\selectfont fvDOM} the full solid angle $4\pi$ is divided into $4N_{\phi}N_{\theta}$ discrete angles. For all simulations, the angular discretization was treated with $N_{\phi}=3$, $N_{\theta}=3$ for the azimuth and inclination, respectively. To keep a moderate computational cost of the additional 36 equations, the RTE was solved only once every 10 flow time steps. This is justifiable by the fact that the radiation field does not change briskly between iterations as other momentum-driven scalars may do \cite{krishnamoorthy2015assessing}. \\

An adequate modeling of the gaseous absorption coefficient is critical for combustion applications because product gases (i.e. CO\textsubscript{2}, H\textsubscript{2}O) are strong selective absorbers and emitters of radiant energy \cite{viskanta1987radiation}. In CFD combustion, $\alpha_{g}$ is often evaluated using the weighted-sum of gray gases model (WSGGM) because it strikes a reasonable compromise between the oversimplified gray gas assumption and a complete model accounting for the entire spectral variations of radiation properties \cite{yeoh2009computational}.\\

The WSGGM postulates that the total gaseous emissivity may be represented by the sum of the emissivities of $N_{g}$ gray gases and one clear gas, weighted by temperature-dependent factors, Eq. (\ref{Eqn:WSGGM}) \cite{smith1982evaluation}. Then Beer’s law, Eq. (\ref{Eqn:Beer_law}) is used to calculate a gray absorption coefficient based on the total emissivity $\varepsilon$, and mean beam length of the chamber $L$, where $L=0.65D$ for spheres \cite{modest2021radiative}.

\begin{align}
    \varepsilon &= \sum_{i=0}^{N_{g}}a_{\varepsilon,i}(T)\left[1-\exp{(-\kappa_{i}p_{a}L)}\right] \label{Eqn:WSGGM}\\
    \alpha_{g} &= \frac{-\log{(1-\varepsilon)}}{L}
    \label{Eqn:Beer_law}
\end{align}

In the present, the WSGGM was implemented into OpenFOAM following the works of Smith et al. \cite{smith1982evaluation} and Kangwanpongpan et al. \cite{kangwanpongpan2012new}. The latter reference provides extended WSGGM correlations that are valid for H\textsubscript{2}O/CO\textsubscript{2} molar ratios between 0.125 to 4.0. These coefficients are valid for the variable molar ratios arising from the combined effect of moisture evaporation and combustion for the entire range of dust concentrations considered here. The WSGGM implementation was validated with benchmark cases from the literature (\textit{see} Appendix A).

\subsection{Solid-phase governing equations}
\label{SubSubSection:Solid-phase_governing_equations}

The combustion of biomass follows a reaction mechanism similar to coal, i.e. it occurs in three consecutive processes: (1) moisture evaporation, (2) devolatilization, and (3) surface reactions.\\

During all stages of biomass combustion, the thermal history of the solid particles is governed by an energy balance which includes the effects of convective and radiative heat transfer, and enthalpy change due to reactions, Eq. (\ref{Eqn:Particle_energy_equation}):

\begin{multline}
    m_{p}C_{p}\frac{dT_{p}}{dt} = \pi d_{p}k_{g}\text{Nu}\left(T_{\infty}-T_{p}\right) + \frac{dm_{p}}{dt}\Delta H \\
    + \pi d_{p}^2\varepsilon_{0}\sigma\left(\theta_{R}^4-T_{p}^4\right)
    \label{Eqn:Particle_energy_equation}
\end{multline}
    
where the Nusselt number is given by the Ranz-Marshall correlation \cite{ranz1952evaporation}. $m_{p}$, $C_{p}$, $T_{p}$, $d_{p}$, $k_{g}$, $T_{\infty}$, $\varepsilon_{0}$, $\sigma$, and $\theta_{R}$ denote the particle mass, particle specific heat, particle temperature, particle diameter, thermal conductivity of the surrounding gas, local temperature of the bulk gas, particle emissivity, Stefan-Boltzmann constant, and the radiation temperature, respectively. \\

Depending on the thermal stage of the particle, $\Delta H$ can denote the latent heat of: (1) evaporation of the moisture, (2) devolatilization, or (3) heat of combustion of the surface reactions. A common practice in CFD is to set the latent heat of evaporation and heat of combustion as 2.25 MJ/kg and 32.9 MJ/kg (in case of C oxidation only), respectively. However, literature review indicates a large scatter for the latent heat of devolatilization \cite{milosavljevic1996thermal}. This is mainly because devolatilization can be driven in either endothermic or exothermic directions by competition between char and tar yields \cite{haseli2012modeling, ragland2011combustion}. Here, the devolatilization is considered as an endothermic reaction \cite{bridgwater2012review}, requiring heat from the surroundings to the particle, which causes the thermal decomposition of biomass into the gaseous species. In all simulations, a value of 100 kJ/kg is adopted based on the most frequent order of magnitude of the values reported by Haseli \cite{haseli2012modeling}.

\subsubsection{Moisture evaporation}
\label{SubSubSection:Moisture_Evaporation}

The moisture evaporation rate is governed by gradient diffusion, with the flux of particle vapor into the gas phase related to the difference in vapor concentration at the particle surface and the gaseous phase, Eq. (\ref{Eqn:Particle_moisture_evaporation}):

\begin{equation}
    \frac{dm_{w}}{dt}=\pi d_{p}D_{0}Sh\left(\frac{p_{sat,T}}{RT_{m}}-X_{w}\frac{p}{RT_{m}}\right)M_{w}
    \label{Eqn:Particle_moisture_evaporation}
\end{equation}

where the Sherwood number is calculated by the equivalent Ranz-Marshall correlation for mass transfer \cite{ranz1952evaporation}. $D_{0}$, $p_{sat, T}$, $R$, $X_{w}$, $p$, and $M_{w}$ denote the vapor diffusion coefficient, the saturation pressure at the bulk temperature, the universal gas constant, the molar fraction of water vapor in the surrounding, the local absolute pressure, and the molar weight of vapor, respectively. In OpenFOAM, the film temperature $T_{m}$ is evaluated using the two thirds rule, $T_{m}=\tfrac{2T_{p}+T_{g}}{3}$.\\

For high rates of vaporization, the heat transfer coefficient should be corrected for both the effect of superheating the vapor as it moves away from the surface, and for the blowing effect of the vapor motion on the boundary layer \cite{ragland2011combustion}. Therefore, in this work Bird's correction \cite{bird2002transport} is applied to the Nu number to account for the reduction of heat transfer, Eq. (\ref{Eqn:Bird_correction}):

\begin{equation}
    \text{Nu*}=\text{Nu}\frac{\beta}{e^{\beta}-1}, \quad \beta=-\frac{C_{p,\text{vap}}\dot{m}_{p}}{\pi d_{p} k_{g}\text{Nu}}
    \label{Eqn:Bird_correction}
\end{equation}

\subsubsection{Devolatilization model}
\label{SubSubSection:Devolatilization_model}

 Here, the devolatilization reaction is described with a single first-order model (SFOM), Eq. (\ref{Eqn:Arrhenius_mass_devolatilization}):
 
 \begin{align}
 \label{Eqn:Arrhenius_mass_devolatilization}
    -\frac{dm_{p}}{dt} &= k\left(T\right)\left(m_{p}-\left(1-f_{\text{VM}_{0}}\right)m_{p_{0}}\right)  \\
    k\left(T\right)&=A\exp{\left(-\frac{E_{a}}{RT}\right)} 
 \end{align}
 
 where $f_{\text{VM}_{0}}$, $m_{p_{0}}$, and $m_{p}$ are the initial mass fraction of volatile matter in the particle, and the initial and instantaneous particle masses, respectively. $k(T)$ is the kinetic rate which takes the form of an Arrhenius expression. \\
 
 In combustion of pulverized biomass, the particles experience very fast heating rates and temperatures \cite{johansen2018high, espekvist2021determination}. For example, in dust explosion testing of carbonaceous dusts, $P_{ex}$ is generally reached in some tens or a few hundreds of milliseconds \cite{ogle2016dust}. Considering that typical adiabatic flame temperatures for biomass lie in the range of 2000 K to 2700 K \cite{jenkins1998combustion} and assuming that this temperature is reached exactly at $P_{ex}$, one can expect heating rates in the order of $10^3$-$10^5$ K/s. These heating rates are comparable to those found in industrial furnaces firing biomass \cite{ma2007modelling,black2013effects}. \\
 
 Although extensive thermogravimetric analysis (TGA) experiments have been conducted to determine devolatilization kinetics, these are mostly valid for low heating rate condition (typically in the order of K/min). On the contrary, entrained flow reactors (EFR) or drop tube reactors (DTR) can operate at the elevated heating rates \cite{wagenaar1993flash}. However, kinetic data from these experiments are very limited, mainly because accurate measurements at such conditions are difficult to perform \cite{dupont2009biomass}, particularly those concerning the particle residence time and thermal history \cite{johansen2016devolatilization}.\\
 
 As an alternative to experiments and given the complexity of the conversion process, the existent advanced network models for coal devolatilization have been extended to predict devolatilization behavior of biomass. Chen at al. \cite{chen1998modeling} adapted the Functional Group-Depolymerization, Vaporization Crosslinking (FG-DVC) model to predict biomass devolatilization at a heating rate $10^3$ K/s. Niksa \cite{niksa2020bio} used the bio-FLASCHAIN model to predict total volatile yields of 13 woods and 22 torrefied woods under heating rates of $10^4$ K/s. Fletcher et al. \cite{fletcher2012prediction,lewis2013prediction} extended the chemical percolation devolatilization (CPD) model to predict pyrolysis yields of biomass at heating rates of $10^3$-$10^4$ K/s, demonstrating good agreement with experiments.\\

  In this study, devolatilization kinetic parameters are determined by employing the CPD model extension for biomass (i.e. the BioCPD model), mainly because is freely available to all researchers \cite{CPDfletcher2020} and has been successfully used as kinetic pre-processor for CFD simulations \cite{zhang2021kinetic}. The BioCPD model characterizes the devolatilization behavior of rapidly heated biomass based on the physical and chemical transformations of lignocellulose structures at a given heating rate \cite{vizzini2008prediction}. The model is used to calculate the conversion curves for cellulose, hemicellulose and lignin independently, while the total volatile yield of biomass is determined by linear superposition of these species \cite{zhang2021kinetic}.\\
  
  \begin{table}[H]
\caption{Estimated lignocellulose composition of the biomass samples}
\label{Table:Lignocellulose_composition}
\begin{tabular}{@{}lll@{}}
\toprule
Label                                & Biomass 1                                                  & Biomass 2                                                    \\ \midrule
Sample                               & \begin{tabular}[c]{@{}l@{}}Pellets\\ Asturias\end{tabular} & \begin{tabular}[c]{@{}l@{}}Cupressus\\ Funebris\end{tabular} \\
\textit{Lignocellulose composition (wt. \%)} &                                                            &                                                              \\
Cellulose                            & 40.37                                                      & 38.41                                                        \\
Hemicellulose                        & 34.70                                                      & 31.62                                                        \\
Lignin                               & 24.93                                                      & 29.97                                                        \\ \bottomrule
\end{tabular}
\end{table}

  Defining the degree of conversion of volatile matter, $\alpha=(m_{p_{0}}-m_{p}) / (m_{p_{0}}-m_{p_{0}}f_{\text{FC}_{0}})$, and assuming devolatilization progresses at a linear heating rate $\beta$, Eq. (\ref{Eqn:Arrhenius_mass_devolatilization}) can be re-written more conveniently as:
 
 \begin{equation}
     \frac{d\alpha}{dT} = \frac{A}{\beta}\exp{\left(-\frac{E_{a}}{RT}\right)}\left(1-\alpha\right)
     \label{Eqn:Arrhenius_conversion_devolatilization}
 \end{equation}

 where $\alpha(T)$ is obtained as output of the BioCPD model. This curve is then used to estimate the kinetic parameters by a fitting procedure using the Coats-Redfern \cite{coats1964kinetic} integral method, Eq. (\ref{Eqn:Coats_Redfern}):
 
 \begin{equation}
     \ln \left[\frac{G(\alpha)}{T^2}\right] = \ln \left(\frac{AR}{\beta E_{a}}\right) - \frac{E_{a}}{R}\frac{1}{T}
     \label{Eqn:Coats_Redfern}
 \end{equation}
 
 where $G\left(\alpha\right)$ denotes the integral function of conversion. The pre-exponential factor $A$ and activation energy $E_{a}$ can be determined from the slope and intercept of the line resulting from plotting the left-hand side of Eq. (\ref{Eqn:Coats_Redfern}) versus $1/T$.\\
 
\begin{figure}
     \centering
     \includegraphics[width=0.43\textwidth]{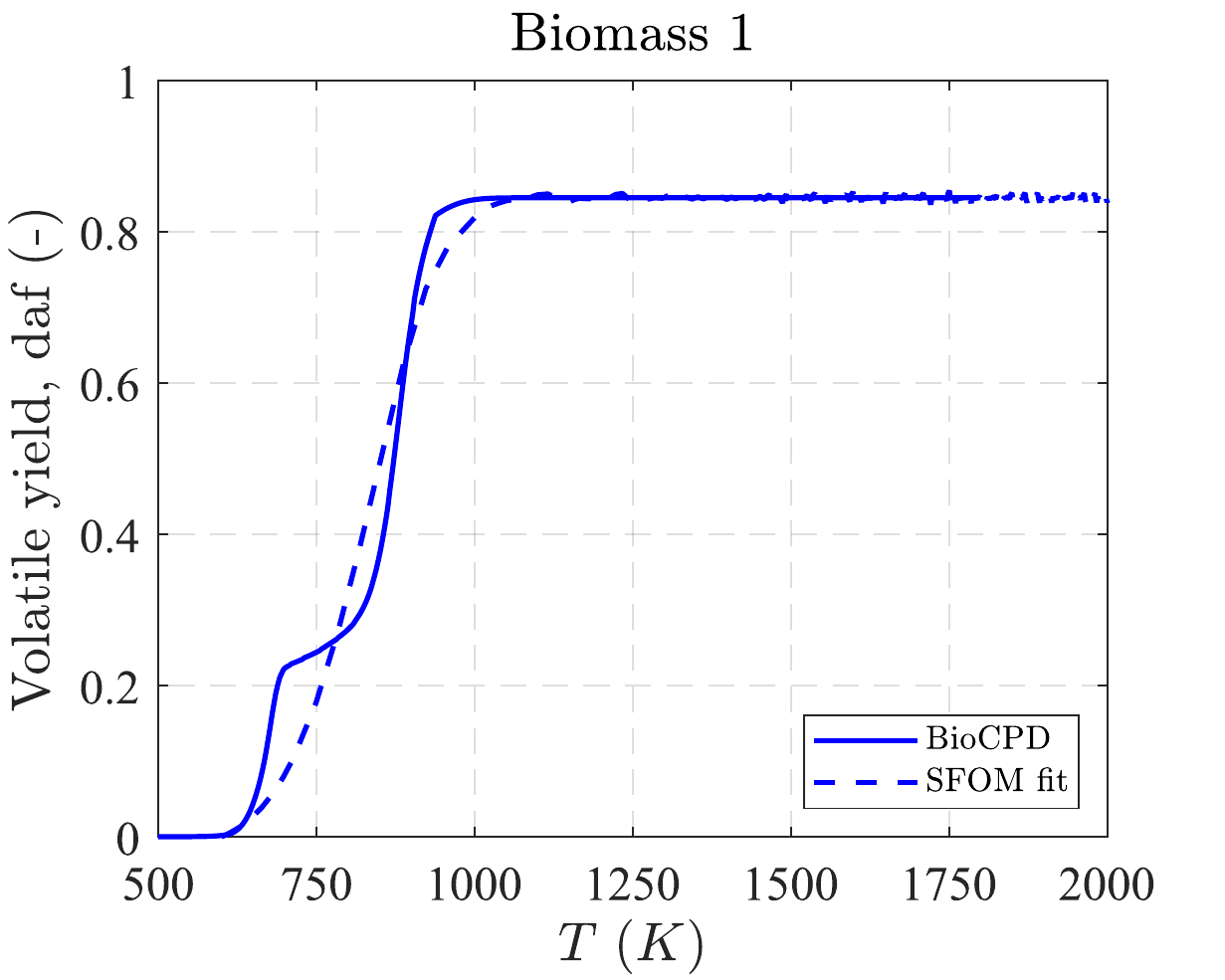}
     \caption{Example of volatile yield prediction using BioCPD model and SFOM fit for \textit{Pellets Asturias}.}
     \label{Fig:BioCPD_Pellets_Asturias}
 \end{figure}

 All BioCPD calculations were computed at a representative heating rate, $\beta=3.7\times 10^4$ K/s, and the lignocellulose composition of each biomass was estimated using the empirical correlations proposed by Sheng and Azevedo \cite{sheng2002modeling}, see Table (\ref{Table:Lignocellulose_composition}). Fig. \ref{Fig:BioCPD_Pellets_Asturias} shows an example of the devolatilization curve obtained with BioCPD model and the corresponding SFOM fit for \textit{Pellets Asturias}. \\
 
\subsubsection{Surface reaction}
\label{SubSubSection:Surface_reaction}

After the volatile gases of the biomass particle are completely evolved, a surface reaction occurs which consumes the char left in the particle. In the present, the char content is assumed as pure C which undergoes complete oxidation:

\begin{equation}
    \text{C}(\text{s}) + \text{O}_{2}\rightarrow{} \text{CO}_{2}
\end{equation}

The rate of char conversion is computed with the kinetic-diffusion limited rate model \cite{baum1971predicting}. It assumes that the overall reaction rate is function of the combined diffusion and kinetic rates, (Eq. \ref{Eqn:Kinetic_diffusion_rate}): 

\begin{align}
    \frac{dm_{p}}{dt}&=-\pi d_{p}^2 p_{o}\frac{R_{diff}R_{kin}}{R_{diff}+R_{kin}} \\
    R_{diff}&= C_{0}\frac{\left[0.5\left(T_{g}+T_{p}\right)\right]^{3/4}}{d_{p}} \\
    R_{kin}&= A\exp{\left(-\frac{E_{a}}{RT_{g}}\right)}
    \label{Eqn:Kinetic_diffusion_rate}
\end{align}

where $R_{diff}$, $R_{kin}$, $C_{0}$, $p_{o}$, $A$, and $E_{a}$ represent the diffusion rate, the kinetic rate, the mass diffusion coefficient, the partial pressure of the oxidizer, and the Arrhenius pre-exponential factor and activation energy, respectively.\\

Because during char burnout part of the heat released may be transferred to the particle rather than the gas phase \cite{boyd1988three}, only during the surface reaction, the second term on the RHS of Eq. (\ref{Eqn:Particle_energy_equation}) is multiplied by a retention coefficient $h_{s}$. In this work $h_{s}$ is assumed to be 0.3 \cite{boyd1988three,ansys2011ansys}.

\subsubsection{Radiative properties of biomass particles}
\label{SubSubSection:Radiative_properties_of_biomass_particles}

In contrast to radiation from hot gases, particulate solids absorb, emit and scatter radiation throughout the wavelength spectrum \cite{mengucc1994determination}. Absorption and scattering of a cloud of particles are proportional to the degree of blockage of incident radiation due to particles. In OpenFOAM these are calculated as \cite{chui1993implementation}:

\begin{align}
    \alpha_{p} &= \sum_{i} \varepsilon_{0} \frac{A_{pp,i}}{V_{i}} \\
    \sigma_{p} &= \sum_{i} \left(1-f_{0}\right)\left(1-\varepsilon_{0}\right) \frac{A_{pp,i}}{V_{i}}
    \label{Eqn:radiation_particles}
\end{align}

where $\alpha_{p}$, $\sigma_{p}$, $\varepsilon_{0}$, and $f_{0}$ denote the particle absorption and scattering coefficients appearing in Eq. (\ref{Eqn:fvDOM}), and the particle emissivity and scattering factors, respectively. $A_{pp,i}$ is the cross-sectional area of all the particles in parcels contained in the $i$-th cell volume $V_{i}$.\\

Contrarily to the limiting case when a size parameter $x=d_{p}/\lambda\rightarrow 0$ (where $d_{p}$: the particle diameter and $\lambda$: the radiation wavelength) for which the simple Rayleigh-scattering formulas are valid for estimating the radiative properties of very small particles (e.g., soot). the complicated Lorenz-Mie scattering theory is applicable to calculate the absorption $Q_{\text{abs}}$ and scattering $Q_{\text{sca}}$ efficiencies of particles clouds with $\mathcal{O}\left(x\right)\sim\left[10^0,10^2\right]$ instead \cite{hulst1981light}. This theory is a formal derivation from Maxwell's equations of electromagnetism \cite{hofgren2015evaluation}, is valid for spherical particles and depends on the complex index of refraction $m = n-ik$, the particle size (i.e., the PSD), the radiation wavelength $\lambda$, and the dust concentration \cite{ogle2016dust}.\\

In this work, the radiative properties of the biomass particles are calculated with the open-source Mie theory code {\fontfamily{qcr}\selectfont mmmie.f} \cite{modest2021radiative}. Along with above variables and a number density (i.e., the number of particles per cm\textsuperscript{3}) this code calculates $Q_{\text{abs}}$ and $Q_{\text{sca}}$ and relates them with the particle cloud absorption and scattering coefficients $\alpha_{p}$ and $\sigma_{p}$, respectively. For these calculations, the complex index of refraction of biomass was assumed to be $m=1.50-0.01i$ \cite{levine1991global}. The particle size distribution effects were resolved by adopting the histogram representation (40 equally spaced bins), the number density in each bin was calculated as the number of physical particles (do not confuse with computational parcels) distributed over a volume of 20,000 cm\textsuperscript{3} (i.e., 20L). The efficiency factors were calculated for each bin, weighted over fractional particle number in each bin, and integrated over the PSD to obtain macroscopic absorption and scattering coefficients of the cloud. The calculations were repeated and averaged for a wavelength interval between $1-10 \mu\text{m}$, with increments of $1\mu \text{m}$. This corresponds to the spectrum of electromagnetic radiation that has a potential range of interaction with combustible dusts \cite{ogle2016dust}. Then $\varepsilon_{0}$ and $f_{0}$ were calculated from Eq. (\ref{Eqn:radiation_particles}) and given as inputs to the CFD code.\\
    
\section{Solution strategy and numerical methods}
\label{Section:Solution_strategy_and_numerical_methods}

Each of the simulation runs is split into two stages: (1) dispersion, and (2) explosion of the dust cloud. The first stage consists of placing the biomass dust in the canister at stagnant conditions. The dust container is pressurized at 21 bar and the sphere is vacuumed to 0.4 bar. The particles are driven from the canister to the sphere by the pressure gradient, while they are dispersed into the chamber by the rebound nozzle. The reader is referred to our previous work for more details on the cold flow simulation \cite{ISLAS2022117033}. After an ignition delay time $t_{d}$ elapses, the cold flow solution is mapped from mesh 1 to mesh 2, where the reactive simulation is resumed.\\ 

\begin{figure}[h]
    \centering
    \includegraphics[width=0.43\textwidth]{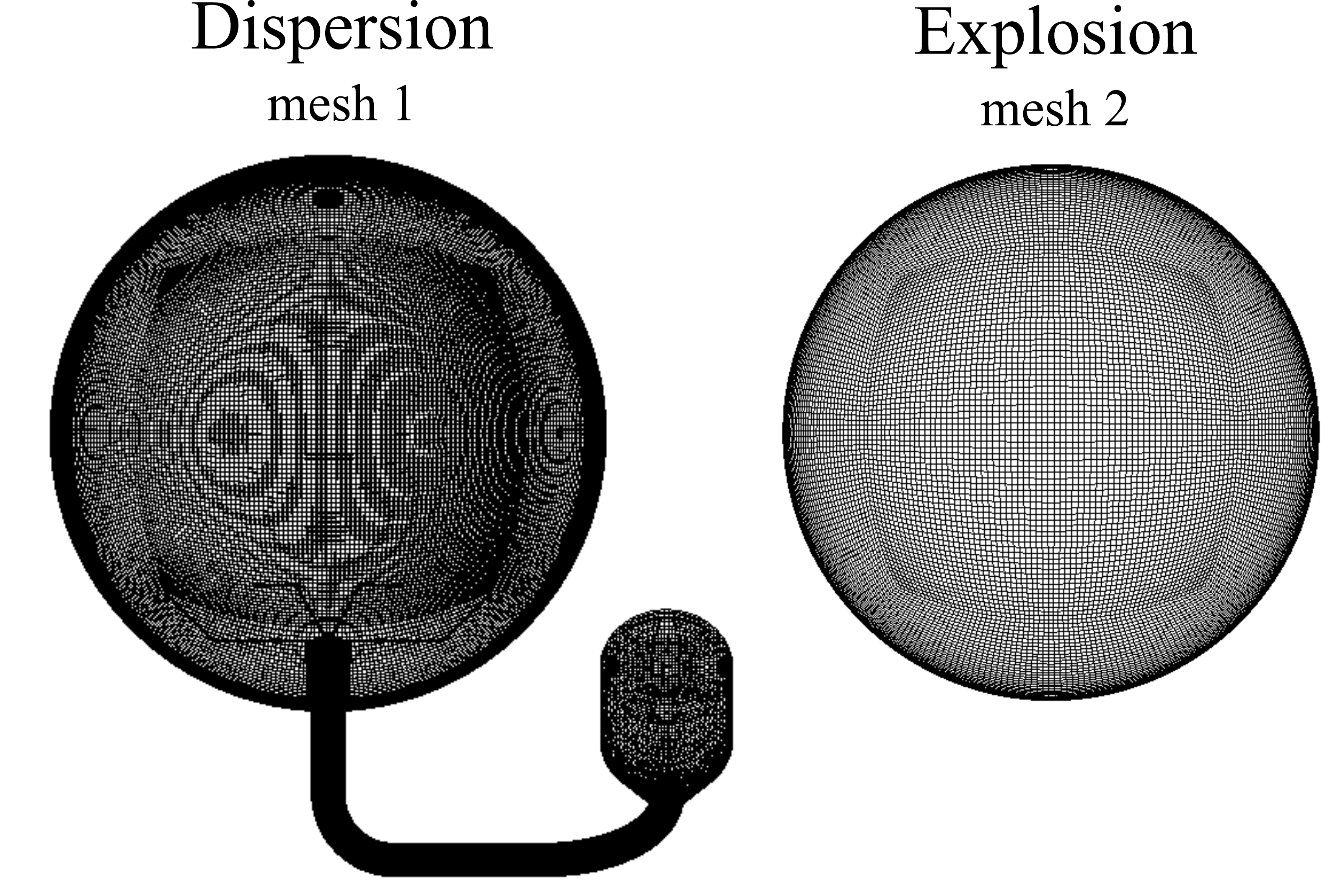}    
    \centering \caption{3D grids employed for the simulation of the dispersion and explosion stages.}
    \label{Figure:mesh}
\end{figure}

During the second stage of the simulation, all the reactive features of the solver are enabled starting with the activation of the ignition mechanism. For all simulations, the pressure-time monitor is reported as a {\fontfamily{qcr}\selectfont patchAverage} value at the walls, whose temperature is fixed at 293 K to represent the cooling effect of the water-jacket in the real apparatus \cite{ASTME1226}.\\

The mapping strategy allows one to preserve the features of the explosion process in a spherical chamber, while the quality metrics of mesh 2 are improved. This facilitates the use of a CFL = 1 condition, leading to better convergence and accuracy of the solution. The 3D grids were generated in ANSYS ICEM\textsuperscript{\textregistered} using a blocking strategy to produce hybrid and structured meshes for mesh 1 and 2, respectively. The corresponding quality metrics are shown in Table (\ref{Table:Mesh_quality_metrics}).\\

\begin{table}[h]
\caption{Mesh quality metrics reported by OpenFOAM's \texttt{checkMesh} utility.}
\label{Table:Mesh_quality_metrics}
\begin{tabular*}{\tblwidth}{@{} LLLL@{} }
\toprule
Parameter              & Mesh 1 & Mesh 2                                                                                   \\ \midrule
Max. Aspect Ratio       & 25.32 & 6.03  \\
Avg. Non-Orthogonality  & 13.72 & 12.42 \\
Max. Non-Orthogonality & 75.36 & 53.35 \\
Min. Angle      & 9.52 & 44.93         \\
Avg. Face interpolation wt. & 0.47 & 0.49 \\
Number of cells        & 1.62M & 2.29M 
\\ \bottomrule
\end{tabular*}
\end{table}

\begin{table}[h]
\caption{Particle properties of the biomass samples}
\label{Tab:Particle_properties}
\adjustbox{width=0.48\textwidth}{%
\begin{tabular}{@{}llll@{}}
\toprule
Particle property                                                          & Biomass 1 & Biomass 2 & Ref. \\ \midrule
Sample & \begin{tabular}[c]{@{}l@{}}Pellets\\ Asturias\end{tabular} & \begin{tabular}[c]{@{}l@{}}Cupressus\\ Funebris\end{tabular} \\
Granulometry & & & \\
particle size distribution (PSD) & general & RR &  \\
$D_{10}$, [$\mu$m] & 99.9 & 8.78\textsuperscript{*} & \textendash , \cite{liu2019explosion}\textsuperscript{*} \\
$D_{50}$, [$\mu$m] & 535.7 & 32.47\textsuperscript{*} & \textendash , \cite{liu2019explosion}\textsuperscript{*}\\
$D_{90}$, [$\mu$m] & 1200.4 & 63.55\textsuperscript{*} & \textendash , \cite{liu2019explosion}\textsuperscript{*}\\ 
 & & & \\
Thermophysical properties & & & \\
density, $\rho_{p}$ [kg/m\textsuperscript{3}] & 1430 & 1430 & measured  \\
specific heat, $c_{p}$ [J/kg K] & 1242 & 1242 & \cite{jenkins1989physical} \\
initial temperature, $T_{0}$ [K] & 300 & 300 & \textendash \\
devolatilization temperature, $T_{\text{dev}}$ [K] & 500 & 500 & \cite{haseli2012modeling} \\
latent heat of devolatilization, $\Delta H_{\text{dev}}$ [J/kg] & $1\times 10^{5}$ & $1\times 10^{5}$ & \cite{haseli2012modeling} \\
 & & & \\
Radiative properties& & & \\
particle emissivity, $\varepsilon_{0}$ [-] & 0.54 & 0.81 & calculated \\
particle scattering factor, $f_{0}$ [-] & 0.91 & 0.52 & calculated \\
 & & & \\
(SFOM) devolatilization parameters & & & \\
pre-exponential factor, $A$ [1/s] & $7.84\times 10^{5}$ & $9.10\times 10^{5}$ & calculated \\
activation energy, $E_{a}$ [J/kmol K] & $5.41\times 10^{7}$ & $5.53\times 10^{7}$ & calculated \\
swelling index & 1.0 & 1.0 & \\
 & & & \\
Surface reaction parameters & & & \\
mass diffusion coefficient $C_{0}$ [kg/m\textsuperscript{2}s Pa] & $5.32\times 10^{-12}$ & $5.32\times 10^{-12}$ & \cite{chen2012oxy}\\
pre-exponential factor, $A$ [1/s] & 0.005 & 0.005 & \cite{chen2012oxy} \\
activation energy, $E_{a}$ [J/kmol K] & $7.4\times 10^{7}$ & $7.4\times 10^{7}$ & \cite{chen2012oxy} \\
retention coefficient, $h_{s}$ & 0.3 & 0.3 & \cite{boyd1988three,ansys2011ansys}\\

\bottomrule
\end{tabular}}
\end{table}

Eqs. (\ref{Equation:mass_transport}-\ref{Equation:species_transport}) were discretized by employing a first order upwind scheme for the convective terms and a second-order central difference scheme for diffusive terms. Gradient terms were evaluated using a cell-limited scheme with cubic interpolation. Transient discretization was calculated using a first-order Euler scheme with an adaptive time-stepping method to satisfy CFL = 5 and CFL = 1, for the cold flow and reactive flow simulations, respectively. The velocity-pressure coupling is solved by the PIMPLE algorithm with 3 correctors per time step. Flow residuals were set to $10^{-8}$ for continuity and pressure, and $10^{-12}$ for momentum, turbulence, and species equations, respectively.\\

The particle velocity and energy equations were solved with Euler and analytical integration schemes, respectively. A limiting Courant number of 0.3 was imposed to guarantee the stability of the coupled solution between Eulerian and Lagrangian phases.\\ 

To obtain statisically significant results of the lagrangian phase, in all simulations the parcel count was set to 1M. A summary of the thermophysical and other particle properties for the two biomass samples considered in this work is presented in Table (\ref{Tab:Particle_properties}).\\

\section{Results and discussion}
\label{Section:Results_and_discussion}

\subsection{Validation of the pressure-time curve}
\label{Subsection:Validation_of_the_pressure-time_curve}

\begin{figure}[h]
    \centering
    \includegraphics[width=0.48\textwidth]{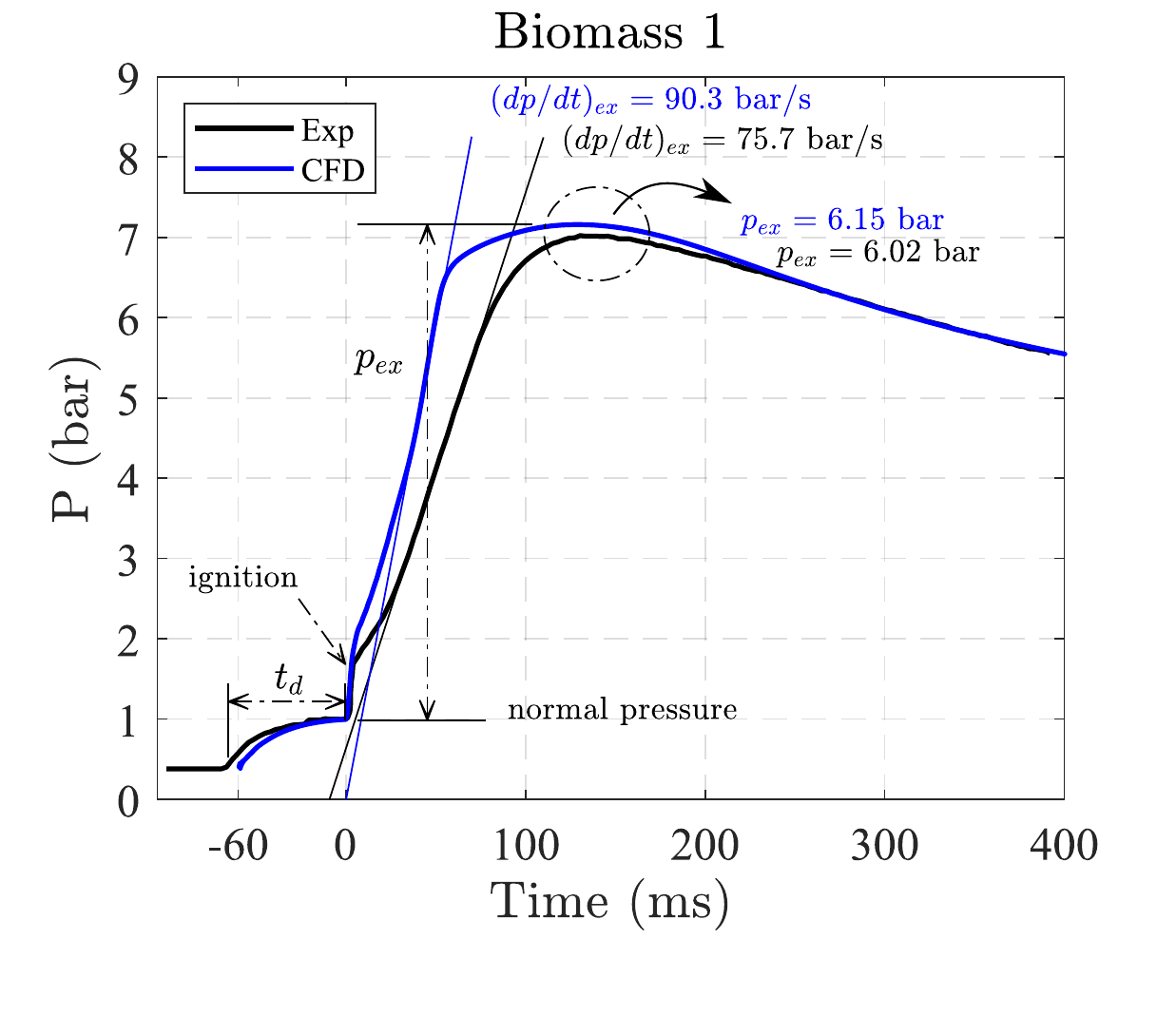}
    \caption{Comparison of the experimental and CFD-predicted pressure-time curve for \textit{Pellets Asturias} (biomass 1).}
    \label{Fig:PA_pressure-time}
\end{figure}

To ensure the accuracy of the numerical modeling and physical considerations described before, a comparison of the experimental and CFD-predicted pressure-time curves of \textit{Pellets Asturias} (biomass 1) is presented in Fig. \ref{Fig:PA_pressure-time}. This curve shows the pressure rise during all stages of the experiment as per the ASTM E1226 standard, i.e. due to injection of air and particles, ignition, and the deflagration itself. This case corresponds to a dust concentration of $C_{0}=750$ g/m\textsuperscript{3}, ignition delay time of $t_{d}=60$ ms, and an ignition energy of 10 kJ.\\

First, during the dispersion stage [-60, 0 ms], the pressure increases from 0.40 to 1 bar, which is the desired normal pressure before the initiation of the deflagration test. Second, the curve is followed by a sharp pressure rise due to the activation of the energetic pyrotechnic ignitors. After their effect is extinguished, the deflagration is self-sustained by the combustion of volatile gases and the particle surface reaction, which increases the over-pressure up to a maximum of 6.15 bar. The relative error between the predicted and experimental explosion pressure $P_{ex}$ is 1.85\%. This error can be considered as an excellent value considering the model assumptions and simplifications of the intricate physics in dust explosions. Although this is not case for the error in the rate of pressure rise $(\text{dP}/\text{dt})_{ex}$, which increases up to 19.2\%, this is still an acceptable value as the ASTM E1226 standard allows a deviation of $\pm30\%$ when $(\text{dP}/\text{dt})_{ex}\lesssim 180\,\text{bar}/\text{s}$. Moreover, it is important to recall that the tangent has to be drawn only after the effect of the ignitors is terminated \cite{ASTME1226}, otherwise the deflagration index $K_{st}$ can be severely over-predicted. Lastly, the pressure drops because the available oxygen is depleted and the cold walls reduce the temperature inside the chamber.\\

\begin{figure*}[h]
    \centering
    \includegraphics[width=0.80\textwidth]{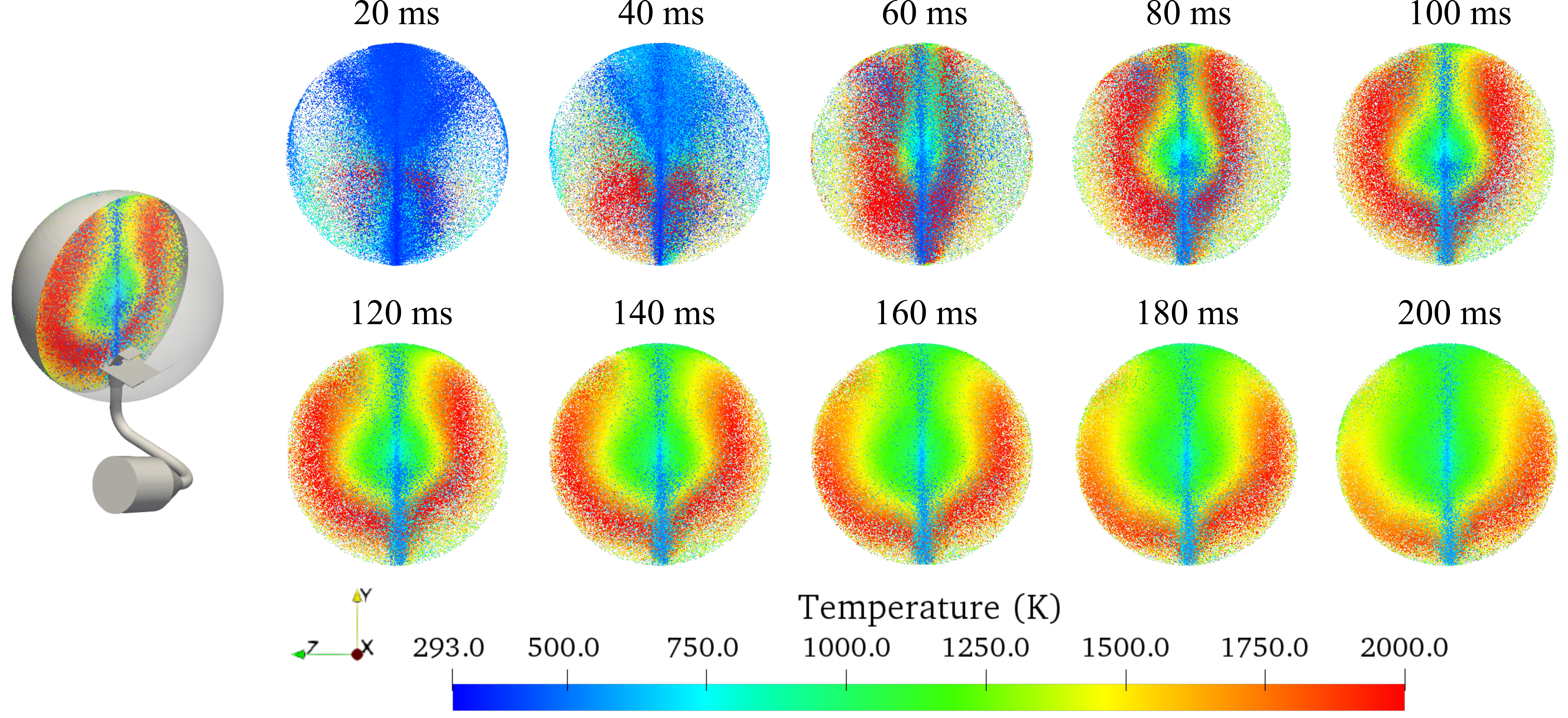}
    \caption{Snapshots of particle tracks colored by particle temperature at selected times during the explosion process of \textit{Pellets Asturias} (biomass 1).}
    \label{Fig:Fig_PA_Particle_temperature}
\end{figure*}

\begin{figure}[h]
    \centering
    
    \begin{tabular}{cc}
    \subfigure[Particle tracks colored by particle diameter during the dispersion process. Snapshot taken at the end of the ignition delay time. \label{Subfigure:PA_particles_transparency}]{\includegraphics[width=0.43\textwidth]{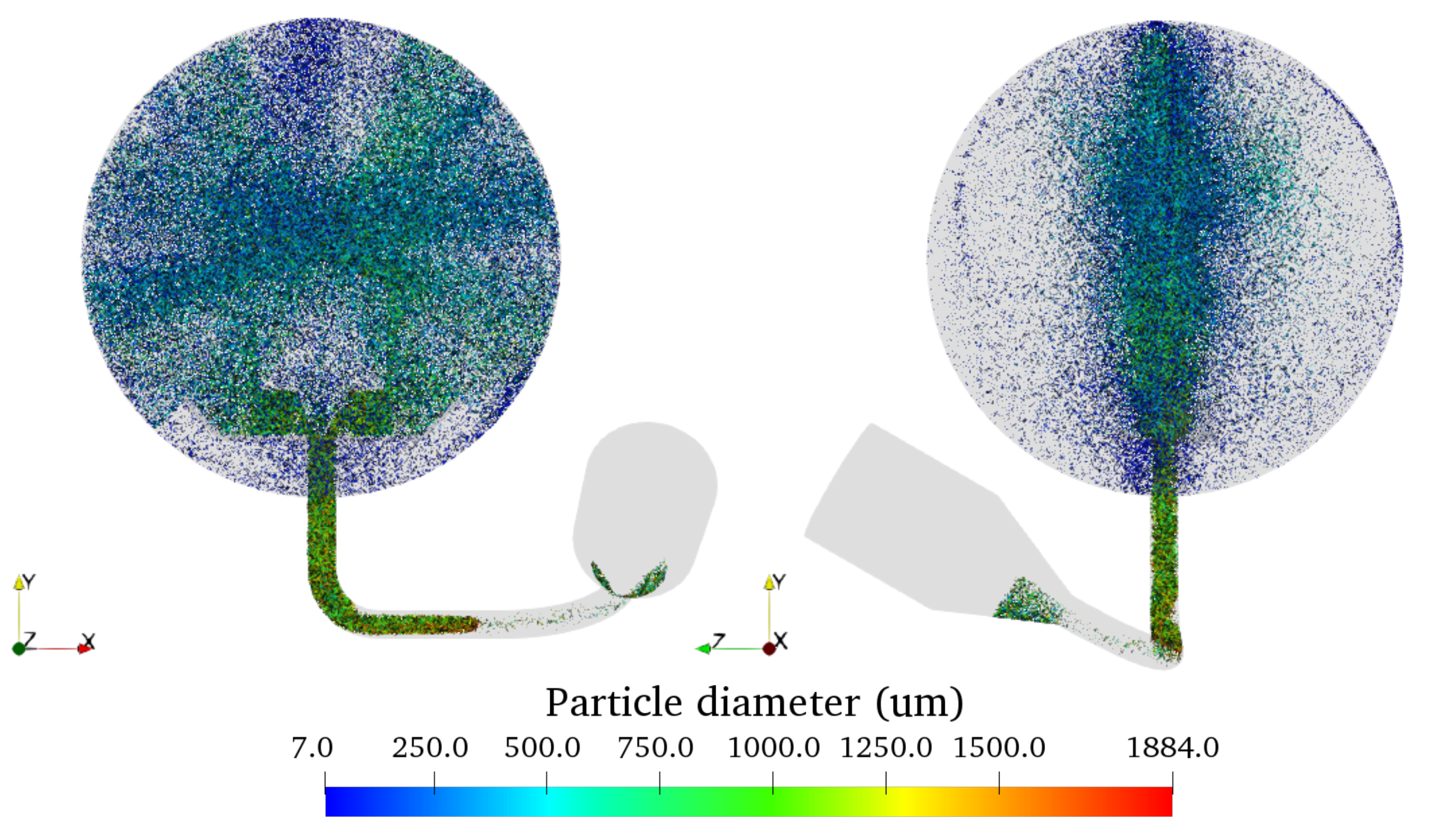}} \\
    \subfigure[Stokes number scatter plot of instantaneous data (blue dots) and profile extracted by moving average (solid line). \label{Subfigure:PA_Stokes_number}]{\includegraphics[width=0.38\textwidth]{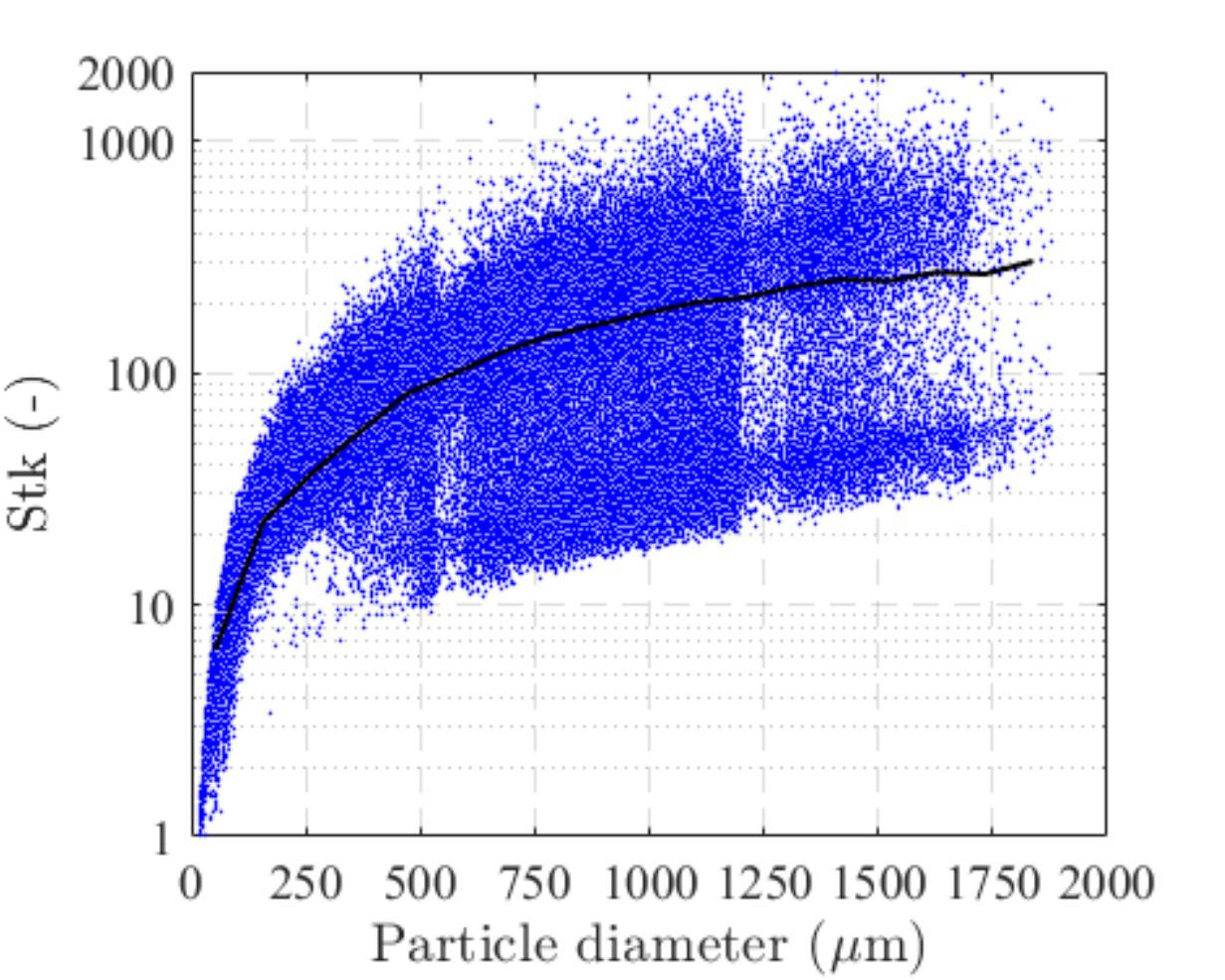}}
    \end{tabular}

    \caption{Kinematic behavior of \textit{Pellets Asturias} (biomass 1) at the onset of the deflagration test.}
    \label{Fig:Fig_PA_Kinematic_behavior_particles}
\end{figure}

Because temperature is closely related to the pressure development inside the vessel, Fig. \ref{Fig:Fig_PA_Particle_temperature} illustrates the particle temperature at different times during the explosion process. These snapshots depict a flame propagating radially from the interior to the walls of the sphere. As the mixture is ignited, the particles at the center heat up, evaporate their moisture content and release volatile gases. The volatile products then ignite and form an attached flame around the particle as oxygen diffuses into the products. The flame, in turn, heats the particles, increasing the rate of devolatilization. The released sensible energy continues heating the neighboring particles and causes a continued chain reaction. Once the volatiles have completely evolved from the particle, the remaining char reacts with the surrounding oxygen, releasing additional energy to the vessel. The figure suggests that the hottest particles are found in the outermost zone of the chamber. This happens because the smaller particles, which dry and react faster, are pushed towards the walls of the chamber during the dispersion process \cite{KALEJAIYE201046,DIBENEDETTO2013cfd,du2015visualization}. Notably, the picture also reveals an agglomeration of cold particles aligned vertically with the y-axis.\\

To explain this, and to further understand the kinematic behavior of the dust cloud prior to the onset of the deflagration test, Fig. \ref{Fig:Fig_PA_Kinematic_behavior_particles} illustrates the particle tracks and Stokes number map classified by particle diameter. At first sight, the front view of Fig. \ref{Subfigure:PA_particles_transparency} suggests the particles are well distributed in a radial direction, however the lateral view evidences that nearly all the particles are concentrated in the XY plane. This is because, although particles above $d_{p}> 200\mu\text{m}$ are less sensitive to the effect of high concentration at the walls caused by the well known two-vortex flow pattern \cite{ISLAS2022117033}, most of the particles in the PSD of \textit{Pellets Asturias} substantially surpass this size ($D_{10}=99.9$, $D_{50}=535.7$, and $D_{90}=1200.4 \, \mu\text{m}$). It is likely that these particles are not distributed homogeneously because their motion is not in equilibrium with the carrier phase.\\

Fig. \ref{Subfigure:PA_Stokes_number} advises that the Stk number rises asymptotically with increasing the particle diameter. A criterion of $\text{Stk}>1$ indicates that the particle momentum response time is larger than the fluid characteristic time scale, thereby suggesting that the surrounding eddies do not deflect the particle trajectories. In other words, most of the particles in the dust cloud adopt a ballistic behavior that is mostly influenced by particle-wall interaction in the XY plane.\\

\begin{figure*}[h]
    \centering
    
    \begin{tabular}{cc}
    \subfigure[Time evolution of the nominal dust concentration in the 20L sphere for \textit{Pellets Asturias}, $C_{0}=750$ g/m\textsuperscript{3}.\label{Subfigure:PA_Mass_time}]{\includegraphics[width=0.38\textwidth]{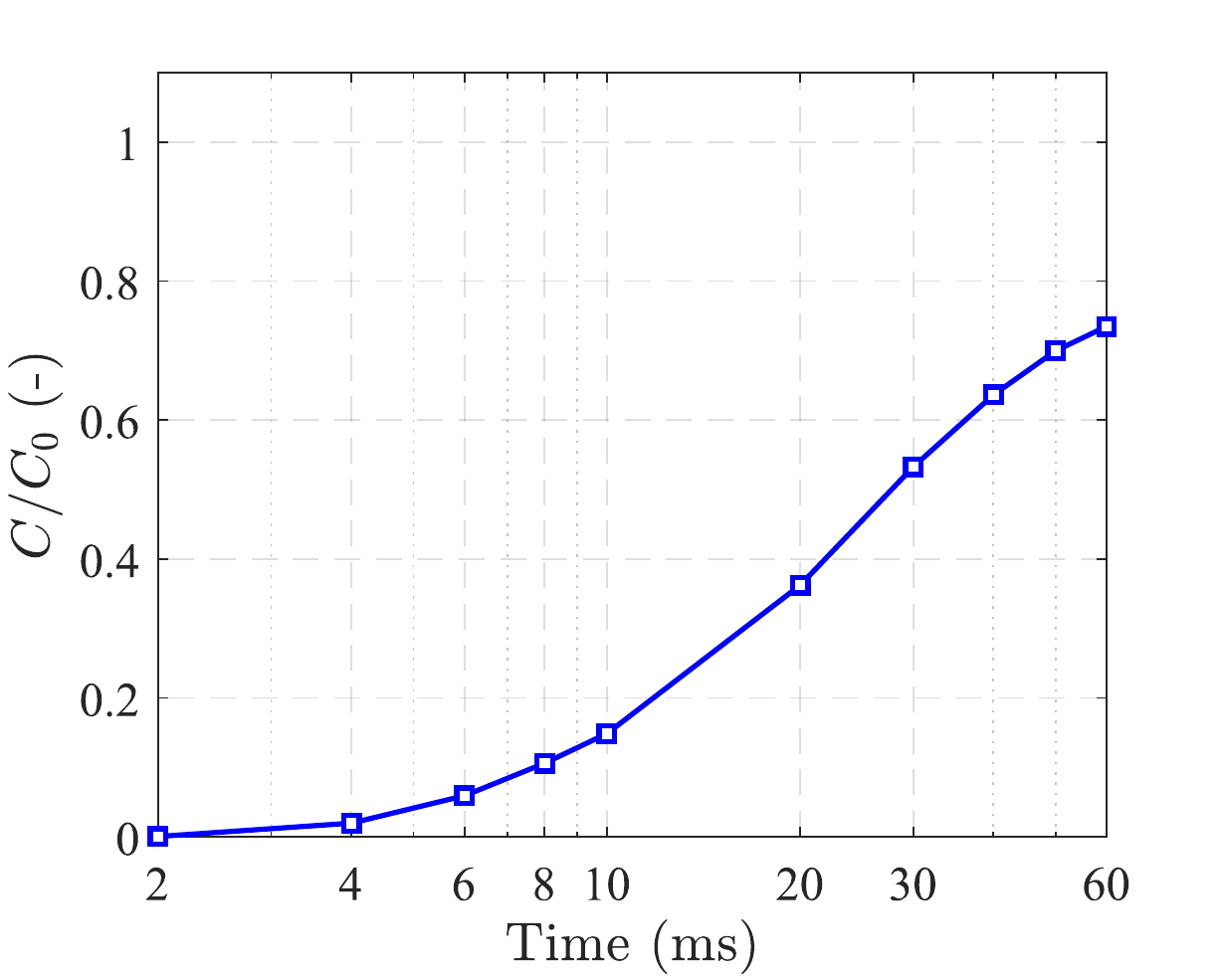}} &
    \subfigure[Comparison of particle size distribution of \textit{Pellets Asturias} measured before and after injection into the 20L sphere.\label{Subfigure:PA_PSD_afterDisp}]{\includegraphics[width=0.38\textwidth]{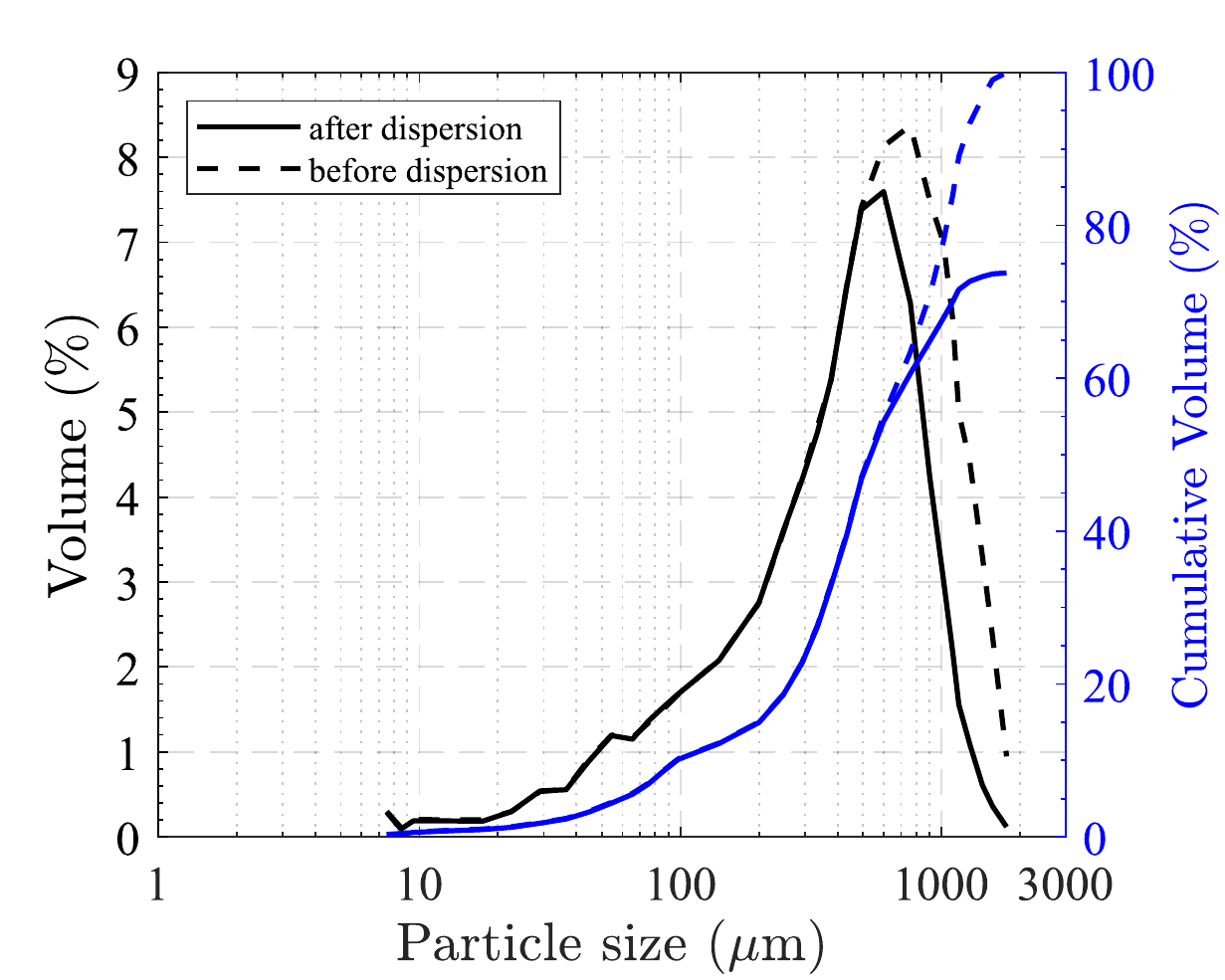}} 

    \end{tabular}
    
    \caption{Time evolution of the dust concentration and particle size distribution measured in the 20L sphere during the dispersion process of \textit{Pellets Asturias} (biomass 1).}
    \label{Fig:Fig_PA_dust_concentration_after_dispersion}
\end{figure*}

Fig. \ref{Subfigure:PA_particles_transparency} depicts that by the end of the ignition delay time, some particles are still on transit through connecting pipe. Given that the PSD considered here is above the recommended limits for dust explosion tests \cite{EN14034}, and that the length of $t_{d}$ determines the concentration of the dust dispersed in the chamber at the moment of ignition \cite{ASTME1226}, Fig. \ref{Fig:Fig_PA_dust_concentration_after_dispersion} quantifies the amount of dust concentration that is attained in the 20L sphere by the end of the dispersion process. First, Fig. \ref{Subfigure:PA_Mass_time} indicates that only about 73\% of the nominal dust concentration is reached in the sphere by the the time the explosive test is initiated. This corresponds to an effective dust concentration of $C_{\text{eff}}=547.5$ g/m\textsuperscript{3}. Second, Fig. \ref{Subfigure:PA_PSD_afterDisp} shows a comparison of the pre-dispersion PSD  and the post-dispersion PSD. While the left halves of both discrete and cumulative curves overlap, there is a clear reduction of particles with diameter above $500 \, \mu\text{m}$. Some of these particles remain adrift within the pipe of the dispersion system once the ignition delay time has been reached, and hence, do not enter the 20L sphere. This behavior can be attributed to various phenomena: (1) the increased inertia of large particles prevents them to follow evenly the motion of the carrier phase in the pipe, (2) by the end of $t_{d}$, the strength of $\nabla p$ has decayed significantly, and (3) the duration of $t_{d}$ is not sufficient to drive the remaining particles into the sphere. Note that this change in the post-dispersion PSD is only associated with re-sampling the particles that managed to enter the 20L sphere, and does not consider size reduction by fragmentation.\\

Although there is no consensus yet on the main mechanism responsible for particle fragmentation in the 20L sphere, particle fragmentation can occur due to a combination of the following mechanisms: (1) mechanical shear caused by the dispersion nozzle \cite{Cesana_manual}, (2) action of the outlet valve \cite{KALEJAIYE201046}, and (3) the baroclininc effect \cite{serrano2020experimental}. Moreover, recent studies suggest that there are other factors that may also play a role on the degree of particle break-up, e.g., the type of nozzle \cite{MURILLO201854}, hardness and fracture toughness of the sample \cite{bagaria2019classification}, and dust concentration \cite{miller2020comminution}. However, according to the breakage classification of Bagaria et al. \cite{bagaria2019classification}, among all the post-dispersion PSD measurements of pharmaceutical, carbonaceous and biomass samples, the latter dusts exhibited the lowest (very little or none) fragmentation during the dispersion process in various closed vessels. After all, findings of this CFD study suggest that particle size also plays a role on the legitimate PSD that enters the 20L sphere, specifically owed to the inertial effects between the gas-solid flow, and therefore the pre-dispersion PSD may not necessarily coincide with the post-dispersion PSD.\\

\begin{figure*}[b]
    \centering
    
    \begin{tabular}{cc}

    \subfigure[Temporal evolution of the particle mass yields and oxygen consumption during the explosion process. \label{Subfigure:PA_Particle_mass_transfer}]{\includegraphics[width=0.38\textwidth]{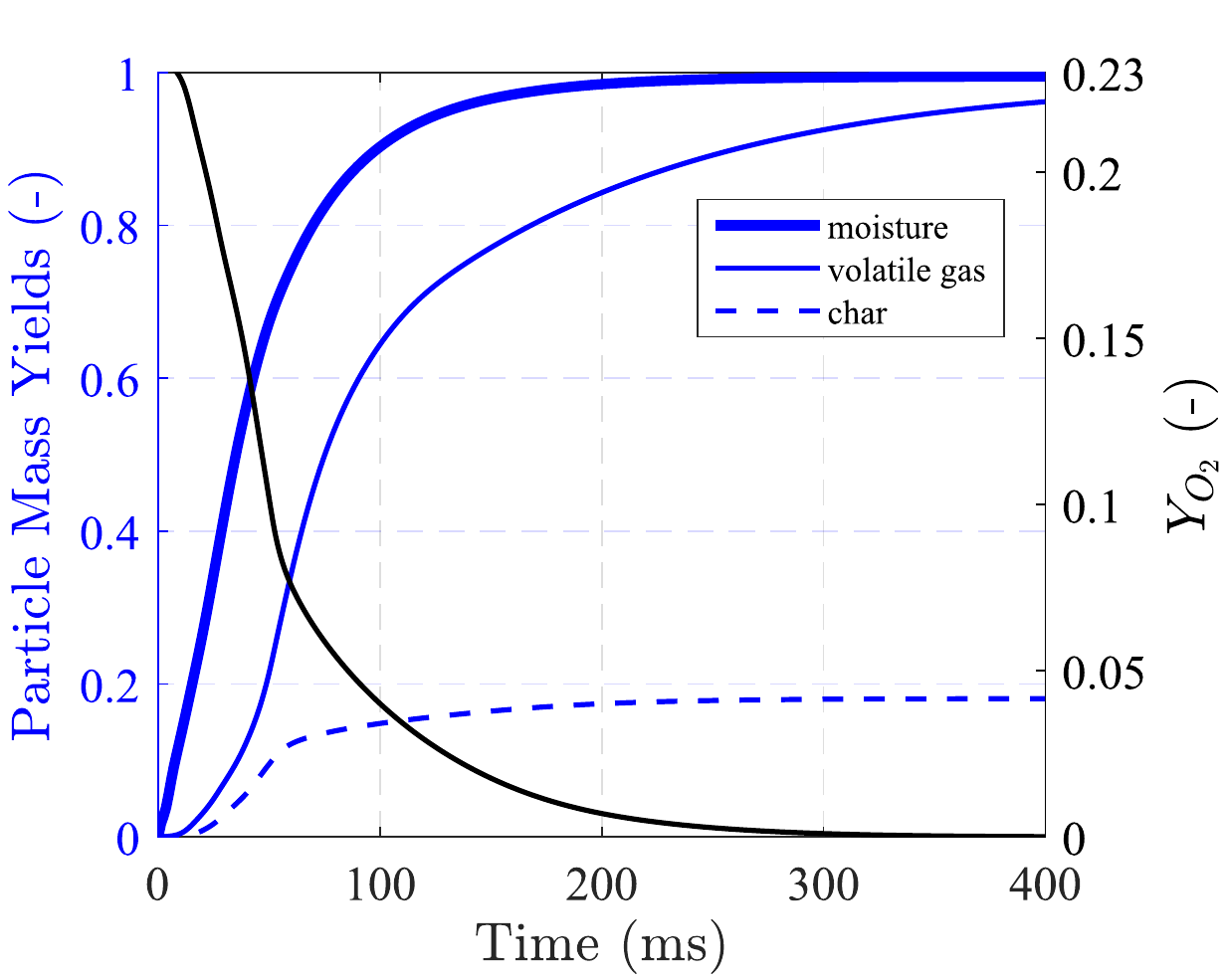}} &
    \subfigure[Char burnout and volatile conversion as function of particle diameter. Instantaneous data (dots) and profiles extracted by moving average (solid lines). \label{Subfigure:PA_Burnout_Volatile_Conversion}]{\includegraphics[width=0.38\textwidth]{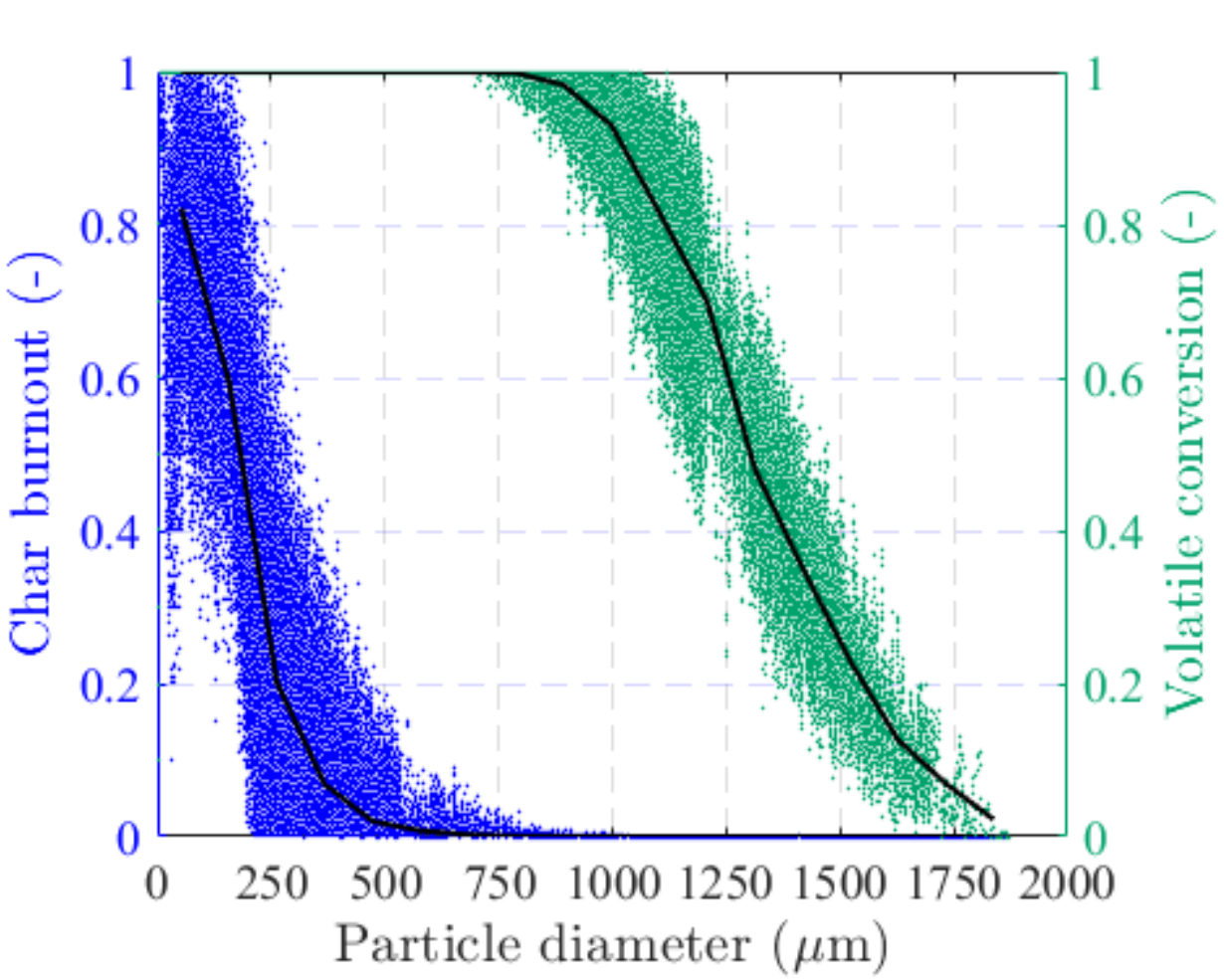}}
    
    \end{tabular}
    
    \caption{Consumption of each component in the biomass particle as function of time and particle diameter for \textit{Pellets Asturias} (biomass 1).}
    \label{Fig:PA_Particle_Mass_Yields_Burnout_Volatile_Conversion}
\end{figure*}

Given that the deflagration test of \textit{Pellets Asturias} (biomass 1) was conducted on the material in an as-received state from a process industry with a somewhat coarse PSD, Fig. \ref{Fig:PA_Particle_Mass_Yields_Burnout_Volatile_Conversion} quantifies the degree of consumption of each component in the biomass particle as a function of time and particle diameter. Fig. \ref{Subfigure:PA_Particle_mass_transfer} shows that the mass transferred from the particle to gas phase due to pyrolysis is dominant over that due to the surface reaction. This is congruent with experiments \cite{JIANG201845,liu2019explosion}, which suggest that once ignited, the overall burning rate of biomass is dominated by the rapid release and combustion of volatile gases.
Indeed, the combustion of volatile matter represents approximately 86 \% of the calorific value of this biomass sample, whereas the role of char oxidation on the energy release of this deflagration test is minor, as only about 20\% of the available char is deployed.\\

In addition, O\textsubscript{2} is completely consumed in about 300 ms, which limits further oxidation of the remaining carbon. Conversely, since devolatilization does not depend on diffusion, the release of volatile gases continues despite the consumption of oxygen, and is limited only by residence time and temperature. Fig. \ref{Subfigure:PA_Burnout_Volatile_Conversion} plots the char burnout and volatile conversion as function of $d_{p}$. Here, it can be seen that the mass loss due to both reactions decreases with increasing particle diameter. Only particles with $d_{p}<750\mu\text{m}$ release all the volatile content, being the ones that exhibit some degree of char consumption. The release of moisture reduces the heat and mass transfer to the particle surface, thus reducing the rate of mass loss of the particle (burning rate). For the largest particles, considerable time is required to heat these to the devolatilization temperature. Finally, since diffusion is generally the rate limiting process for the surface reaction of large particles \cite{ragland2011combustion}, burnout times scale with $\sim d_{p}^2$.  Note that due to modeling assumptions, the char contained within a parcel can not burn until its volatile gases content has been totally released.\\

The CFD results suggest that, in order to test a similar sample under the same concentration condition, and for a more explosive scenario a smaller particle size ($d_{p}<750 \mu\text{m}$) would be needed. Furthermore, since some particles exceeding a diameter of 500 microns may not enter the sphere during the dispersion process see Fig. \ref{Subfigure:PA_PSD_afterDisp}, basically owed to the increased particle inertial effects (i.e., $\mathcal{O}(\text{Stk})\sim 10^2$) which reduces the interaction between the carrier and the particles, the latter threshold value prevails as the closing recommendation. Overall, since the CFD model predicted reasonably well the transient behavior of the explosion process, the models and physical considerations described earlier can be assumed to reveal fairly well the explosion pressure of biomass dust.\\

\subsection{Validation of the peak pressures as function of dust concentration}
\label{Subsection:Validation_of_the_peak_pressures_as_function_of_dust_concentration}

\begin{figure}[h]
    \centering
    \includegraphics[width=0.48\textwidth]{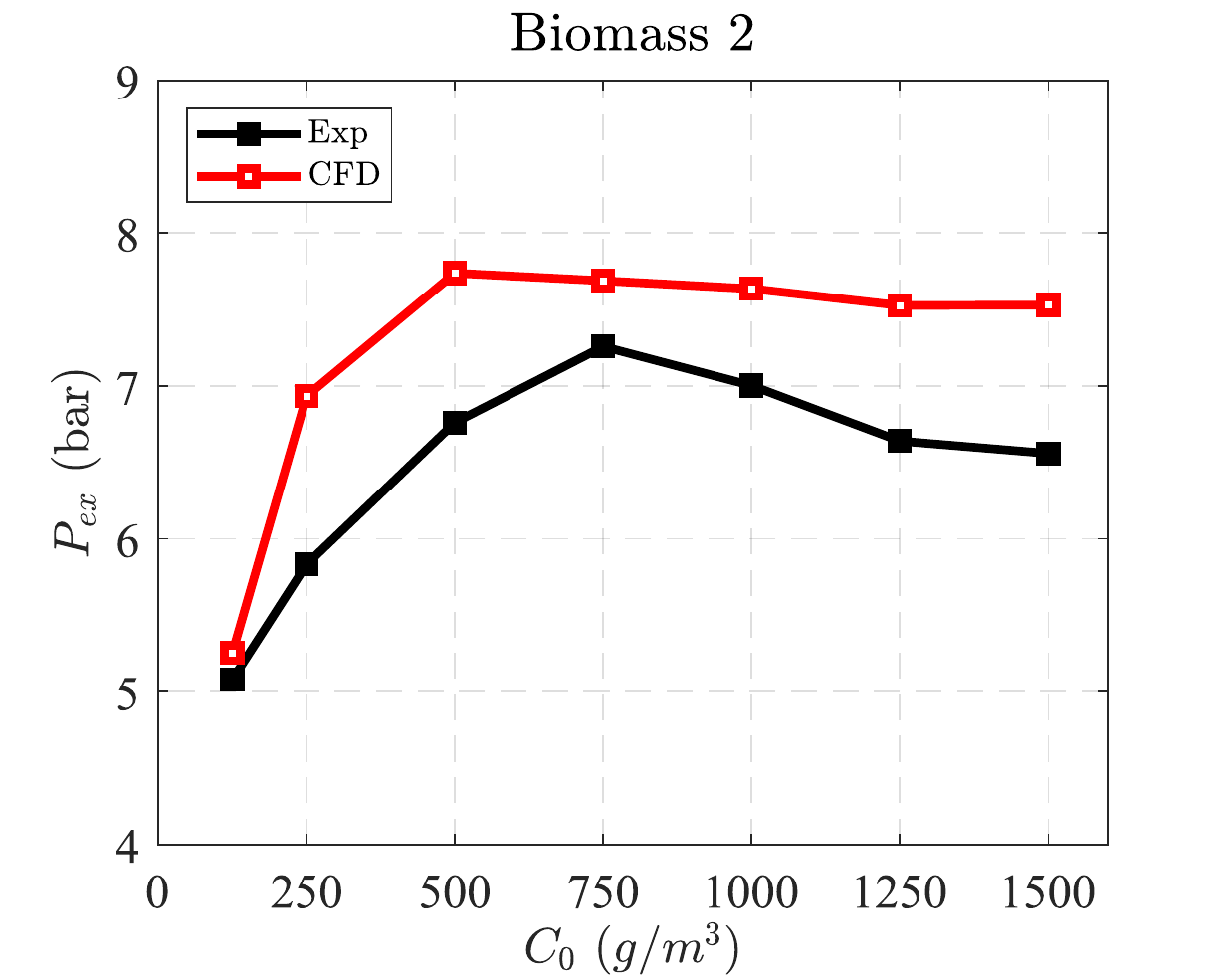}    
    \centering \caption{Comparison of experimental (Liu et al. \cite{liu2019explosion}) and CFD-predicted explosion pressures $P_{ex}$ as function of dust concentration for \textit{Cupressus Funebris} (biomass 2).}
    \label{Figure:CF_Exp_pts}
\end{figure}

Next, the assessment of model predictions over a wide range of dust concentrations is conducted. For this task, the biomass sample \textit{Cupressus Funebris} (biomass 2) was simulated following the same solution strategy. Fig. \ref{Figure:CF_Exp_pts} shows a comparison of the explosion pressures predicted by the model and the  experiments of Liu et al. \cite{liu2019explosion} for a range of concentrations between 125 to 1500 g/m\textsuperscript{3}.\\

In this case, $P_{ex}$ calculated by the CFD model moderately overestimate the experimental measurements. The minimum relative error in the explosion pressure is 2.86\% for the lowest concentration, while the maximum error is 16\% for $C_{0}=250 \, \text{g}/\text{m}\textsuperscript{3}$. This represents in a worst-case scenario, a 1 bar difference in the explosion pressure reported by Liu et al. \cite{liu2019explosion} and the one determined by the model. The increase in error can be attributed mainly to the fact that there is uncertainty in both the chemical composition of this biomass and the particle size distribution. On one hand, although the proximate and ultimate analyses of this same biomass species are reported in the literature by Shen \cite{shen2014emission}, the moisture content may vary 
depending on the process where the sample was collected. 
Moisture in the dust reduces both ignition sensitivity and the explosion violence of dust clouds markedly \cite{eckhoff2019dust}. Besides, given that only 3 percentiles of the size distribution were known, the rest of the distribution was assumed to follow a Rosin Rammler profile, which might not be necessarily true. On the other hand, in the same work, Liu et al. \cite{liu2019explosion} reported the explosion parameters in the 20L sphere for other two samples with no further details on the operating conditions of the experiments, specifically about the testing frequency.
According to the ASTM E1226 standard, a high testing frequency (20 to 40 explosions per day) can increase the chamber temperature by approximately 40 to 50$^{\circ}$C, which can reduce the explosion pressure by up to 15\% \cite{ASTME1226}.\\

\begin{figure}[h]
    \centering
    \includegraphics[width=0.48\textwidth]{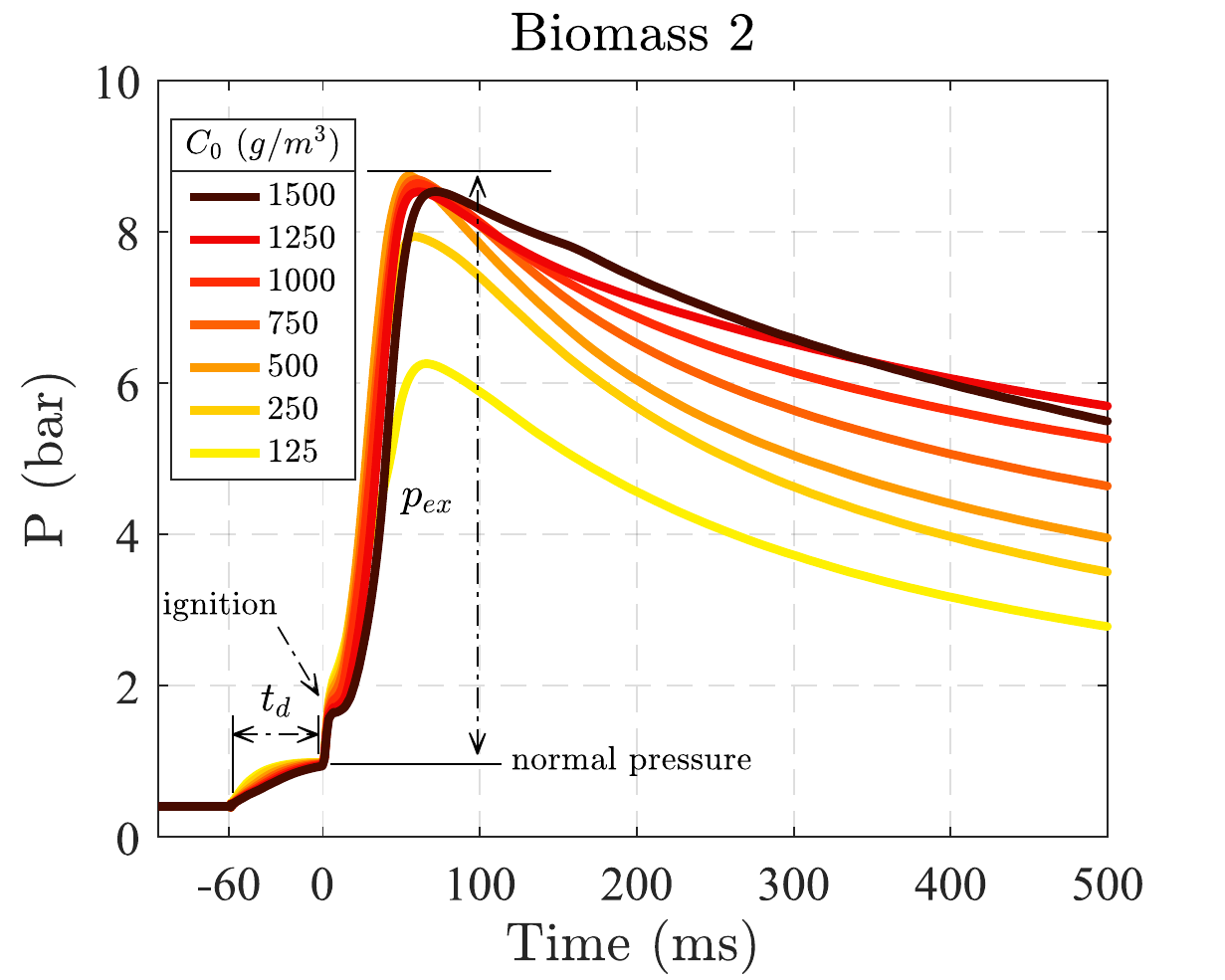}    
    \centering \caption{Comparison of the CFD-predicted pressure-time curves for the various dust concentrations of \textit{Cupresus Funebris} (biomass 2).}
    \label{Figure:CF_Exp_time}
\end{figure}

Despite these uncertainties, the model predicts that the maximum explosion pressure is $P_{max}=7.73$ bar, while the experimental test indicates $P_{max}=7.26$ bar. This is a relative error of 5.81\%, which is a reasonable deviation. Moreover, it is seen that the trend of the curve holds reasonably well, with an increase in the explosion pressure for the first 3 concentrations, and then a continuous decrease with successive concentrations. \\ 

Fig. \ref{Figure:CF_Exp_time} shows the pressure-time evolution during the explosion process of the various dust concentrations simulated in this study. Again, the pressure curves rise sharply during the time the ignition mechanism is active (0-25 ms). From here, the pressure increases almost linearly up to reaching $P_{ex}$ (40-60 ms), while decreases non-linearly at different rates. At low dust concentrations, the pressure decays faster than at high concentrations. This is because, for dense clouds the cooling effect of the walls cannot dissipate the thermal energy out of the chamber at the same rate as for dilute clouds.\\

\begin{figure*}[h]
    \centering
    \includegraphics[width=0.70\textwidth]{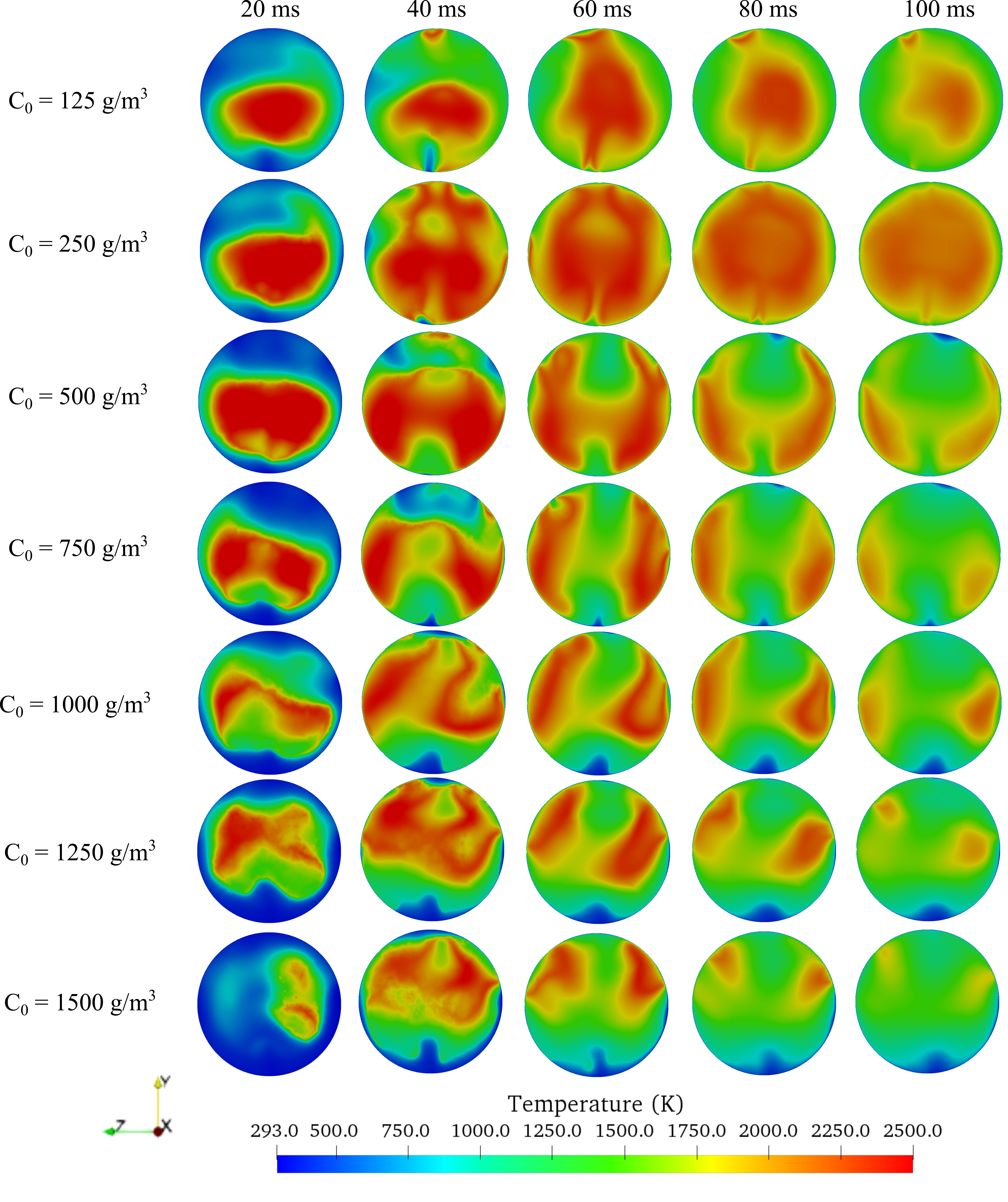}
    \caption{Snapshots of the flame evolution during the first 100 ms of the explosion process of \textit{Cupressus Funebris} (biomass 2).}
    \label{Fig:Fig_CF_Gas_Temperature}
\end{figure*}

As in the case of the first biomass, Fig. \ref{Fig:Fig_CF_Gas_Temperature} depicts the flame evolution during the first 100 ms of the explosion process for the different dust concentrations simulated. First, there is an apparent reduction of the flame temperature with increasing dust concentration beyond $C_{0}=500 \, \text{g}/\text{m}\textsuperscript{3}$. This is because as dust concentration increases, the particles act as heat sink consuming the available energy to heat up and get dried. Regardless of dust concentration, the energy release is limited by oxygen concentration, which is always 23\% (w.t.). This is not the case of single-phase mixtures (gases), where oxygen concentration is reduced by increasing the fuel concentration. \\

Second, the figure suggests that the flame spreads radially with an irregular morphology. This can be attributed to differences in the dust cloud patterns produced during the dispersion process. Only for the first two concentrations, the flame exhibits a somewhat uniform radial propagation, while for concentrations above 500 g/m\textsuperscript{3}, the flame stretches from an initial spheroid shape to a hatchet profile \cite{LI2020cornstarch}. In all cases, these snapshots advise that the maximum flame temperatures are reached between 40 to 60 ms, similar to the times when the explosion pressures are reached.\\

\begin{figure}[h]
    \centering
    \includegraphics[width=0.48\textwidth]{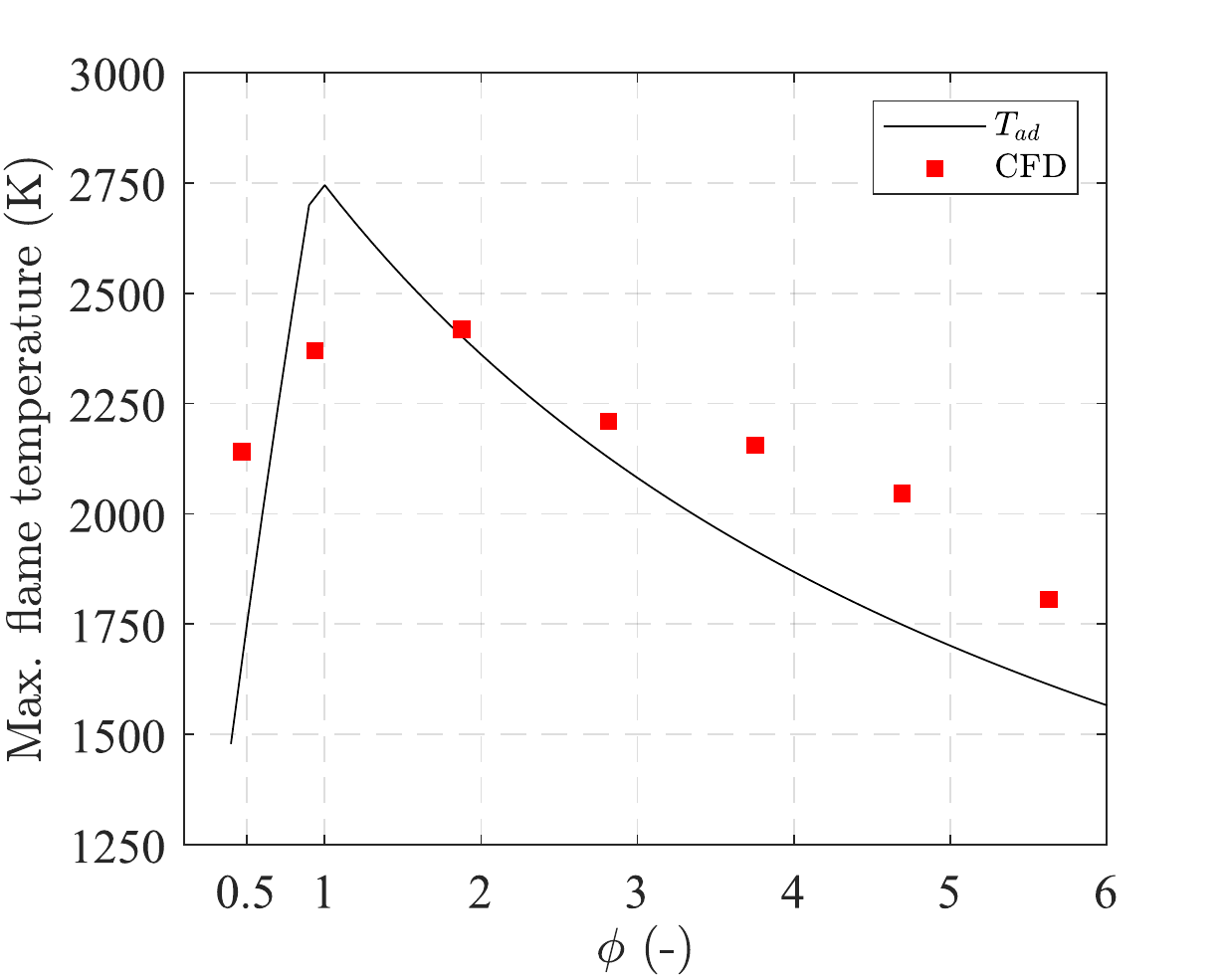}
    \caption{Maximum flame temperature as function of equivalence ratio for \textit{Cupressus Funebris} (biomass 2).}
    \label{Fig:Fig_CF_Adiabatic_Flame_Temperature}
\end{figure}

To further illustrate the role of dust concentration on the maximum flame temperature obtained during the deflagration process, Fig. \ref{Fig:Fig_CF_Adiabatic_Flame_Temperature} presents the maximum flame temperatures calculated by the CFD model as function of equivalence ratio $\phi$. Assuming that the postulate substance C\textsubscript{x}H\textsubscript{y}O\textsubscript{z} reacts to completion with oxygen, the equivalence ratio (ratio of the actual fuel-to-air (F/A) ratio to the stoichiometric (F/A)\textsubscript{st} ratio) can estimated from the following balance \cite{mcallister2011fundamentals, Lee2016}:

\begin{multline}
    \phi \text{C}\textsubscript{x}\text{H}\textsubscript{y}\text{O}\textsubscript{z} + \frac{4\text{x}+\text{y}-2\text{z}}{4}\left(\text{O}\textsubscript{2}+3.76\text{N}\textsubscript{2}\right) \\
    \rightarrow \text{x}\text{CO}\textsubscript{2}+0.5\text{y}\text{H}\textsubscript{2}\text{O} + 3.76\frac{4\text{x}+\text{y}-2\text{z}}{4}\text{N}\textsubscript{2} \\
    + \left(\phi-1\right)\text{C}\textsubscript{x}\text{H}\textsubscript{y}\text{O}\textsubscript{z}
    \label{Eqn:Stoich_reaction}
\end{multline}

Following the procedure illustrated by Ogle \cite{ogle2016dust}, the dust concentration that corresponds to $\phi=1$ was calculated at $C_{\text{st}}=266.39 \, \text{g}/\text{m}\textsuperscript{3}$.\\
 
The adiabatic flame temperature for a combustible dust is a function of the equivalence ratio. In the present, a pseudo-adiabatic flame temperature is included in Fig. \ref{Fig:Fig_CF_Adiabatic_Flame_Temperature} for the sake of comparison with the maximum flame temperatures predicted by the model. The pseudo-adiabatic flame temperature was calculated using a constant specific heat approach, and neglecting dissociation effects \cite{mcallister2011fundamentals,ogle2016dust}.\\

\begin{figure}[h]
    \centering
    \includegraphics[width=0.34\textwidth]{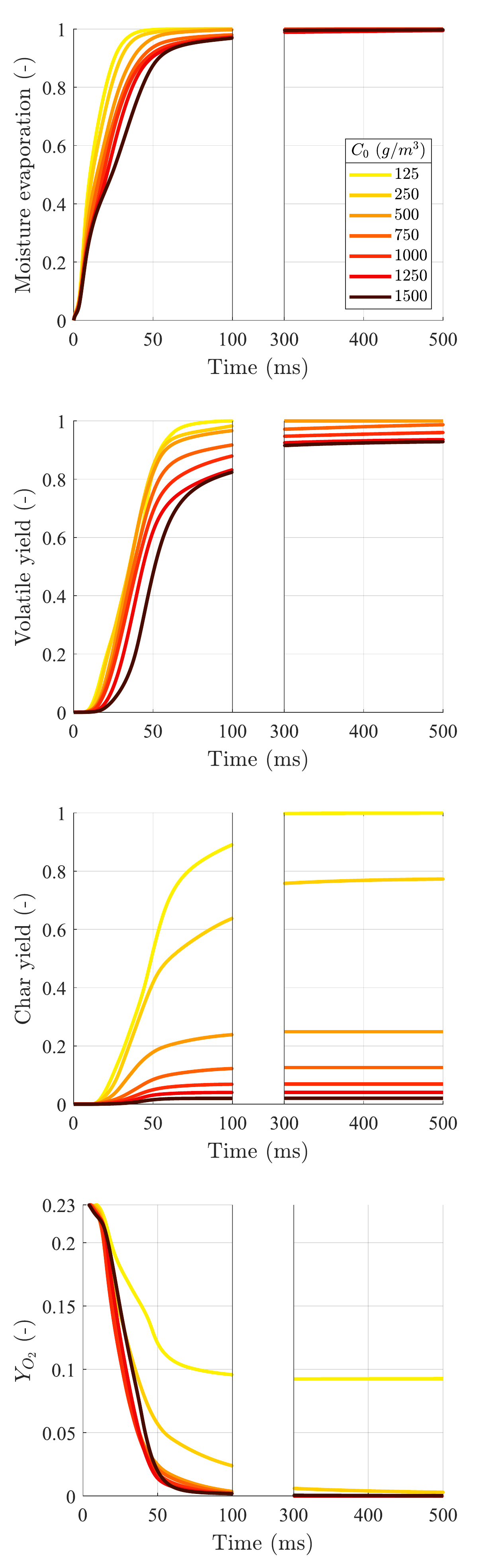}    
    \centering \caption{Comparison of the temporal evolution of the particle mass yields and oxygen consumption during the explosion process of \textit{Cupressus Funebris} (biomass 2). Note the break and change of scale in the x-axis.}
    \label{Fig:Fig_CF_Particle_mass_transfer}
\end{figure}

Here it is observed that flame temperature dependency on dust concentration behaves in a similar fashion as $P_{ex}$ does. At fuel-lean conditions ($\phi<1$), the maximum flame temperature increases with equivalence ratio, while at fuel-rich conditions ($\phi>1$) decreases nonlinear with successive values of $\phi$ \cite{ogle2016dust}. However, for the CFD case the absolute max. flame temperature is given under a slightly fuel-rich condition, corresponding to $1<\phi<1.8$, in a similar trend to the maximum explosion pressure ($P_{max}$). Following this approach, it can be assured that the optimum concentration for $P_{max}$ will be somewhere between 250 and 500 g/m$^{3}$, which cannot be precisely calculated without conducting intermediate simulations. This is in agreement with the explosion experiments of biomass reported by Huéscar Medina et al. \cite{HUESCARMEDINA201591} and Lee et al. \cite{Lee2016}.\\

This can be explained by the fact that under fuel-rich combustion of multi-phase (gas-solid) mixtures, the particles may react only partially, where the thermal histories depend strongly on size effects. Moreover, at such conditions the combustion process tends to create products of incomplete combustion. For organic fuels, this means the production of carbon monoxide and other species, therefore the optimal dust concentration will be larger than the stoichiometric concentration. These results reaffirm the fact that the criterion of fuel lean and rich conditions in dust explosions shall not be same as the criterion for gas combustion \cite{Lee2016}.\\

Next, Fig. \ref{Fig:Fig_CF_Particle_mass_transfer} shows a comparison of the time evolution of the mass transferred from the discrete to the gaseous phase for the full range of concentrations. It can be seen that in all cases, 100\% of the moisture is evaporated, while for concentrations above or equal to 250 g/m\textsuperscript{3} both the volatile matter and char content react partially. This is because biomass combustion is limited by the availability of oxygen, which, as explained above, is always 23 \% (by weight). Note that the rate of the volatile yield curves resemble the corresponding rates of oxygen consumption. Again, this behavior suggests that biomass combustion is dominated by the rapid release and combustion of volatile gases. Contrarily, since the rates of the char yield are smoother, the CFD model suggests that the surface reaction play a secondary role in the deflagration process. As dust concentration increases, the char yield decreases drastically from 1 (100\% burnout state) to a minimal value of 0.02 for the highest concentration. Moreover, from these curves it can be said then, that after 300 ms the particles no longer interact chemically with the fluid flow, and experience inert heating only.\\

\begin{figure*}[h]
    \centering
    \includegraphics[width=0.65\textwidth]{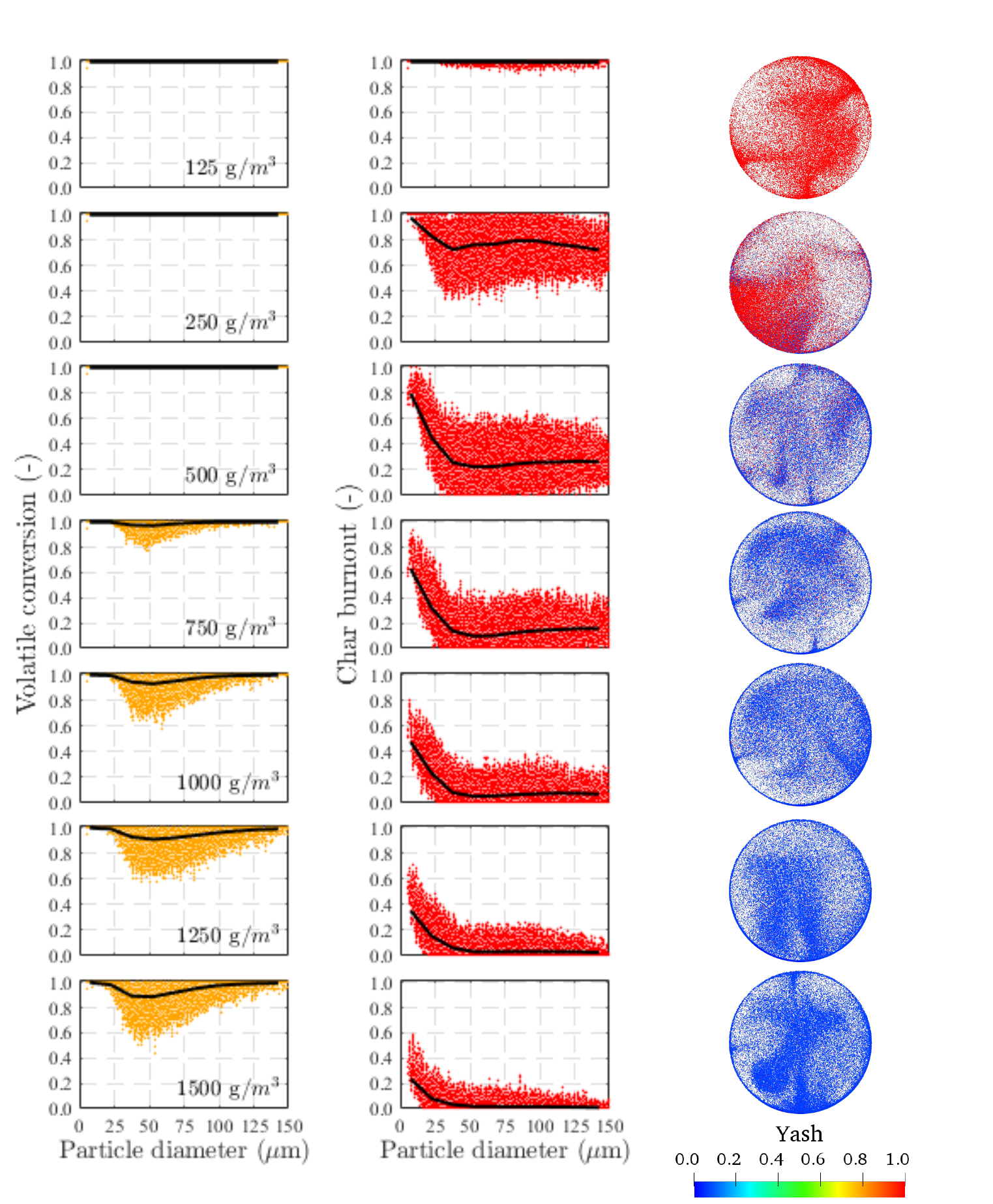}    
    \centering \caption{Comparison of volatile conversion and char burnout as function of particle diameter for the various dust concentrations of \textit{Cupressus Funebris} (biomass 2). Instantaneous data (dots) and profiles extracted by moving average (solid lines). On the right, particle tracks colored by ash fraction.}
    \label{Fig:CF_fuel_consumption}
\end{figure*}

Lastly,  Fig. \ref{Fig:CF_fuel_consumption} plots the char burnout and volatile conversion as function of particle diameter. It can be seen that, starting with a concentration of 750 g/m\textsuperscript{3}, the volatile conversion decreases slightly for a particle size range between 25 and 75 $\mu$m. Although devolatilization does not depend on particle size, this occurs because, according to the size distribution, it is in this range where the greatest amount of dust is concentrated. Therefore, it can be stated that the volatile conversion is not 100\% complete not because the particle size is too large, but because the cloud is dense in this range. However, even for the case with the highest dust concentration, a significant amount of volatile content is released into the fluid phase, which in all cases is capable of igniting the gaseous flame that sustains the deflagration.\\

On the other hand, only for the case with a concentration of 125 g/m\textsuperscript{3}, a state of complete char burnout is reached. This is because it is the only concentration with enough oxygen to consume all the fuel, while for concentrations above 250 g/m\textsuperscript{3}, the char burnout is partial, with particles smaller than 50 microns being the most reactive. Note that above this threshold value, the char depletion stagnates at specific values for each concentration, independent of particle size. This suggests that char content is consumed equally for particles in the range of 50 to 150 $\mu$m. \\

The particle tracks included in the figure give a qualitative impression of the degree of total fuel consumption of the dust cloud. The red color represents that the particles have consumed all the char content, reaching an ash fraction of 1.0, while the particles in blue have not yet experienced the surface reaction, thus the ash fraction is 0. Note that for the concentration of 250 g/m\textsuperscript{3}, there is a slight layer of blue particles in the vicinity of the walls. Contrarily to the PSD of \textit{Pellets Asturias}, the size distribution of \textit{Cupressus Funebris} is substantially smaller than the former, thus more prone to be affected by the two-vortex flow pattern during the dispersion process. This confirms that the effect of the two-vortex flow pattern caused by the rebound nozzle prevails even during the explosion process. Therefore, when performing explosivity tests, the degree of mixing during the formation of the dust cloud should be carefully scrutinized to assess the further behavior of the dust explosion in other geometries.\\

\subsection{Role of ignition delay time on the explosion behavior of biomass}
\label{Subsection:Role_of_ignition_delay_time_on_the_explosion_behavior_of_biomass}

\subsubsection{Aspects to consider during the dispersion process}
\label{SubSubsection:Aspects_to_consider_during_the_dispersion_process}

Turbulence is generally accepted to play an important role in the propagation of dust explosions \cite{amyotte1988effects,pu1991turbulence,bradley1989burning,SONG2020429}. Pre-ignition turbulence is caused by the air blast which disperses the dust particles into the chamber. In dust explosion testing in the 20L sphere, turbulence can be adjusted by varying the ignition delay time, which has been agreed to $t_{d}=60 \,\text{ms}$ since the establishment of the ASTM E1226 or EN 14034 standards \cite{ASTME1226,EN14034}. This value is meant to reproduce the same turbulence levels found in the 1m\textsuperscript{3} explosion chamber after a dispersion time of 600ms. However, latter experimental studies found that the turbulence levels between the two vessels were indeed different. Pu et al. \cite{pu1991turbulence} used hot wire anemometer (HA) to determine that an ignition delay time of 200 ms should be used in the 20L sphere instead. Similarly, van der Wel et al. \cite{van1992interpretation} used HA to suggest that turbulence levels between the two vessels was equal when $t_{d}$ was adjusted to 165 ms in the 20L sphere. More recently, Dahoe et al. \cite{dahoe2001transient} used two-dimensional laser Doppler anemometer (LDA) to report that $t_{d}$ should be modified to about 200 ms.\\

Although an ignition delay time of 60 ms introduces higher turbulence levels, dust explosion results obtained by the 20L sphere are usually unassailable because they are on the "safe side" \cite{van1992interpretation}. Experiments demonstrate that at higher turbulence levels, the severity of the explosion parameters increases \cite{bartknecht1989dust,eckhoff2003dust}. In practice, dust explosions in the process industries occur in very different geometries than the standardized vessels and under a wide range of turbulence conditions. Therefore, the last section of this work is devoted to study the effect of the ignition delay time on the explosion pressure of \textit{Cupressus Funebris} (biomass 2). Namely, the ignition delay times of 30, 90 and 120 ms are considered, while results are compared to those obtained under the standard value of $t_{d} = 60 \,\text{ms}$.\\

\begin{figure}[h]
    \centering
    \includegraphics[width=0.43\textwidth]{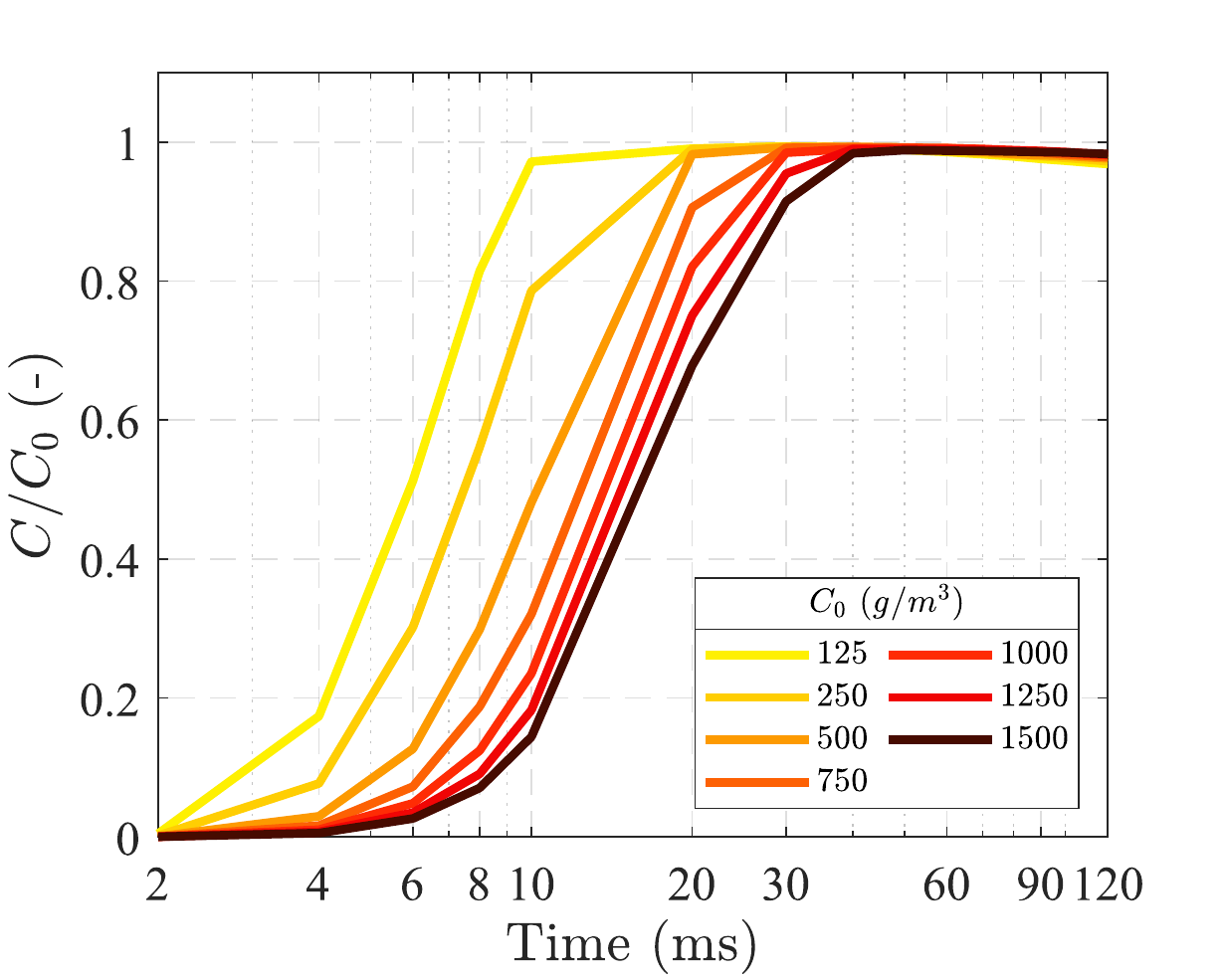}    
    \centering \caption{Time evolution of the nominal dust concentration attained in the 20L sphere during the dispersion process of \textit{Cupressus Funebris} (biomass 2).}
    \label{Figure:CF_mass_time}
\end{figure}

\begin{figure}
    \centering
    \includegraphics[width=0.43\textwidth]{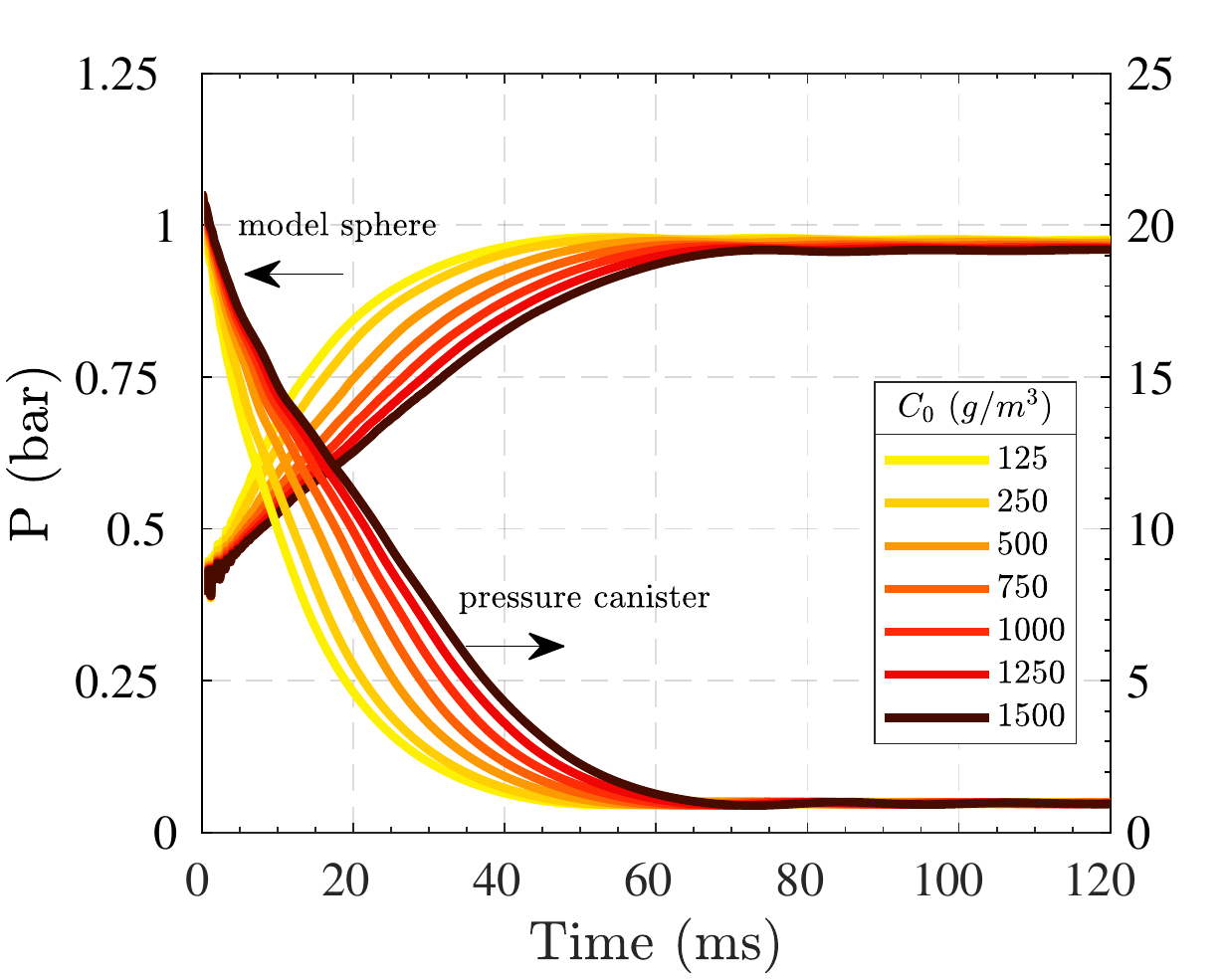}
    \caption{Comparison of the time-evolution of the pressure rise in the 20L sphere and pressure decrease in the canister for the various dust concentrations of \textit{Cupressus Funebris} (biomass 2).}
    \label{Figure:Fig_CF_dispersion_pressure}
\end{figure}

First, considering that the amount of dust concentration that is attained in the 20L sphere during the dispersion process is function of $t_{d}$, Fig. \ref{Figure:CF_mass_time} shows the time-evolution of the normalized dust concentration for the various concentrations considered in the previous section. For all cases, the higher the dust concentration, the longer it takes for the particles to enter from the canister to the sphere. The greatest difference in the mass filling occurs for a time $t=10 \,\text{ms}$, where almost 95\% of the mass for $C_{0}=125 \,\text{g}/\text{m}\textsuperscript{3}$ has entered the sphere, while for $C_{0}=1500 \,\text{g}/\text{m}\textsuperscript{3}$, this percentage is only $\sim16\%$. From here, particles continue entering progressively until 40 ms, time at which all nominal concentrations are reached. However, note that at 30 ms, the two highest concentrations $C_{0}=1250 \,\text{g}/\text{m}\textsuperscript{3}$ and $C_{0}=1500 \,\text{g}/\text{m}\textsuperscript{3}$ are not fully reached, as there are particles still on transit from the canister and the tube. Contrarily to the case of \textit{Pellets Asturias} (biomass 1), where the nominal dust concentration was not reached because large particles did not enter the sphere ($d_{p}>500 \,\mu\text{m}$) at $t_{d}=60 \,\text{ms}$, these results confirm that when performing explosion tests at an ignition delay time of 30 ms, concentrations above 1000 g/m\textsuperscript{3} may not be fully discharged into the 20L sphere.\\ 

\begin{figure}
    \centering
    \includegraphics[width=0.43\textwidth]{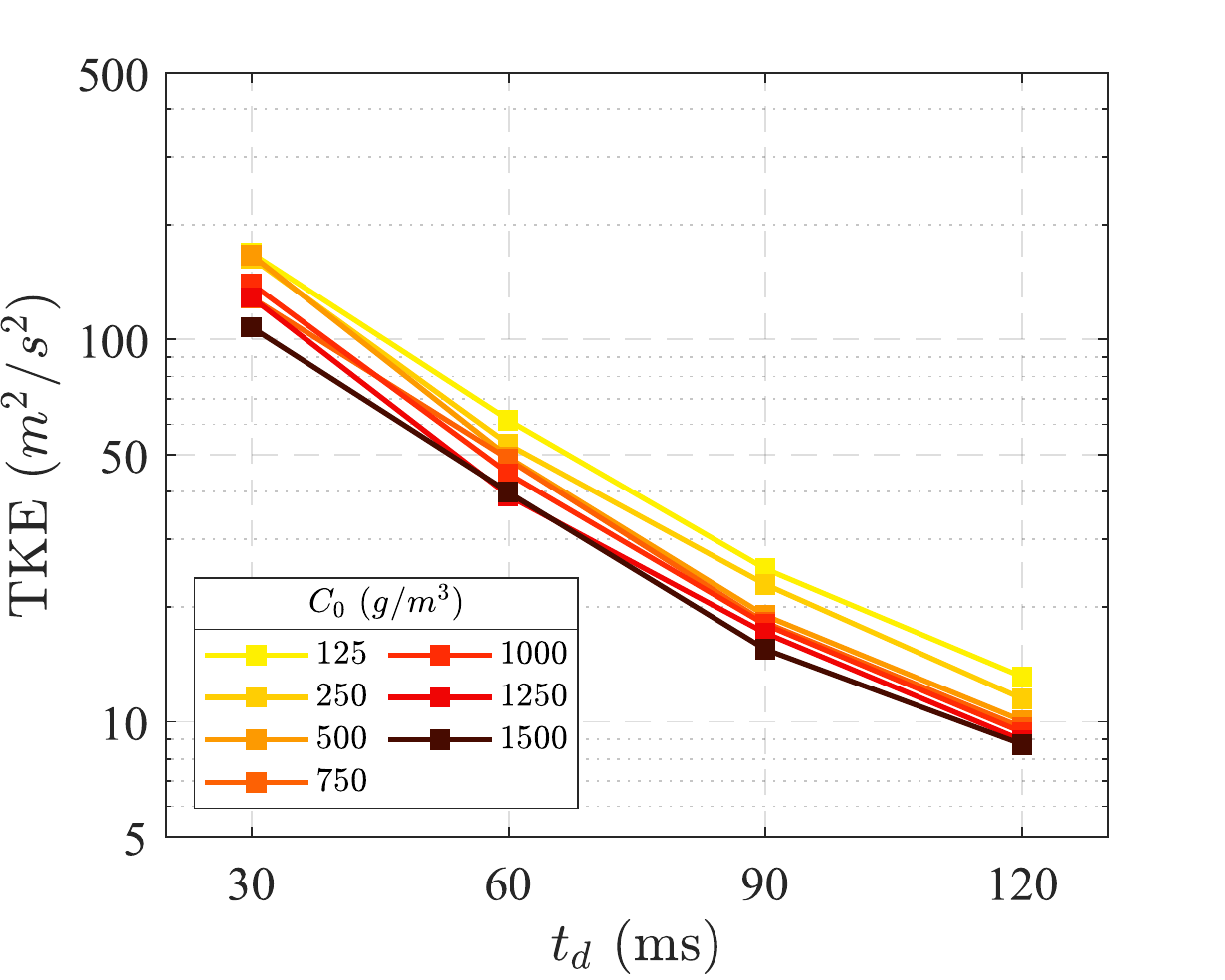}    
    \centering \caption{Comparison of TKE at the end of various ignition delay times $t_{d}$ for the full range of concentrations of \textit{Cupressus Funebris} (biomass 2).}
    \label{Figure:CF_TKE_ign_delay_time}
\end{figure}

Second, the time-evolution of the pressure rise in the 20L sphere and the pressure decrease in the dust container during the dispersion process is shown in Fig. \ref{Figure:Fig_CF_dispersion_pressure}. Again, there is a apparent dependence of the rates of pressure rise and pressure decrease on the dust concentration. This happens because the higher the concentration, the more particles hinder the flow from the canister to the sphere, thus slowing down the rate of pressure change in both reservoirs. This is a critical aspect to consider when performing experiments at $t_{d}<60 \,\text{ms}$ because the pressure in the sphere may not be 1 bar exactly. For instance, when $t_{d}=30 \,\text{ms}$ the pressure at the time of ignition is $ p_{0} = 0.92 \,\text{bar}$ for the lowest concentration, while $p_{0} = 0.73 \,\text{bar}$ for the highest dust concentration. This condition advises that the explosion pressure $P_{ex}$ should be measured from different reference points. Moreover, igniting the dust-air mixtures at pressures below 1 bar may affect the mass transfer rates, especially due to evaporation, as phase change of moisture from liquid to gaseous state depends strongly on pressure. At $p_{0} < 1 \,\text{bar}$ the evaporation point of moisture is reduced, so at least for the time in which $p_{0}$ keeps below atmospheric pressure, the biomass conversion will proceed faster as compared to dust explosions tests performed at $t_{d} = 60\,\text{ms}$.\\

Continuing with the analysis of the conditions prior to any ignition, Fig. \ref{Figure:CF_TKE_ign_delay_time} shows the turbulent kinetic energy (TKE) in the sphere as function of the ignition delay time for the various dust concentrations. Here it can be seen that the TKE reduces log-linearly with increasing $t_{d}$. This observation is consistent with the experiments of Dahoe et al. \cite{dahoe2001transient}, who proposed an exponential correlation for the decay of the pre-ignition turbulence. For more details on the time-evolution of the TKE predicted by our CFD model, refer to our previous work \cite{ISLAS2022117033}. In addition, this plot suggests that for all cases the turbulent kinetic energy decreases with increasing dust concentration, similar to the observations of Di Sarli et al. \cite{di2014cfd}. This can be explained by the fact that, the higher dust concentration, the increased dissipation due to particle drag \cite{balachandar2010turbulent}, which along with the increased inertial effects of dense particle-laden flows, both contribute to the turbulence modulation phenomenon.\\ 

\begin{figure}[h]
    \centering
    \includegraphics[width=0.49\textwidth]{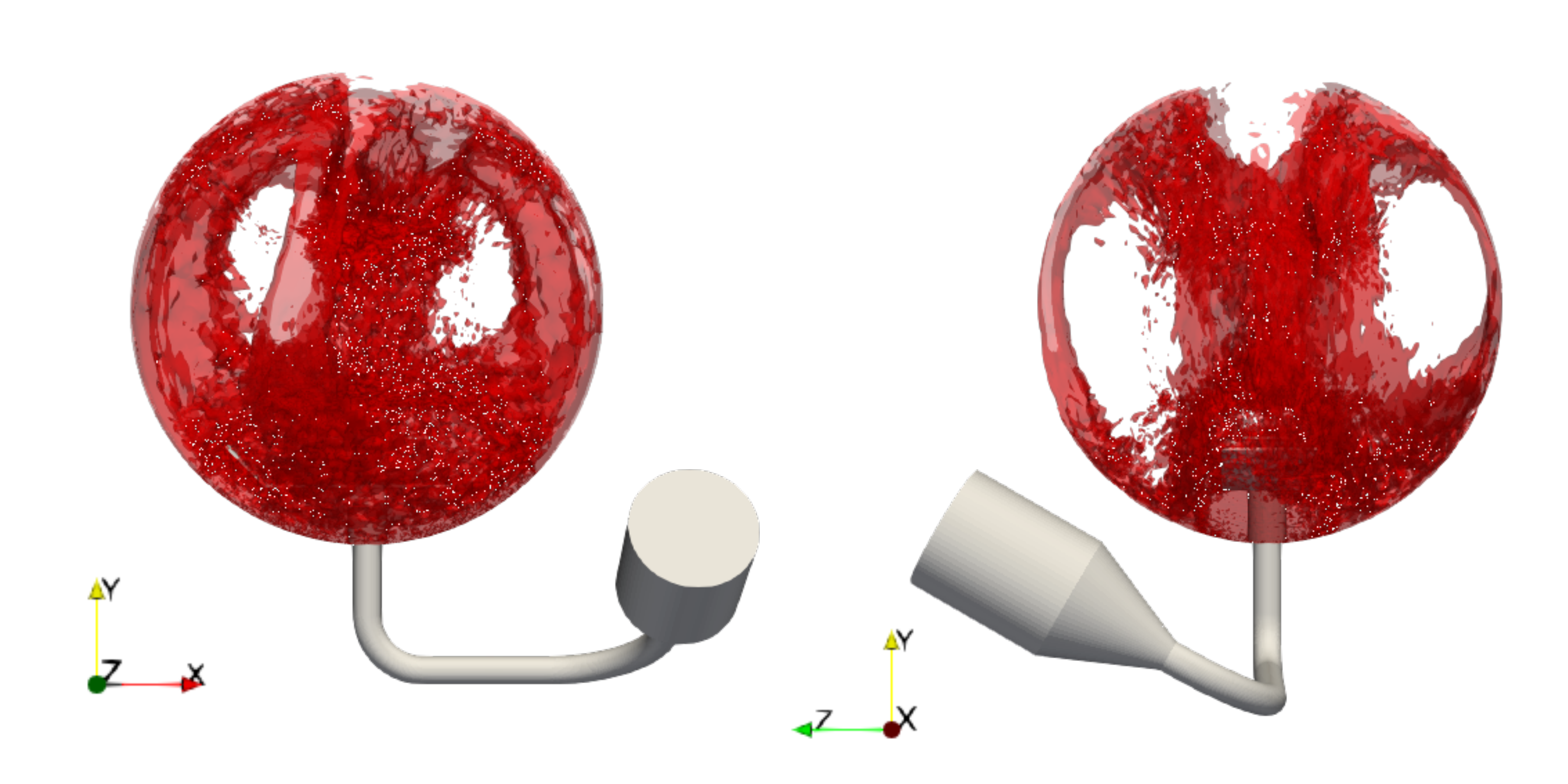}
    \caption{Snapshot of the spatial distribution of the dust cloud ($C_{0}=250 \,\text{g}/\text{m}^3$) in the 20L sphere at end of the dispersion process of \textit{Cupressus Funebris} (biomass 2).}
    \label{Fig:Fig_CF_distribution_transparency}
\end{figure}

Since Sh and Nu numbers scale with $\text{Re}\textsubscript{p}^{1/2}$, turbulence speeds-up both the mass and heat transfer rates between the reactive particles and the flow. This implies that at shorter ignition delay times the diffusion-controlled reactions (moisture evaporation and char burning) will proceed faster, thus increasing the rate of pressure rise during the course of the explosion process. Yet, there is another way in which turbulence may influence the combustion of the dust particles in the 20L sphere. Traditionally, turbulence is thought to disperse particles and to act as a source of increased particle diffusion that smooths sharp gradients in the particle concentration fields \cite{swaminathan2022advanced}. However, during the dispersion process in the 20L chamber, the two-vortex flow pattern has already been recognized as a mechanism that promotes non-homogeneous mixing of the dust cloud. Many CFD studies have revealed a preferential dust concentration towards the wall \cite{DIBENEDETTO2013cfd,murillo2016cfd,portarapillo2020cfd,ISLAS2022117033}. Fig. \ref{Fig:Fig_CF_distribution_transparency} gives a qualitative impression of the spatial distribution of the dust cloud at the end of the standard ignition delay time $t_{d}=60\,\text{ms}$.\\ 

From Fig.\ref{Fig:Fig_CF_distribution_transparency} it can be clearly seen that there are regions with practically no particles present. These hollow regions coincide with the zones where the two-vortex flow pattern develops. The formation of these vortices is due to the design of the rebound nozzle and the spherical shape of the 20L vessel, which create the two capsules of recirculating flow. In our previous work \cite{ISLAS2022117033}
we have shown that, depending on the particle inertial effects (particle Reynolds and Stokes numbers) these vortices may promote the increased particle concentration at the wall or not. Consequently, as the most of the particles in the PSD of \textit{Cupressus Funebris} are smaller than $100 \,\mu\text{m}$ (low particle inertia), the dust cloud is considerably affected by this circumstance. \\

\begin{figure}[h]
    \centering
    \includegraphics[width=0.33\textwidth]{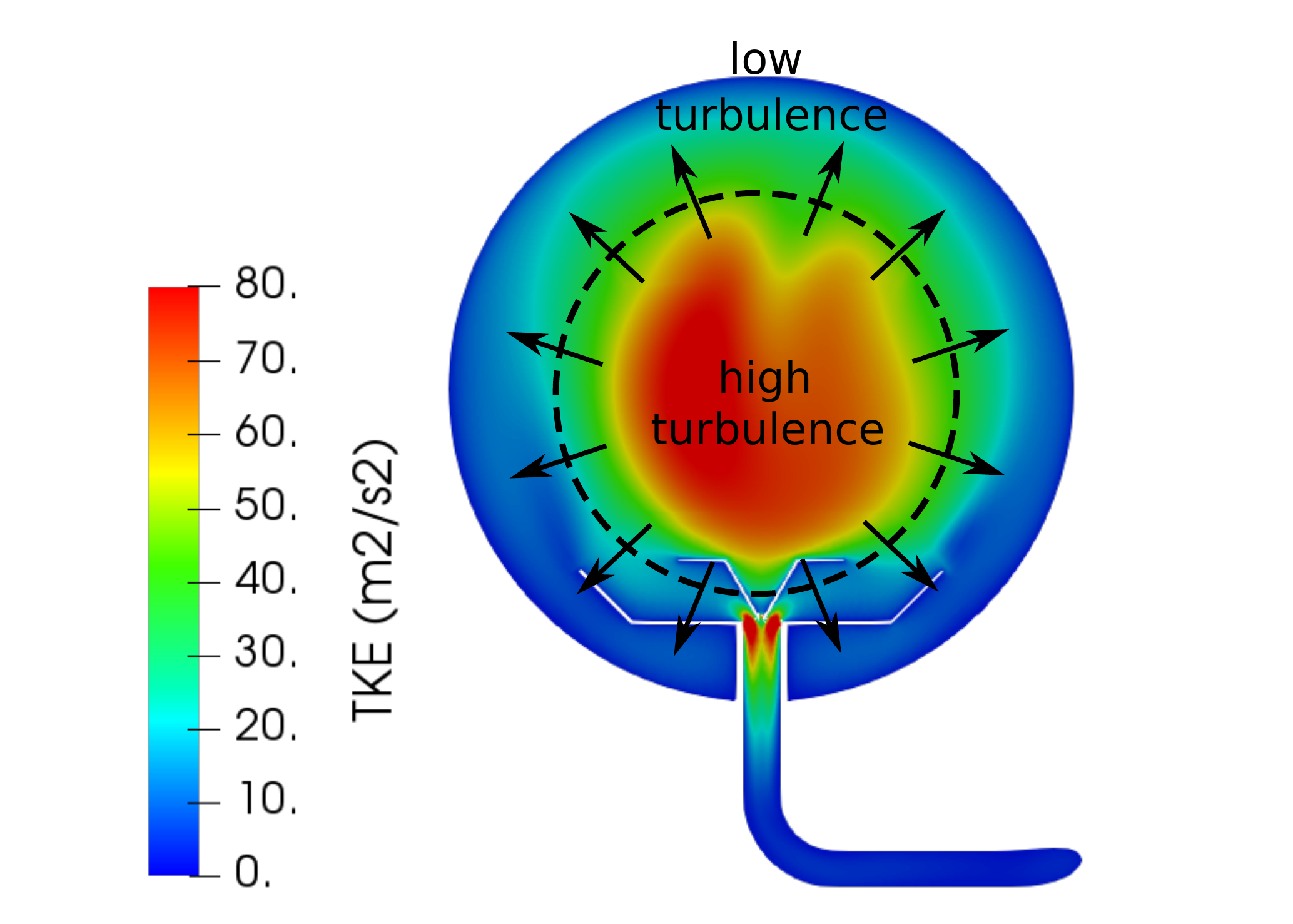}
    \caption{Illustration of the turbophoresis effect during the dispersion process ($C_{0}=250 \,\text{g}/\text{m}^3$) of \textit{Cupressus Funebris} (biomass 2).}
    \label{Fig:Fig_CF_TKE_contour}
\end{figure}

One of the mechanisms for transport of particles towards a wall is caused by the in-homogeneity of the turbulent velocity fluctuations and is called turbophoresis \cite{reeks1983transport}. This phenomenon is driven by a differential in turbulent dispersion rates between different regions of a flow. Particles in regions with higher turbulent intensity disperse more quickly than those in more quiescent regions, causing particles to accumulate with longer residence times and higher concentrations in regions of lower turbulence intensity \cite{johnson2020turbophoresis}. Since the velocity fluctuations are directly related to the turbulent kinetic energy, Fig. \ref{Fig:Fig_CF_TKE_contour} illustrates the turbophoresis effect by depicting the TKE contour at the end of $t_{d}=60 \,\text{ms}$. The contour shows that there is a region of high TKE located at the center of the sphere. This is result of the vigorous activity of the recirculating flow in this zone. Then, the TKE slowly fades out as it propagates radially in an outward direction. In wall-bounded turbulent flows, the no-slip condition cause turbulence intensity to vanish at solids boundaries, resulting in sharp gradients of turbulence intensity and turbulent kinetic energy in the viscous sublayer and buffer region \cite{marchioli2002mechanisms}. Turbophoresis then, may increase the mean particle concentration at the wall even up to a thousand times the bulk value \cite{swaminathan2022advanced}. \\

\begin{figure*}[h]
    \centering
    \includegraphics[width=0.85\textwidth]{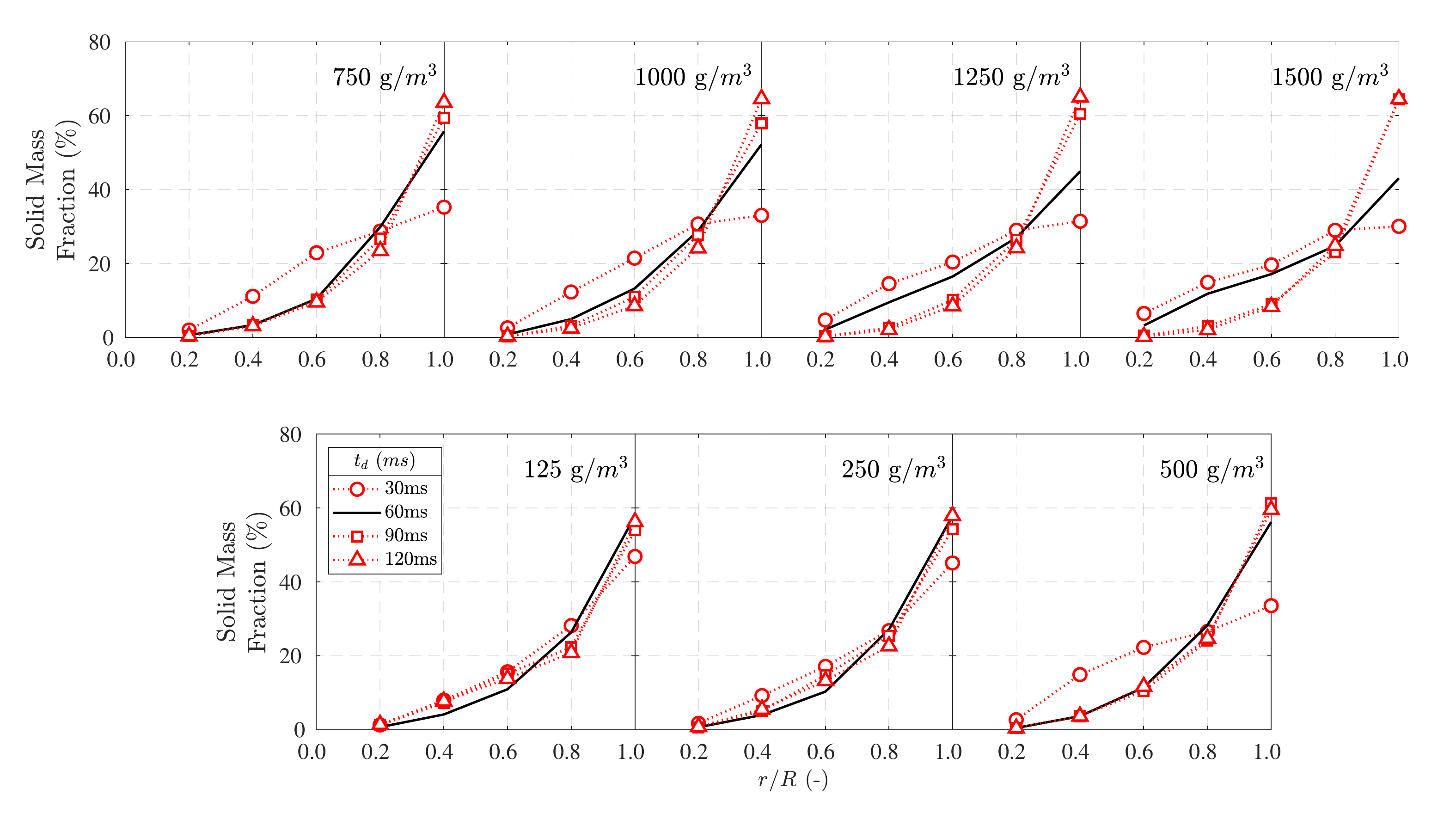}    
    \centering \caption{Distribution of the dust particles versus radial position for the various dust concentrations during the dispersion process of \textit{Cupressus Funebris} (biomass 2).}
    \label{Fig:Fig_CF_solidMassFraction_versus_radius_rows}
\end{figure*}

To understand the effect of varying the ignition delay time on the spatial distribution of the dust cloud at the time of ignition, Fig. \ref{Fig:Fig_CF_solidMassFraction_versus_radius_rows} shows the mass fraction of particulates versus radial position for the various dust concentrations. These calculations were performed in five spherical shells at different radii ratios. In all cases, the solid mass fraction increases as the particles approximate the vicinity of the wall, so it is likely that the turbophoresis effect is always present regardless dust concentration, and depends only on particle size effects. Moreover, note that when $t_{d}=30 \,\text{ms}$ the particle accumulation at the wall is slightly attenuated with respect to the case when $t_{d}=60 \,\text{ms}$. When $t_{d}>60 \,\text{ms}$ it is likely that particle concentration at the near-wall region will increase with ignition delay time. Only for the two most diluted cases, $C_{0}=125$ and $C_{0}=250 \,\text{g}/\text{m}\textsuperscript{3}$ the concentration at the outermost spherical shell remain almost equal. However, because $P_{max}$ is usually registered under a slightly fuel-rich condition, the main inference is that, when performing dust explosion tests at longer ignition delay times, the spatial distribution of the dust cloud will exhibit an increased concentration at the near-wall region by increasing $t_{d}$.\\

\subsubsection{Influence of ignition delay time on the explosion pressure}
\label{Subsection:Influence_of_ignition_delay_time_on_the_explosion_pressure}

\begin{figure}[h]
    \centering
    \includegraphics[width=0.43\textwidth]{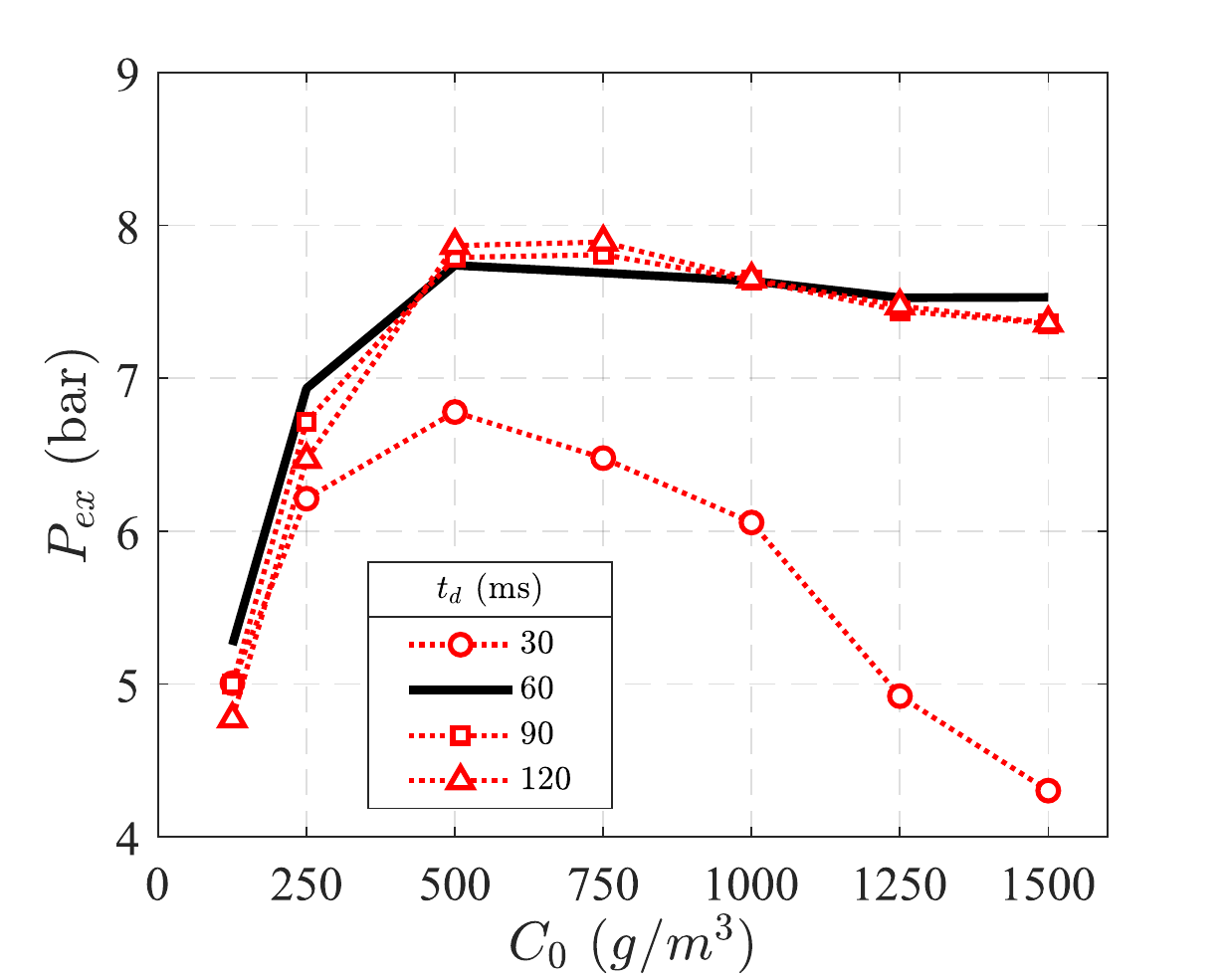}    
    \centering \caption{Comparison of the CFD-predicted explosion pressures $P_{ex}$ of \textit{Cupressus Funebris} (biomass 2) for various ignition delay times $t_{d}$.}
    \label{Figure:CF_Exp_pts_ign_delay_time}
\end{figure}

Once acknowledging that the characteristics of the cold-flow are specific for each condition of dust concentration and ignition delay time, a set of 21 additional reactive simulations were conducted. As usual, the cold-flow solution was mapped from mesh 1 to mesh 2 at the corresponding ignition delay times, and results were compared to the CFD cases of the Section \ref{Subsection:Validation_of_the_peak_pressures_as_function_of_dust_concentration}. Fig. \ref{Figure:CF_Exp_pts_ign_delay_time} shows the explosion pressures $P_{ex}$ for all dust concentrations and ignition delay times. For the dilute concentrations ($C_{0}=125$ and $C_{0}=250\,\text{g}/\text{m}^3$), the explosion pressures obtained at $t_{d}\neq60\,\text{ms}$ are lower than those obtained at $t_{d}=60\,\text{ms}$. This is consistent with other experimental works that studied the effect of varying $t_{d}$ on the explosion behavior of coal particles at a dilute concentration of $C_{0}=250\,\text{g}/\text{m}^3$ \cite{WANG2019509,li2020influence}. Moreover, the present CFD work reports a similar behavior for the dense concentrations ($C_{0}=1250$ and $C_{0}=1500\,\text{g}/\text{m}^3$). Only for the intermediate concentrations ($C_{0}=500-1000\,\text{g}/\text{m}^3$) the explosion pressure slightly increased when $t_{d}\neq60\,\text{ms}$.\\

Note that in all cases, the increasing trend of $P_{ex}$ with respect to $C_{0}$ holds for the first three concentrations, while the decreasing trend is markedly accentuated when $t_{d}=30 \,\text{ms}$. This suggest that regardless the increased velocity-induced mass transfer rates associated with higher TKE levels at shorter $t_{d}$, the dust-air mixtures ignited at $t_{d}=30 \,\text{ms}$ are less reactive than those ignited at the standard (or longer) ignition delay time.\\

To understand why the dust cloud is not burning to completion, this work endorses the idea of interpreting dust explosions on the basis of time scales \cite{van1992interpretation}. The interaction mechanism between turbulence and the combustion zone is examined through a Karlovitz number. A Karlovitz number (Ka) can be defined as the quotient of the chemical reaction time scale $\tau_{c}$ to the mixing time scale $\tau_{m}$. It can be related to the reactive volume fraction in cell $\kappa$ appearing in Eq. (\ref{Equation:PaSR}), as $\text{Ka}=\tfrac{\kappa}{1-\kappa}$.\\

\begin{figure}[h]
    \centering
    \includegraphics[width=0.43\textwidth]{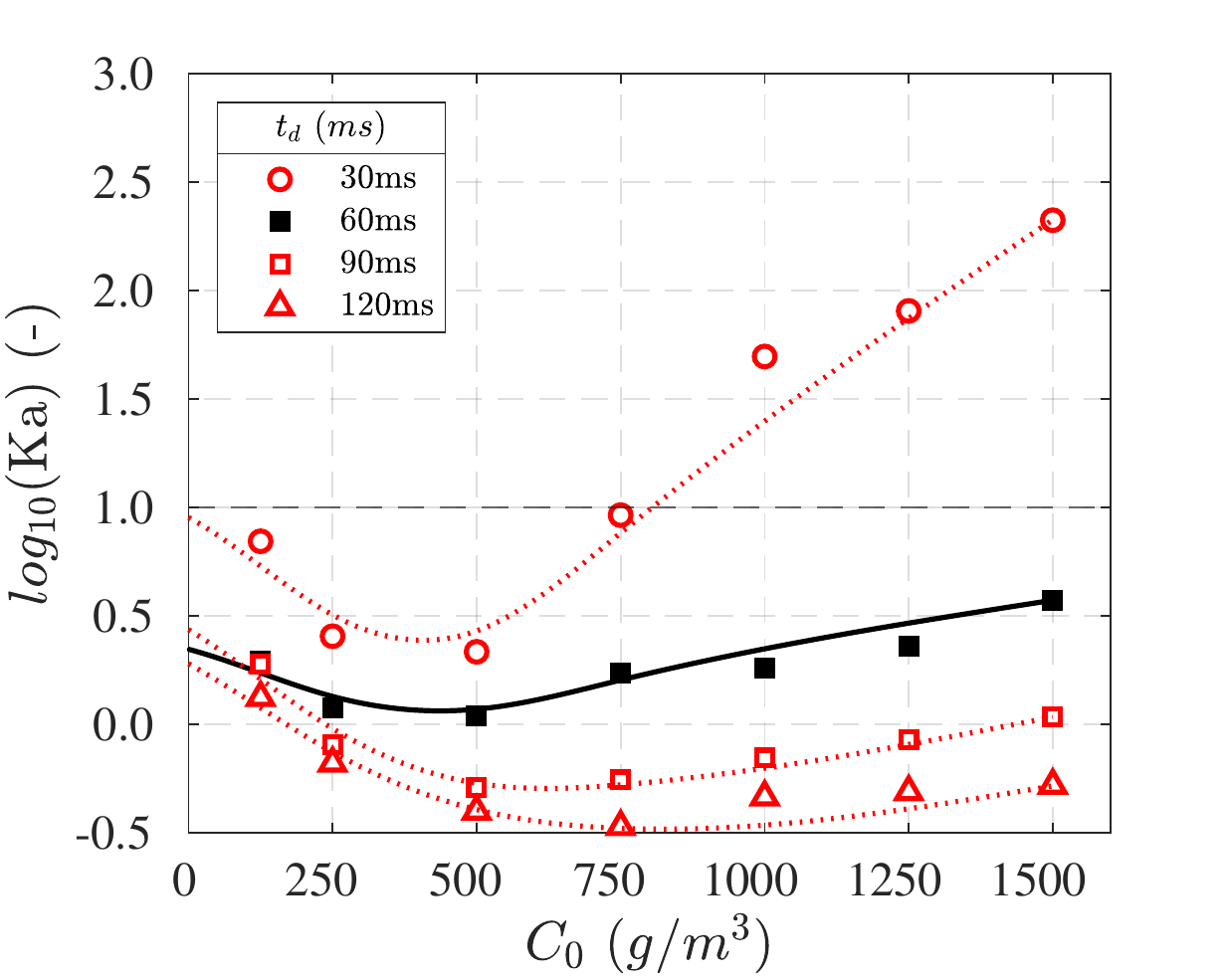}
    \centering \caption{Comparison of the time-averaged values of the Karlovitz number during the first 100 ms of the explosion process of \textit{Cupresuss Funebris} (biomass 2).}
    \label{Figure:CF_Karlovitz_number}
\end{figure}

Fig. \ref{Figure:CF_Karlovitz_number} shows the time-averaged Ka number as function of dust concentration. The burning rate depends on the evolution of the Karlovitz and Reynolds number that embody the competition between mixing and chemistry \cite{swaminathan2022advanced}. Both quantities evolve locally in the flow and depend on the flame propagation pattern and its physical overlap with sources of turbulence generation, e.g. the two-vortex flow pattern, shear layers, etc. Note that a decreasing trend of the Karlovitz number is maintained for the first three concentrations, while the trend is increasing for the successive concentrations. This suggests that for all ignition delay times, the Ka number peaks its minimum for $C_{0}=500\,\text{g}/\text{m}^3$ so that the explosion is governed by fast-chemistry rather than by turbulent diffusion effects.\\

\begin{figure*}[h]
    \centering
    \includegraphics[width=0.95\textwidth]{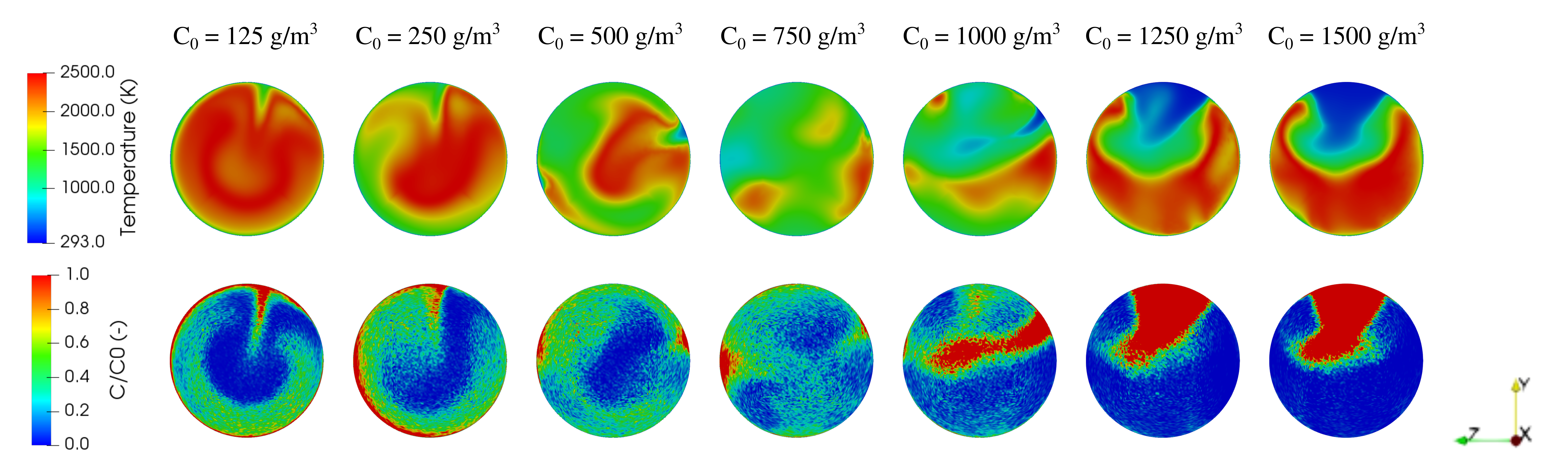}    
    \centering \caption{Contours of flame temperature and normalized dust concentration when employing an ignition delay time $t_{d}=30\,\text{ms}$. Snapshots taken at the time $P_{ex}$ was reached.}
    \label{Fig:Fig_CF_Exp_30ms_T_rhoEffLagrangian_zNormal}
\end{figure*}

\begin{figure*}[h]
    \centering
    \includegraphics[width=0.95\textwidth]{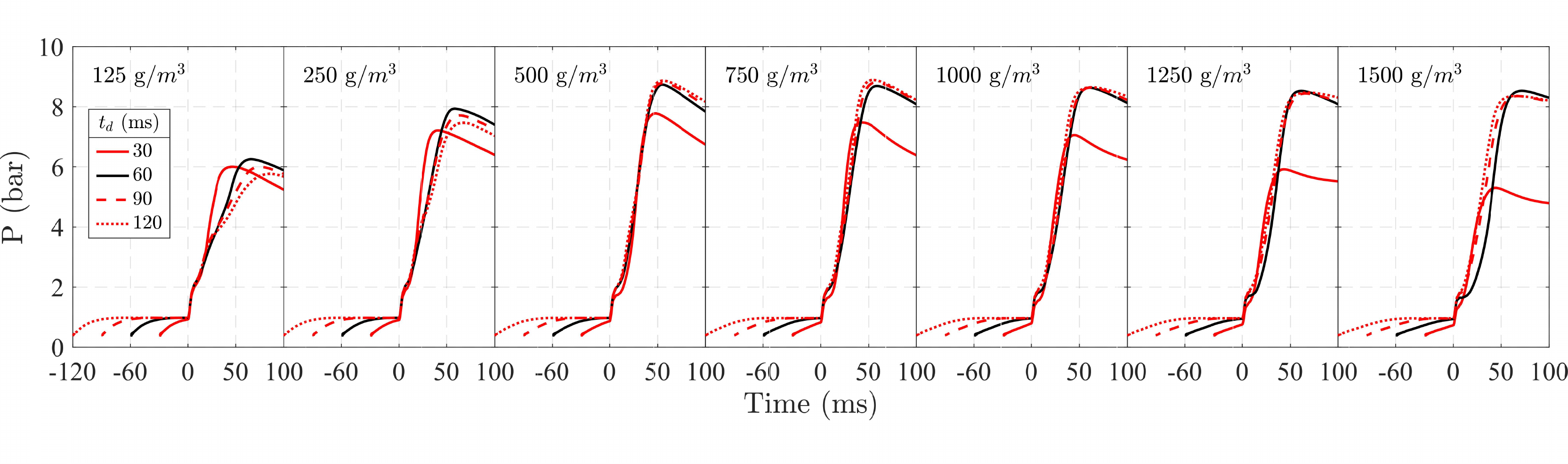}    
    \centering \caption{Comparison of the different pressure-time curves obtained by varying the ignition delay time in the explosion process of \textit{Cupressus Funebris} (biomass 2).}
    \label{Figure:CF_Standard_Explosion_all_ign_delay_time}
\end{figure*}

Fig. \ref{Fig:Fig_CF_Exp_30ms_T_rhoEffLagrangian_zNormal} illustrates the contours of flame temperature and corresponding dust distribution when the explosion pressures were reached in each simulation. These contours reveal a self-evident dependency between the flame propagation patterns and the distribution of the dust cloud. The dust agglomeration at high dust concentrations is responsible for the noticed flame discontinuities and irregular morphologies. Dust agglomeration phenomenon is important in turbulent combustion of solid powders because it strongly affects the local fuel concentration \cite{jenny2012modeling}. This phenomenon has been observed in other CFD studies on dust explosions in the 20L sphere \cite{LI2020118028,LI2020cornstarch}. In the present, the existence of particle clusters may have an impact on the heterogeneous reactions. This is due to the fact that particles that are concentrated in particle clusters will soon consume most of the reactant species (oxygen) within the cluster. In this way, a particle inside a cluster will have access to less reactant species than a particle outside the cluster \cite{haugen2018effect}. Therefore, the conversion of these particles will be slower than for a similar fluid-particle realization that is not clustered. Moreover, if turbulent transport conveys such agglomerates in low temperature regions, their devolatilization and oxidation will be abruptly interrupted. It is important to realize that for particle clustering to have an effect on the conversion rate of the particles, the lifetime of a typical cluster cannot be much shorter than the time it takes for the particles to consume a significant fraction of the surrounding reactants \cite{swaminathan2022advanced}. However, in these simulations the particle clusters observed in the cases of dense dust concentrations persisted during the pressure drop after $P_{ex}$ was reached (contours not included for the sake of brevity). \\

Based on the apparent reduced reactivity of the dust cloud when $t_{d}=30 \,\text{ms}$, the following criterion for the Ka number is proposed:

\begin{itemize}
    \item For $\text{Ka}\lessapprox\mathcal{O}\left(10^{1}\right)$ the chemical reactions are fast compared to the small scale turbulence characteristic time-scale. Therefore, the chemical reaction will dominate over the small scale flow phenomena.
    \item For $\text{Ka}\gtrapprox\mathcal{O}\left(10^{1}\right)$ the chemical reactions are slow compared to the small scale turbulence characteristic time-scale. Strong turbulence-chemistry interaction (TCI) effects are involved, therefore dust agglomeration phenomenon may be responsible for attenuating the explosion pressure due to uneven burning of the dust cloud.
\end{itemize}

Finally, Fig. \ref{Figure:CF_Standard_Explosion_all_ign_delay_time} shows the pressure-time curves for all dust concentrations. In these curves, the pressure trace appearing on the negative range of the x-axis corresponds to the pressure increase due to the air-blast, while the pressure trace on the positive x-axis plots the course of the explosion process. It is recalled that, in all cases the air-dust mixtures were ignited with an energy source of 10 kJ. \\

For dilute concentrations ($C_{0}=125$ and $C_{0}=250\,\text{g}/\text{m}^3$), the pressure increases more sharply when $t_{d}=30\,\text{ms}$ due to the increase in TKE which enhances the heating and mass transfer rates from the particle to the gas-phase. Contrarily, pressure increases more slowly as $t_{d}$ increases. This is congruent with statement that higher pre-ignition turbulence levels increase the rate of pressure rise in the explosion process. However, this trend is less notable for the subsequent concentrations. This can be attributed to competition of the various $C_{0}$-dependent factors mentioned above: (1) the pre-ignition pressure, (2) the turbulent kinetic energy of the cold-flow, (3) the turbophoresis effect, and (4) the dust agglomeration phenomenon. For concentrations exceeding $C_{0}\geq500\,\text{g}/\text{m}^3$ the pressure-time curve almost overlaps for all cases, except when igniting the air-dust mixtures at $t_{d}=30\,\text{ms}$, in which dust agglomeration seems to be responsible of pressure attenuation, as previously discussed.

\section{Conclusions}

In this work, the dust explosion behavior of biomass in the 20L sphere was assessed via numerical simulations conducted with the open-source CFD code OpenFOAM 8. A total of 29 simulations (dispersion and explosion) were performed, in which an in-depth analysis of the elements affecting the dynamics of the cold-flow and the reactivity of the dust explosion is presented for each biomass sample. The CFD results were validated with experimental data of two biomass samples, namely from: (1) the pressure-time curve of \textit{Pellets Asturias} (biomass 1) and (2) the explosion pressures versus dust concentration graph of \textit{Cupressus Funebris} (biomass 2). Results showed good agreement with experimental data, reporting relative errors of 1.85\% and 5.81\% for $P_{ex}$ of biomass 1 and $P_{max}$ of biomass 2, respectively. Furthermore, when comparing the pressure-time curve of biomass 1, although $(\text{dP}/\text{dt})_{ex}$ is moderately overestimated by the CFD model, the relative error of 19.2\% is still compliant with the ASTM E1226 standard. Then, the model was used to appraise the influence of the ignition delay time on the explosion pressure of biomass 2.\\

On one hand, the results suggest that the burning rate is dominated by the combustion of volatile gases and that particles exceeding $d_{p}>500\,\mu\text{m}$ may not enter the 20L sphere during the dispersion process. This value agrees with the EN 14034 standard and experimental researchers are advised to be cautious when running tests under such particle size conditions as the nominal dust concentration $C_{0}$ may not be fully discharged into the chamber. Moreover, those large particles that managed to enter the 20L sphere and that exceed $d_{p}>750\mu\text{m}$ are likely to react partially due to the increased heating times necessary for devolatilization. On the other hand, both the explosion pressure and maximum flame temperature increase with $C_{0}$ up to reaching $P_{max}$ in the fuel-rich region ($1<\phi<1.8$). Similarly, other variables like (1) the TKE of the cold flow, (2) the rates of oxygen depletion and (3) the rates of biomass conversion (evaporation, devolatilization and char oxidation) depend on $C_{0}$. When running tests at modified ignition delay times, there are important implications that experimental researchers should acknowledge before conducting such tests. For instance, if $t_{d}<60\,\text{ms}$, neither the pre-ignition pressure nor the dust concentration may reach 1 bar or $C_{0}$. In such conditions, although the TKE of the flow is higher and presumably, the rate of pressure rise should increase as well, the simulations evidence that this does not apply for all cases as only the dilute concentrations $C_{0}\leq 250\,\text{g}/\text{m}^3$ follow this pattern. The strong turbulence chemistry interactions promote dust agglomeration and as the dust concentration increases the explosion pressure is significantly reduced. Contrarily, if $t_{d}>60\,\text{ms}$, although the turbophoresis effect is responsible of promoting a non-uniform mixing (increased dust concentration at the walls), the resulting explosion pressures are comparable to those obtained under the standard ignition delay time $t_{d}=60\,\text{ms}$.\\

In short, these CFD results are aimed to: (1) help practitioners who conduct dust explosion testing to broaden the interpretation of test results in the 20L sphere experiment, and (2) to emphasize that the course of a dust explosion is strongly coupled to the dispersion process of a dust cloud and its dynamic behavior. Thereafter, when employing CFD methods to estimate the explosion parameters in the process of conducting dust explosion risk assessments, a reactive simulation should always be pre-assessed with the corresponding dust dispersion simulation, despite if the geometry is a standardized vessel or a large industrial enclosure.

\section*{Declaration of Competing Interest}
The authors declare that they have no known competing financial interests or personal relationships that could have appeared to influence the work reported in this paper. \\

\section*{CRediT Authorship Contribution Statement 
}
\textbf{A. Islas:} Conceptualization, Formal analysis, Data curation, Methodology, Software, Validation, Investigation,  Resources, Writing - original draft, Writing - review \& editing, Visualization. \textbf{A. Rodríguez Fernández:}  Methodology, Software, Validation, Investigation,  Resources, \linebreak Writing - original draft, Writing - review \& editing. \linebreak \textbf{E. Martínez-Pañeda:} Conceptualization, Writing - review \& editing, Funding acquisition. \textbf{C. Betegón:} Writing - review \& editing, Supervision, Project administration, Funding acquisition. \textbf{A. Pandal:} Conceptualization, Methodology, Software, Investigation, Resources, Writing - review \& editing, Supervision, Funding acquisition. \\

\section*{Acknowledgements}

Authors acknowledge that this work was partially funded by CDTI (Centro para el Desarrollo Tecnológico Industrial de España, IDI-20191151), Universidad de Oviedo and \linebreak PHB WESERHÜTTE, S.A., under the project "FUO-047-20: Desarrollo de silo metálico de grandes dimensiones ante los condicionantes de explosividad de la biomasa". A. Islas acknowledges support from the research grant \#BP20-124 under the 2020 Severo Ochoa Pre (Doctoral) Program of the Principality of Asturias.\\


\bibliographystyle{unsrt_abbrv_custom}

\bibliography{cas-refs}




\section*{Appendix A: WSGGM validation}

The weighted sum of gray gas model (WSGGM) was first developed by Hottel and Sarofim \cite{hottel1967radiative}. 
It replaces the spectrum with few gray gases and transparent windows according to

\begin{equation}
    \varepsilon = \sum_{i=0}^{N_{g}}a_{\varepsilon,i}\left(T\right)\left[1-\exp{(-\kappa_{i}p_{a}L)}\right]
    \label{Eqn:WSGGM_emissivity}
\end{equation}

where $N_{g}$ is the number of gray gases and $a_{i}$ are the emissivity weighting factors.
The bracketed quantity in Eq. (\ref{Eqn:WSGGM_emissivity}) is the $i$-th gray gas emissivity with
banded absorption coefficient $\kappa_{i}$ and pressure-path length $p_{a}L$. The pressure $p_{a}$ is expressed by summing the partial pressures of the participating gases, namely H\textsubscript{2}O and CO\textsubscript{2}

\begin{equation}
    p_{a}=(X_{\text{CO}_{2}}+X_{\text{H}_{2}\text{O}})p
    \label{Eqn:WSGGM_pressure}
\end{equation}

where $X_{i}$ denotes the molar fraction of each species and $p$ is the total pressure in atm. To represent the transparent parts of the spectrum, the banded absorption coefficient $\kappa_{i=0}=0$. Since total emissivity approaches unity in the limit of the pressure-path length, the emissivity weighting factors must sum unity, and all adopt positive values. This implies that $a_{\varepsilon,0}=1-\sum_{1}^{N_{g}}a_{\varepsilon,i}$, such that only $N_{g}$ weighting factors need to be determined.\\

Commonly, the emissivity weighting factors are assumed to be a temperature dependent polynomial function of order $(N_{g}-1)$ \cite{smith1982evaluation,modest2021radiative}, i.e.

\begin{equation}
    a_{\varepsilon,i}\left(T\right)=\sum_{j=1}^{N_{g}}b_{\varepsilon,i,j}T^{j-1}
    \label{Eqn:WSGGM_weighting_factor}
\end{equation}

where $b_{\varepsilon,i,j}$ are the polynomial coefficients. However, this expression does not allow to consider variations in the composition of the gas mixture, so that coefficients must be determined for specific molar ratios, $\text{MR}=X_{\text{H}_{2}\text{O}}/X_{\text{CO}_{2}}$. Because in explosion testing of biomass or carbonaceous dust in the 20L experiment, composition of the combustion products may be not uniform in the chamber (e.g., due to uneven burning of the dust cloud owed to non-uniform particle mixing) a more versatile model is advisable.\\

Alternatively, Kangwanpongpan et al. \cite{kangwanpongpan2012new} derived a new set of correlations for WSGGM from fitting total emittances
generated by line-by-line (LBL) calculations from the HITEMP 2010 database \cite{rothman2010hitemp}. In their work, the emissivity weighting factors express each of the polynomial coefficients $b_{\varepsilon,i,j}$ in Eq. (\ref{Eqn:WSGGM_weighting_factor}) as an independent polynomial function of the molar ratio, leading to

\begin{equation}
    a_{\varepsilon,i}\left(T\right)=\sum_{j=1}^{N_{g}}\left(\sum_{k=0}^{2}c_{\varepsilon,i,j,k}\text{MR}^{k}\right)\left(\frac{T}{T_{ref}}\right)^{j-1}
    \label{Eqn:WSGGM_molar_ratio_weighting_factor}
\end{equation}

Moreover, to keep the same level of precision among the polynomial coefficients, the temperature-dependent relation is normalized by a reference temperature $T_{ref}$. In the same way, the banded absorption coefficients are expressed as another polynomial function of the molar ratio (i.e. $\kappa_{i}=\sum_{k=0}^{2}d_{\varepsilon,i,j,k}\text{MR}^{k}$). These new correlations are valid for a continuous range of dry ($0.125<\text{MR}<1.0$) and wet conditions ($1.0<\text{MR}<4.0$).\\

In this work, two versions of the WSGGM were implemented into OpenFOAM 8, namely: (1) WSGGM-SMITH82 based on the model coefficients of Smith et al. \cite{smith1982evaluation} and (2) WSGGM-KANGWANPONGPAN2012 based on the correlations of Kangwanpongpan et al. \cite{kangwanpongpan2012new}. The numerical calculations were performed with the {\fontfamily{qcr}\selectfont fvDOM} and an angular discretization of $N_{\phi}=3$, $N_{\theta}=3$. Results are compared with benchmark data from statistical narrow band (SNB) and LBL models of the literature.

\subsection{Benchmark case 1}

\begin{figure}[h]
    \centering
    \includegraphics[width=0.45\textwidth]{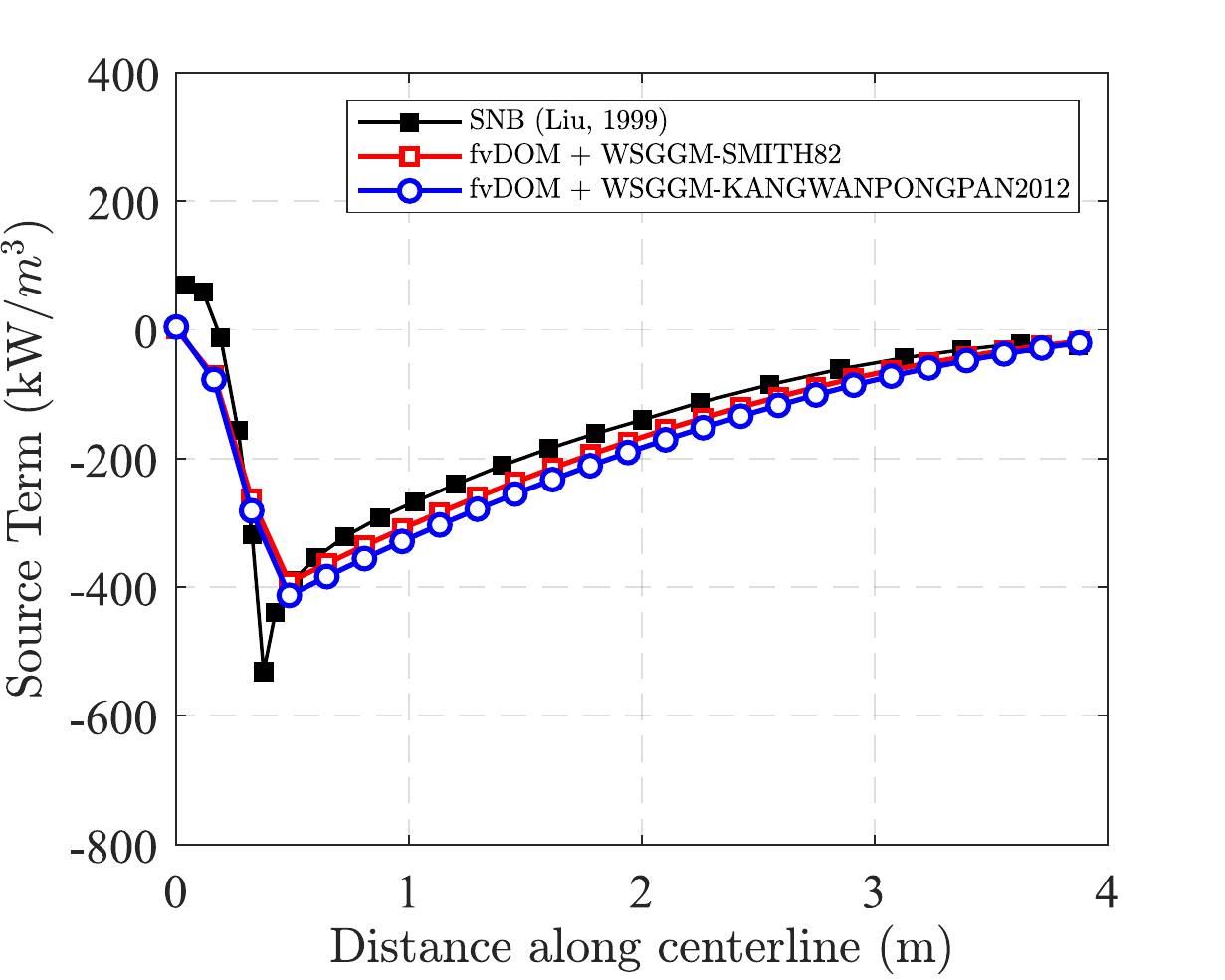}
    \caption{(Benchmark case 1) Comparison of radiative source term along the centerline between benchmark data \cite{liu1999numerical} and WSGGM implementation in OpenFOAM using {\fontfamily{qcr}\selectfont fvDOM} with Smith \cite{smith1982evaluation} and Kangwanpongpan et al. \cite{kangwanpongpan2012new} model coefficients.}
    \label{Fig:Fig_WSGGM_Box}
\end{figure}

The geometry for test 1 is a 3D rectangular enclosure of dimensions 2m x 2m x 4m with the walls being black ($\varepsilon$ = 1.0) at 300 K. The gas temperature is non-uniform but symmetrical about the centerline of the enclosure and specified according to Liu \cite{liu1999numerical}. This profile simulates roughly the temperature distribution of a flame. The medium is assumed to be $0.1\text{CO}_{2}+0.2\text{H}_{2}\text{O}+0.7\text{N}_{2}$ (mole basis), which results in a constant molar ratio $\text{MR}=2.0$.\\

Fig. \ref{Fig:Fig_WSGGM_Box} shows a comparison of the calculated radiative source term along the centerline for the two WSGGM implementations and the benchmark data by Liu \cite{liu1999numerical}. The radiative source term is the link between the radiative transfer equation (RTE) and the energy equation, so prediction of this term is necessary to correctly calculate flame temperature in combustion applications.\\

As shown, both models are in good agreement with the SNB data, with the WSGGM-SMITH82 implementation slightly more accurate in the downstream region of the flame tip ($>0.375\,\text{m}$). This is because the coefficients of Smith et al. \cite{smith1982evaluation} are determined exactly for a $\text{MR}=2.0$ condition, while the correlations of Kangwanpongpan et al. \cite{kangwanpongpan2012new} introduce subtle rounding errors during the interpolation. However, this difference is almost negligible and it is recalled that the same set of Kangwanpongpan's coefficients are valid for a significantly wider range of molar ratios. Overall, the largest errors for both implementations take place in the upstream side of the flame tip where the temperature increases abruptly from 400 K ($0\,\text{m}$) to 1800 K ($0.375\,\text{m}$). This can be improved by either increasing the grid resolution along the centerline or by increasing the angular discretization of the {\fontfamily{qcr}\selectfont fvDOM}.

\subsection{Benchmark case 2}

\begin{figure}[h]
    \centering
    \includegraphics[width=0.45\textwidth]{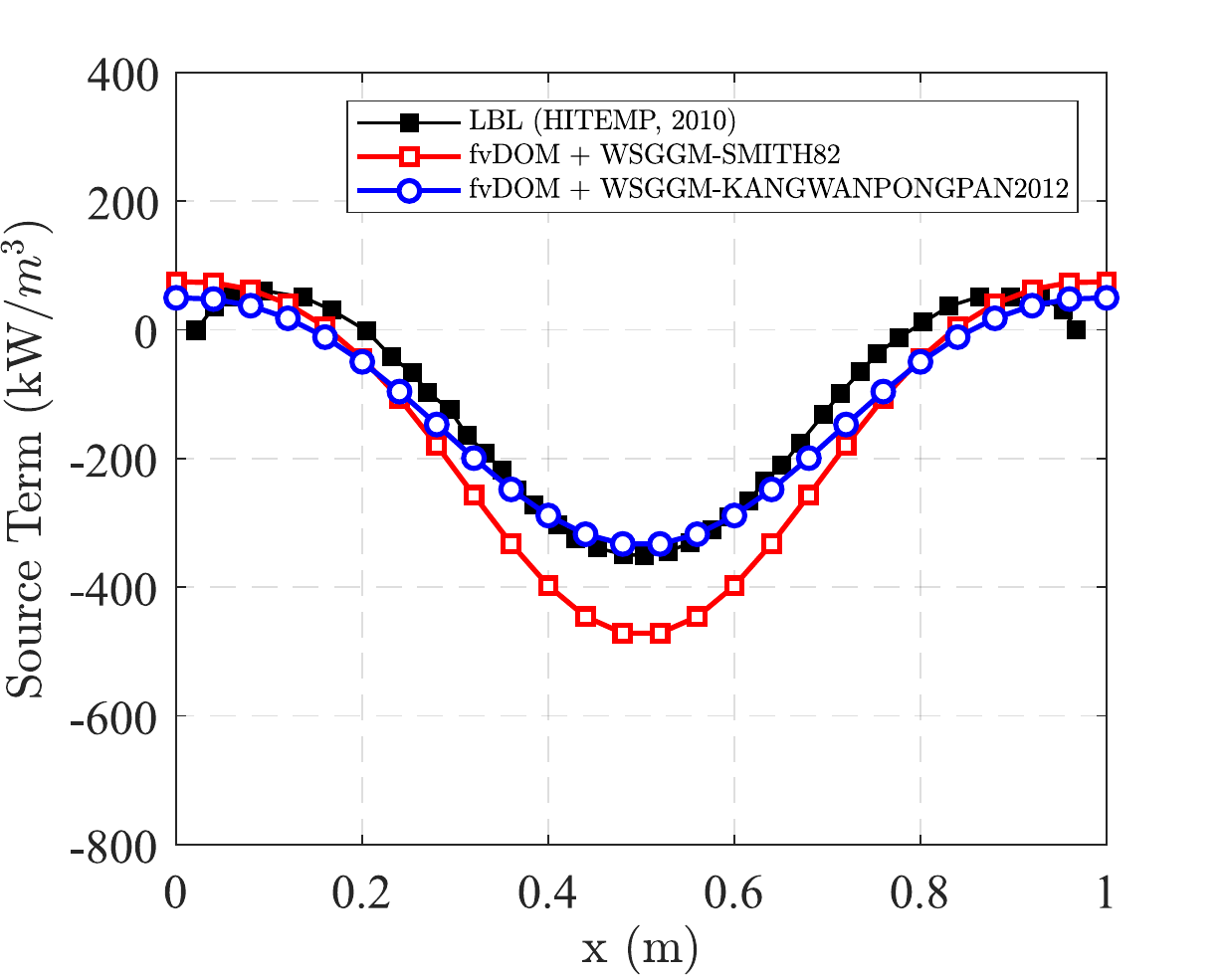}
    \caption{(Benchmark case 2) Comparison of radiative source term along the length coordinate between benchmark data \cite{rothman2010hitemp} and WSGGM implementation in OpenFOAM using {\fontfamily{qcr}\selectfont fvDOM} with Smith \cite{smith1982evaluation} and Kangwanpongpan et al. \cite{kangwanpongpan2012new} model coefficients.}
    \label{Fig:Fig_WSGGM_Slab}
\end{figure}

The geometry for test 2 is a 1D infinite slab separated by a distance $L=1\,\text{m}$. The walls are black ($\varepsilon=1.0$) and the temperature between the plates follows a cosine profile $T=1400\,\text{K}-(400\,\text{K})\cos{\left(\frac{2\pi x}{L}\right)}$. The gas is comprised of a mixture of CO\textsubscript{2}, H\textsubscript{2}O, and N\textsubscript{2}. The molar fraction of carbon dioxide is fixed at $X_{\text{CO}_{2}}=0.8$, and the molar fraction of water follows the profile $X_{\text{H}_{2}\text{O}}=0.12+0.04\cos{\left(\frac{2\pi x}{L}\right)}$, which results in a variation of $0.1<\text{MR}<0.2$.\\

Similarly, Fig. \ref{Fig:Fig_WSGGM_Slab} shows a comparison of the radiative source term for both implementations. This time, the correlations of Kangwanpongpan et al. \cite{kangwanpongpan2012new} exhibit an increased agreement with the LBL benchmark data, particularly at the interval $0.3\leq x\leq0.7\,\text{m}$. Contrarily, model coefficients of Smith et al. \cite{smith1982evaluation} show a gross underestimation around the minimum radiative source term, $\varepsilon_{rel}\sim34\%$. Therefore, the WSGGM-KANGWANPONGPAN2012 implementation was used for all the simulations in this paper.

\end{document}